\documentclass[useAMS,usenatbib]{mn2e}

\usepackage{graphicx}
\usepackage{amstext}
\usepackage{amsmath}
\usepackage{amssymb}
\usepackage{color}
\usepackage[normalem]{ulem}
\usepackage{multirow}
\usepackage{longtable,lscape}
\usepackage{bm}
\usepackage{rotating}

\begin{document}

\title[The structure and fate of WD merger remnants]{The structure and fate of white dwarf merger remnants}

\author[M. Dan, S. Rosswog, M. Br\" uggen, P. Podsiadlowski]{Marius
  Dan$^{1,2}$\thanks{E-mail: marius.dan@hs.uni-hamburg.de}, Stephan Rosswog$^{2,3}$,
  Marcus  Br\" uggen$^{1}$, Philipp Podsiadlowski$^{4}$\\
$^{1}$Hamburger Sternwarte, Universit\"at Hamburg,
  Gojenbergsweg 112, 21029 Hamburg, Germany\\
$^{2}$School of Engineering and Science, Jacobs University
Bremen, Campus Ring 1, 28759 Bremen, Germany\\
$^{3}$The Oskar Klein Centre, Department of Astronomy, 
AlbaNova, Stockholm University, SE-106 91 Stockholm, Sweden\\
$^4$Department of Astronomy, Oxford University, Oxford OX1 3RH, UK}

\date{Accepted ?. Received ?; in original form ?}

\pagerange{\pageref{firstpage}--\pageref{lastpage}} \pubyear{2013}

\maketitle

\label{firstpage}

\def\msun{$M_{\odot}$}
\def\Msun{$M_{\odot}$ }

\begin{abstract}
We present a large parameter study where we investigate the
structure of white dwarf (WD) merger remnants 
after the dynamical phase. A wide range of WD
masses and compositions are explored and we also probe the effect of different initial
conditions. We investigated the degree of mixing between the WDs, the
conditions for detonations as well as the amount of gas ejected. 
We find that systems with lower mass ratios have more total angular
momentum and as a result more mass is flung out in a tidal
tail. Nuclear burning can affect the amount of mass ejected.  
Many WD binaries that contain a helium-rich WD achieve the conditions
to trigger a detonation. In contrast, for carbon-oxygen transferring
systems only the most massive mergers with a total mass $M \ga 2.1 M_\odot$
detonate. Even systems with lower mass may detonate long after the
merger if the remnant remains above the Chandrasekhar mass and carbon
is ignited at the centre. 
Finally, our findings are discussed in the context of several possible
observed astrophysical events and stellar systems, such as hot
subdwarfs, R Coronae Borealis stars, single massive white dwarfs,
supernovae of type Ia and other transient events. 
A large database containing 225 white dwarf merger remnants is made 
available via a  dedicated web page.
\end{abstract}

\begin{keywords}
white dwarfs -- accretion, accretion disks -- nuclear reactions, 
nucleosynthesis, abundances -- hydrodynamics
\end{keywords}

\section{Introduction}

White dwarfs are the most common outcomes of stellar evolution. It is estimated that 
there are about $10^{10}$ WDs in our Galaxy \citep{napiwotzki09} out of which $2.5\times 10^8$ 
reside in binaries consisting
of two white dwarfs \citep{nelemans01a}. About half of these are close enough (typical orbital 
periods $\lesssim 8$ hr) that gravitational wave emission can drive them to a phase of mass 
transfer within a Hubble time. In recent years, surveys such as the ESO SN Ia Progenitor 
Survey (SPY) and Sloan Digital Sky Survey (SDSS) have drastically increased the number of 
known detached and interacting WD binaries \citep[e.g.][]{napiwotzki04,nelemans05,kilic11}. 
Among the about 50 known WD binaries, there are four eclipsing WD systems that allow for 
a very precise measurement of the masses and radii of both components 
\citep{steinfadt10,parsons11,brown11b,vennes11}. The gravitational waves (GWs) produced 
by some of these systems will be within the detection limit and used as verification 
sources for the future space-based gravitational wave facilities, such
as eLISA/NGO \citep{nelemans09,marsh11,amaro12}. 

Mass transfer critically affects the further evolution of the binary. Depending on whether 
it is stable or not, the WD binary may either survive the initial
mass-transfer phase to become a semi-detached 
system or else end up merging completely. The stability of mass transfer depends sensitively on (i) the 
response of the two WDs to the mass transfer, (ii) on the stars' masses and (iii) on various 
angular momentum transport mechanisms \citep[e.g.][]{marsh04,gokhale07}. 

\begin{figure*}
\centerline{
 \includegraphics[height=5.5in]{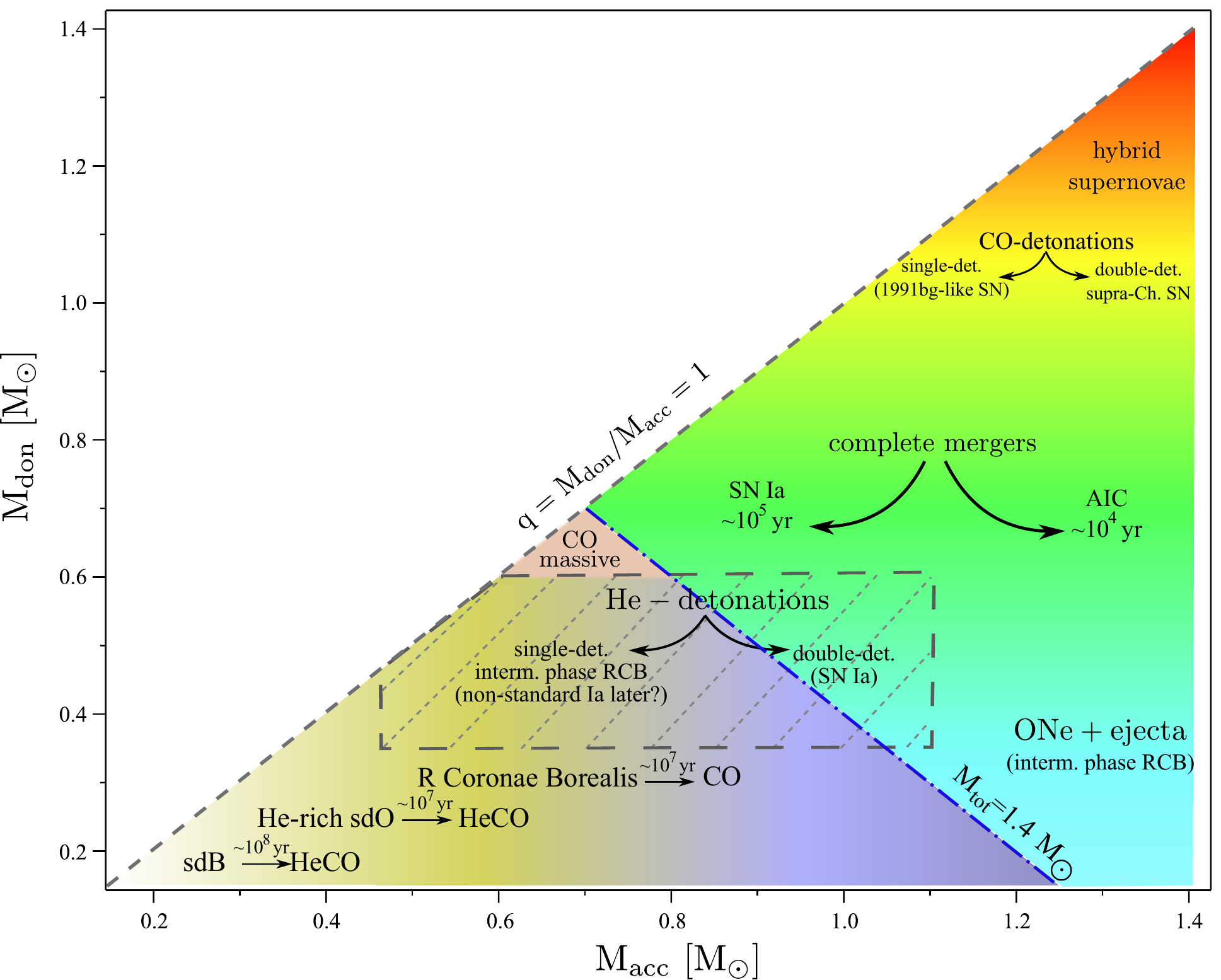}
}
\caption{Binary evolution predicts that a good fraction of the total number of double white 
dwarf systems will come into contact within a Hubble time and, via merging, they could lead 
to a variety of outcomes, depending mainly on the masses of the donor and the accretor, 
$M_{\rm don}$ and $M_{\rm acc}$, respectively. The potential WD merger products are 
helium-rich subdwarfs (sdB and sdO stars), R Coronae Borealis stars, high-mass CO/ONe white 
dwarfs, neutron stars, Type Ia supernovae and low-mass black holes, see text for additional 
details. 
The hashed region roughly marks the region where we found helium-detonations during 
the mass transfer and at the merger moment \citep[see][]{dan11,dan12}. As we will show in 
Section \ref{sec:results}, minutes after the merger a good fraction of these systems can 
still undergo dynamical burning.}
\label{fig:DDO}
\end{figure*}

Once the WDs have merged, the subsequent evolution may lead to a flurry of different outcomes
including He-rich hot subdwarfs of spectral type B \citep[sdB;][]{saio00,han02,heber09} and O (sdO), 
R Coronae Borealis (RCB) stars \citep{webbink84,iben96,justham11,jeffery11,clayton12}, 
massive carbon/oxygen (CO) or oxygen/neon (ONe) WDs \citep{bergeron91,segretain97}, 
explosions as supernova Type Ia \citep[SN Ia;][]{webbink84,iben84}, an accretion-induced collapse (AIC) 
to a neutron star (NS) \citep{saio85,nomoto91,saio04} or perhaps even  to a 
black hole (BH).
For a chart that summarises the possible outcomes of WD mergers see Figure \ref{fig:DDO}. 

Type Ia SN find ample use as cosmological distance indicators, but,
despite the existence of well-established correlations, the identity
of the companion that donates the mass and the exact nature of the
explosion mechanism have remained unclear.  The idea that WD mergers
near the Chandrasekhar mass could cause SNe Ia has been formulated
decades ago \citep{iben84,webbink84} and particularly massive cases
would be natural candidates to explain particularly bright events
\citep[for example SNLS-03D3bb is more than twice as bright as a
normal SN Ia; see][]{howell06}. More recently, it has been realized
that also WD-WD systems with a total mass below the Chandrasekhar mass
could be responsible for (at least a fraction) of SN Ia
\citep{rosswog09c,sim10,vankerkwijk10,woosley11}.

A He, or hybrid He-CO, plus a CO WD could possibly lead to a thermonuclear runaway and hence 
a nova or a SN explosion prior or at the merger time \citep{guillochon10,dan12}. 
\cite{guillochon10} found that Kelvin-Helmholtz instabilities are able to trigger He explosions 
at the surface of the accretor if mass transfer is highly unstable (i.e. large $\dot M$) and 
occurs in the direct impact regime. A second detonation could be triggered by shock compression 
in the CO core leading to spectra and light-curves similar to ``normal'' SNe Ia \citep[e.g.][]{sim10}. 
If the core fails to detonate, these events may resemble sub-luminous SNe 
Ia/Ib \citep{shen10,waldman11}. 
Alternatively, if the explosion during 
mass transfer is avoided, they could undergo hydrodynamical burning when the donor is tidally 
disrupted and plunges onto the more massive WD. 

In order to understand the wealth of possible outcomes of WD--WD mergers, an exhaustive parameter 
study is required. Here we study these properties minutes after the merger of 225 WD-WD systems 
covering a large range of WD masses and chemical compositions.  Our study is complementary to the 
work of \cite{guerrero04,aguilar09,raskin11,zhu13}. These studies investigated a smaller parameter 
space, used different initial conditions \citep[with the exception of][]{raskin11} and \cite{zhu13} 
did not included the feedback from nuclear burning.

We investigate whether minutes after the WD merger the conditions for hydrodynamical burning (i.e. 
characteristic time for heating by nuclear reactions becomes faster than the dynamic response time) 
that could trigger a supernova or supernova-like explosion are
reached. This is complementary to 
the study of \cite{guillochon10,dan11,dan12} where we investigated whether hydrodynamical burning occurs during 
mass transfer or, later, during the final coalescence.

We provide a large database containing the thermodynamic and rotational profiles via the web 
page \verb+www.hs.uni-hamburg.de/DE/Ins/Per/Dan/wdwd_remnants+, so
that they can be used for subsequent stellar evolution studies. 

This paper is organized as follows: in Section \ref{sec:methods} we briefly describe our
numerical methods and in Section \ref{sec:results} we present our
results in detail. Specifically,  
\S \ref{sec:profiles} is devoted to the structure of the merger
remnants, \S~\ref{sec:spin_dep} to a comparison between the results
obtained with corotating and non-rotating initial conditions,
\S~\ref{sec:mixing} to an analysis of mixing between the binary
components, \S~\ref{sec:burning} to whether conditions for
hydrodynamical burning occur, \S~\ref{sec:ejecta} to  
a discussion of how the unbound mass and angular momentum depend on
the stars' masses, \S~\ref{sec:convergence}  
to a resolution dependence study and \S~\ref{sec:comparison} to a
detailed comparison to previous work. In Section
\ref{sec:applications} we discuss the applications of our simulations
to astrophysical systems. In Section \ref{sec:summary} we summarize
our results.

\section{Numerical methods and initial conditions}
\label{sec:methods}

\begin{figure*}
  \centerline{
    \includegraphics[height=1.9in]{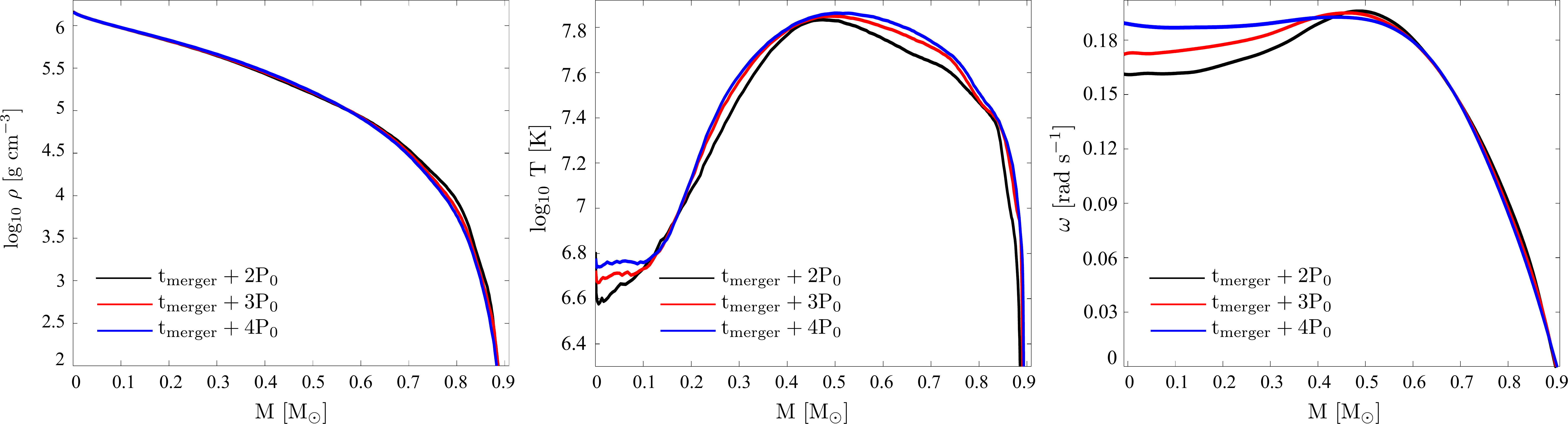}
  }
  \caption{Remnant evolution for the $0.45+0.45\,M_\odot$ system between
    2 (red line), 3 (blue line) and 4 (black line) times the initial period since the merger moment, $t_{\rm merger}$.
   Shown are density (left), temperature
    (center) and angular velocity (right) profiles, averaged over equipotential surfaces, as a
    function of enclosed mass from the center of the accretor.
    The evolution of the remnant at this stage is driven by the 
  artificial viscosity. The core is spun up and evolves towards
  rigid rotation, the temperature increases overall and the density is
  increasing in the core and decreasing in the outer regions. Overall the properties
  do not change much and as the remnant evolves there is even less variation in its properties.
 }  
  \label{fig:234P0wd045wd045}
\end{figure*}

Our numerical code and  initial conditions have been described in detail in the literature
\citep{rosswog04,rosswog08,dan11,dan12}, therefore we only briefly summarize the 
main ingredients of the adopted approach. We use a 3D  smoothed particle hydrodynamics
\citep[SPH; for recent reviews see][]{monaghan05,rosswog09} code that treats shocks by
means of an artificial viscosity scheme with time dependent parameters  \citep{morris97,rosswog00} 
so that viscosity is essentially absent unless it is triggered. Moreover, a switch is
applied to suppress the viscous forces in pure shear flows \citep{Balsara95}. 
We use the binary tree  of \cite{benz90} to search the SPH neighbors and to calculate the 
gravitational forces. The code uses the Helmholtz equation of state \citep{timmes00}
which is coupled to the minimal nuclear reaction network of \cite{hix98} to include the 
energetic feedback onto the gas and to obtain (approximate) information on the 
nuclear composition. 

To construct the initial stars the SPH particles are placed on a stretched hexagonal lattice 
 \citep{rosswog09b} and provided with the properties derived from solving the Lane-Emden 
equations. The particle distribution is subsequently relaxed in isolation at $T=10^5$ K and 
a velocity-dependent force is applied to drive them to equilibrium. As before \citep{dan12},
we use exemplary chemical compositions, in nature also alternative compositions may be 
realized that depend on the exact evolutionary history of the stars. Below 0.45 $M_\odot$ our 
WDs are made of pure helium, between 0.45 and 0.6 $M_\odot$ 
they have a $0.1\,M_\odot$ helium mantle and a pure carbon-oxygen core (mass 
fractions X(${}^{12}{\rm C}$) = 0.5 and X(${}^{16}{\rm O}$) = 0.5); between 0.6 and 1.05 $M_\odot$ 
WDs are made entirely of carbon-oxygen with mass fractions X(${}^{12}{\rm C}$) = 0.4 
and X(${}^{16}{\rm O}$) = 0.6 uniformly distributed through the star and above 
1.05 \Msun they are made of oxygen, neon and magnesium with 
$X({}^{16}{\rm O})= 0.60$, $X({}^{20}{\rm Ne})= 0.35$ and $X({}^{24}{\rm Mg})= 0.05$, respectively.

It is not well known what the WD spins are at the moment when numerically resolved mass 
transfer sets in (our initial mass transfer rates are well above the Eddington limit).
The work of \cite{fuller12,burkart12} suggests that a coupling between 
tides and stellar pulsations may drive the binary to a corotating state prior to mass transfer initiation.
Since it is physically plausible and computationally convenient, we start our simulations from 
corotating initial conditions. The technical procedure how to construct an equilibrium fluid
configuration has been explained in detail before \citep{rosswog04,dan11} and we refer the 
interested reader to this work and the references cited therein.  
In \cite{dan11} we had demonstrated the sensitivity of the orbital and mass transfer evolution
on the initial conditions. In particular we found that approximate initial conditions can underestimate
the duration of the phase of (numerically resolvable) mass transfer by orders of magnitude and 
that they, containing the  wrong amount of angular momentum, also impact on the resulting
remnant structure. Generally, approximate initial conditions lead to a too fast inspiral and to too 
large densities and temperatures. How serious this is for the further evolution depends on the
system under consideration.

We scan the parameter space with as many as 225 simulations which are summarized
in Table \ref{tab:bigtable1} in the appendix. The masses of the donor stars range from 0.2 to 
1.05 $M_\odot$, those of the accretors from 0.2 to 1.2 $M_\odot$, both ranges are probed
in steps of 0.05 $M_\odot$. In this broad parameter scan, each binary system is modeled with 
only 40 000 SPH particles. While this does certainly not yield fully converged results, our 
analysis in  \S \ref{sec:convergence} suggests that the qualitative results and trends are robust.
This study is intended to identify interesting systems, they should be studied in more detail at higher 
resolution in the future.

\section{Results}
\label{sec:results}

First results from the WD systems that we study have been presented  in \cite{dan12}. Here
we give a more complete overview over the parameter space. We determine the remnant
properties at a time of three initial orbital periods after the moment when 
the donor is fully disrupted and its center of mass cannot be identified anymore. At this point 
we stop the calculations since the dynamical evolution is essentially over and the changes occur 
at a much more moderate pace. We illustrate this in Figure \ref{fig:234P0wd045wd045} for a  
system of  two 0.45 $M_\odot$ WDs. Density, temperature and 
angular velocity are averaged over equipotential surfaces and are shown as a function of the 
enclosed mass at two, three and 
four initial orbital periods after the merger moment. The density profile hardly changes during this
time span, but numerical dissipation damps out the differential rotation in the core which goes 
along with a secular increase in the temperature. We repeated this analysis for systems of 
$0.2+0.8\,M_\odot$, $0.45+1.0\,M_\odot$ and $0.8+0.9\,M_\odot$ and found very similar
results.
 
\begin{figure*}
  \begin{center}
    \centerline{
      \includegraphics[height=6.0in]{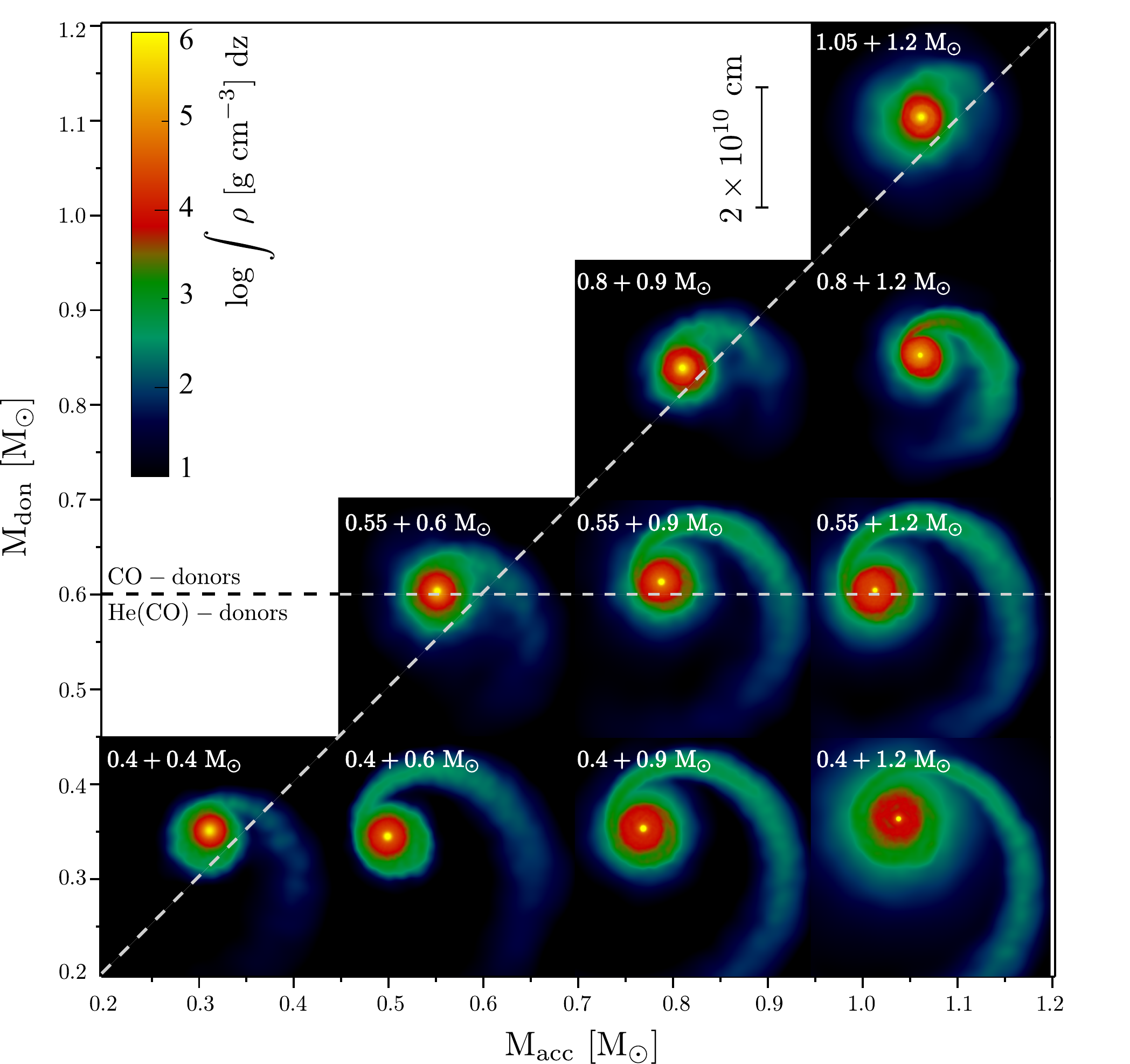}}
  \end{center}
  \caption{Column density snapshots taken after a timescale of three
    initial periods after the merger moment. This corresponds to the 
    moment at which we analyze all merger remnants presented in 
    Table \ref{tab:bigtable1} in the Appendix.
    The 10 snapshots shown here are
    representative of all white dwarf chemical composition and mass
    combinations. The horizontal dashed line divides the CO- and 
    He-mass-transferring systems and the diagonal dashed line shows 
    where the masses of the binary's components are equal, ie. the mass 
    ratio $q=M_{\rm don}/M_{\rm acc}=1$.  During the merger, because the 
    matter coming from the former donor has a high angular momentum and cannot be
    accreted directly by the accretor, a disk and a tidal tail will form. 
    For the systems with a lower mass ratio
    $q$ the matter has a larger specific angular momentum causing the
    disk to spread over a larger distance and a more extended tail. 
}  
  \label{fig:remnants}
\end{figure*}

\begin{figure*}
    \centerline{
      \includegraphics[height=3.35in]{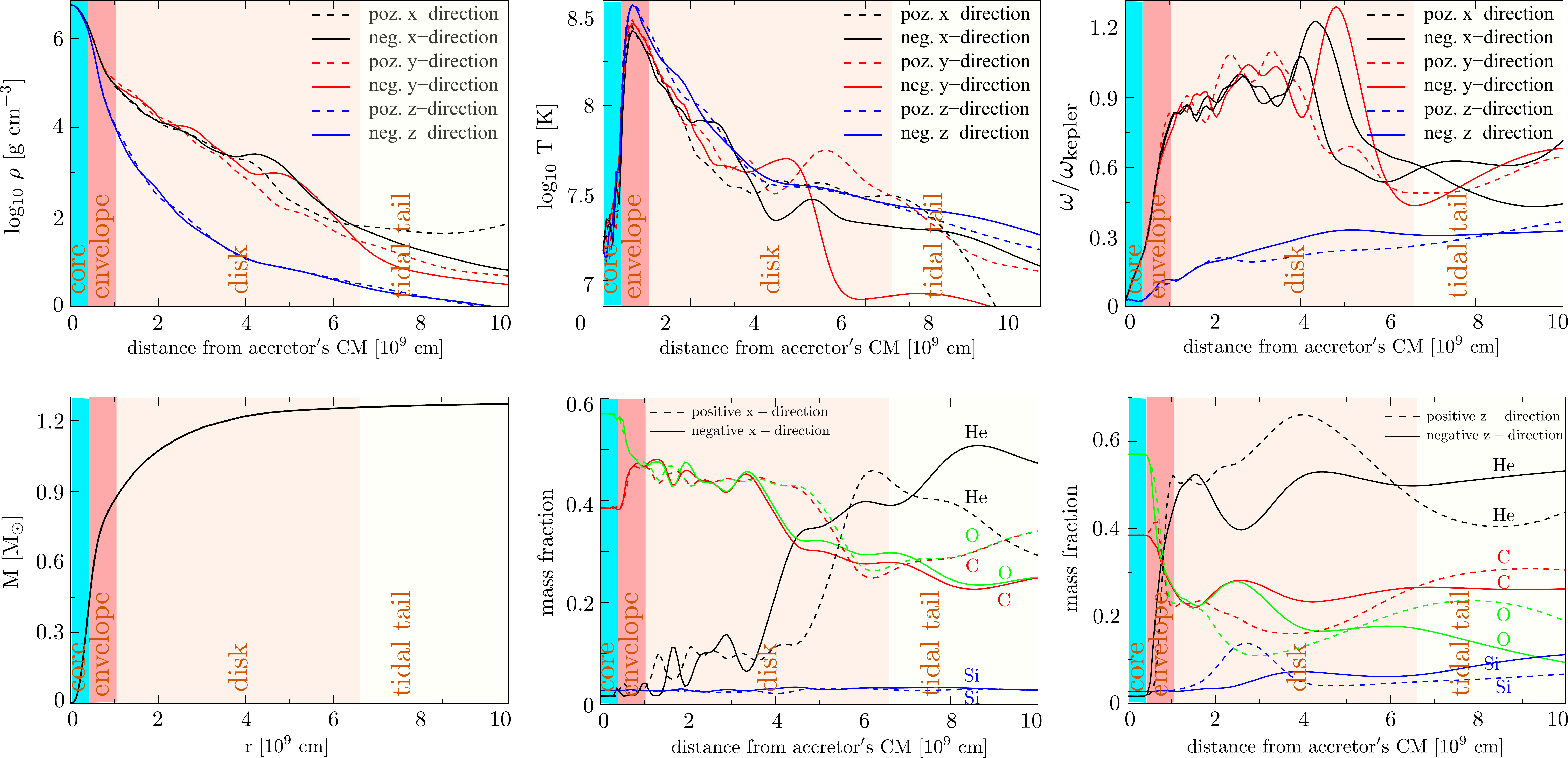}
    }
    \caption{The different regions of the remnant are shown here in the density (top left), temperature
      (top center), angular velocity (top right), enclosed mass
      (bottom left) and mass fractions along X- (bottom center) 
      and Z-axes (bottom right) profiles for the system with $0.6+0.7$ \msun. Here the 
      quantities are SPH-interpolated values along different positive and negative directions with respect
      to the former accretor's center of mass. We distinguish between the different regions based on the
      temperature, density and angular velocity profiles. 
      The central region of the former accretor is nearly
      isothermal and forms the core (cyan region). The outer edge of the core is taken where there is a steep increase in
      temperature. 
      Its outer regions have been
      heated up during the mass transfer and at the merger
      moment and together with a fraction of the matter coming from
      the donor it forms the hot, thermally supported envelope (light red region).
      The hot envelope is surrounded by a (nearly) Keplerian disk (light orange region) and a fraction of the total 
      mass is flung 
      out in a tidal tail (light yellow region) that will fallback and reshape the outer region of the disk.
 }
    \label{fig:structure}
  \end{figure*}

\subsection{Remnant profiles}
\label{sec:profiles}

Figure \ref{fig:remnants} gives an overview over the morphology for the whole range of mass 
distributions at the moment when the simulations are stopped ($t_{\rm merger} + 3 P_0$). This is
different from Figure 7 in \cite{dan12} where the configuration was shown at the moment when
the conditions are most promising for explosion.  At the merger point about $1\%$ of the system's total mass 
is flung out into a tidal tail. Most of this material is still gravitationally bound to the central remnant 
and it will fall back, interact with and shape the outer disk. Systems with a mass ratio 
close to unity (those near the diagonal dashed line) lead to an almost spherical remnant with 
little material in the tidal tail, while smaller mass ratio systems (lower right corner of the digram) 
have a more extended tail. 

Based on temperature, density and rotation profiles, we can distinguish different 
regions in the remnant as shown in Figure \ref{fig:structure} for a  
$0.6+0.7\,M_\odot$ binary system. 
We identify four main regions of the remnant: a cold (nearly) isothermal core; a hot 
envelope; a centrifugally supported Keplerian disk and a tidal tail. Fits for the masses
of these regions can be found in the Appendix (Eqs. \ref{eq:mcore} - \ref{eq:mfb}).

\begin{figure*}
\centerline{
 \includegraphics[height=2.2in]{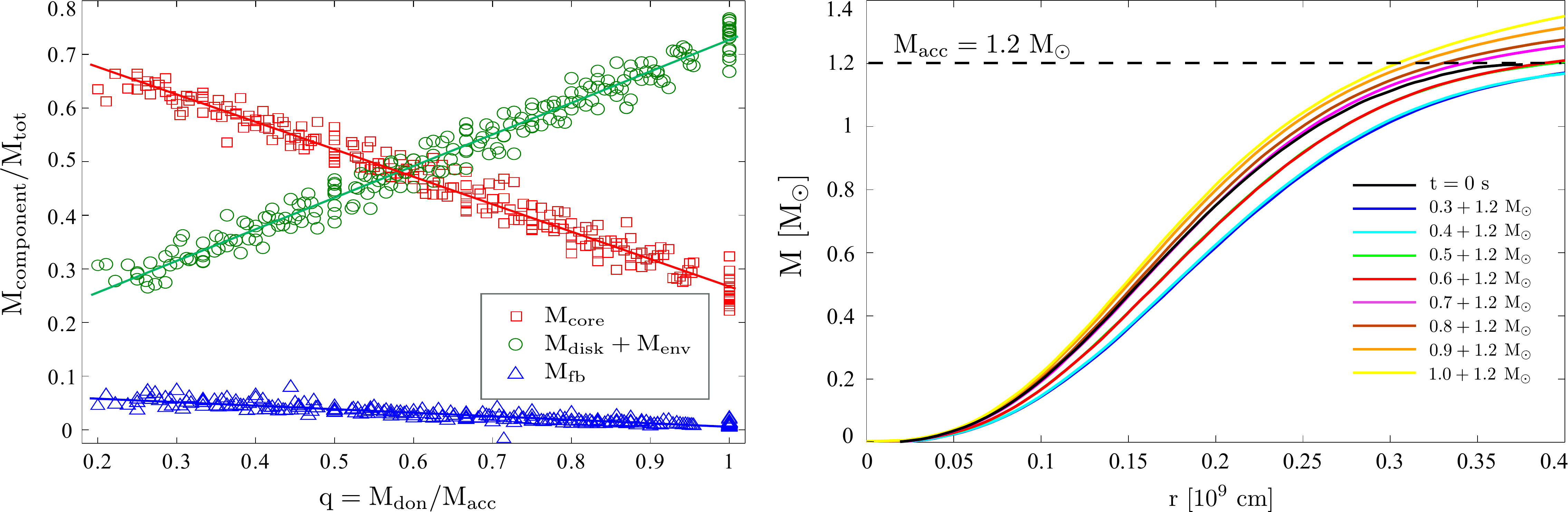}
}
\caption{(left) Mass of the core (red squares), hot envelope plus disk (green circles) 
and tidal tail material (blue triangles) as a function of mass ratio 
$q=M_{\rm don}/M_{\rm acc}$. The degree of heating experienced by the core is increasing with the mass ratio. 
The mass of the core and the 
envelope plus disk are equal at around $q=0.7$ with the mass of the core decreasing down to $30\%$ 
of the total mass for a mass ratio of unity. The mass of the fallback
material $M_{\rm fb}$ increases with decreasing mass ratio.
Continuous lines are fits to the data, see
Eq. \ref{eq:mcore}-\ref{eq:mfb} in the Appendix. (right) Enclosed mass
as a function of radius with  
respect to the WD center of mass for the remnants with $1.2\,M_\odot$ accretor 
and donor masses ranging from 0.3 to 1.0 $M_\odot$. For low mass ratios the 
core experiences an expansion as the accretor is heated through shocks 
while for higher mass ratios the core is compressed by the massive outer material. 
}   
\label{fig:masstrends}
\end{figure*}

\begin{figure*}
\centerline{
 \includegraphics[height=4.6in]{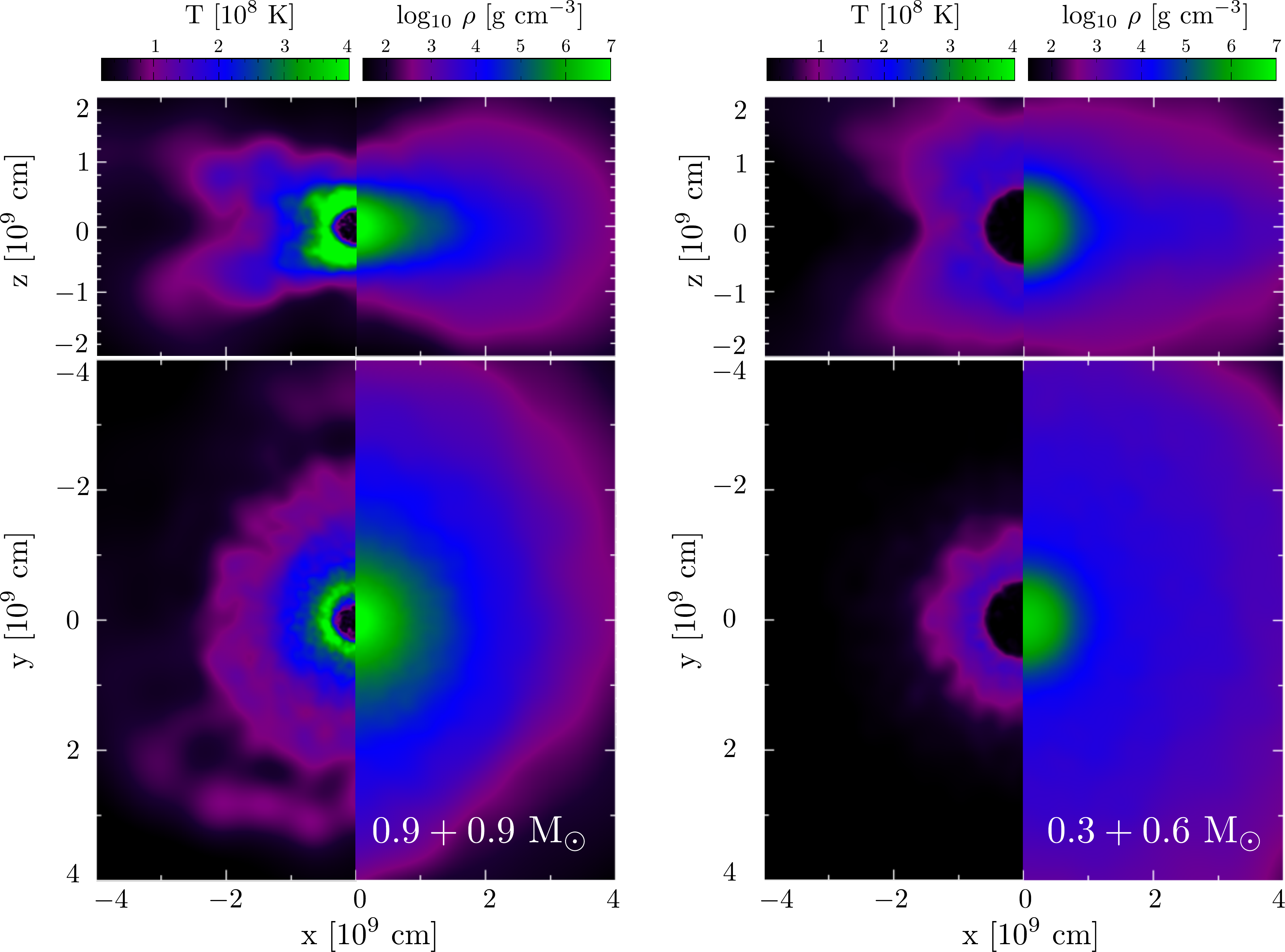}
}
\caption{Comparison of the final remnants resulting from the mergers of $0.9+0.9\,M_\odot$ 
(left) and $0.3+0.6\,M_\odot$ (right) in both XZ (upper panels) and XY (lower panels) planes. 
Within each panel, the left subpanel shows the temperature and the right panel the density. 
The degree of heating experienced by the core increases with the mass ratio. However, the 
central region (core)  is always colder than the surrounding region (envelope), even for 
systems with a mass ratio $q$ of unity. 
}   
\label{fig:Trho}
\end{figure*}

The degree of heating experienced by the core is closely related to the degree of mixing of
the two stars (discussed in more detail in Sec.~\ref{sec:mixing}) and increases with the mass ratio. 
For a mass ratio of $q\sim 0.7$ the mass of the core and the disk plus envelope are equal, with the mass of the core decreasing 
down to $30\%$ of the total mass for a mass ratio of unity, see left panel of Figure \ref{fig:masstrends}.
In our simulations, even equal mass cases do not lead to perfectly symmetric remnants. This is 
because the systems reacts sensitively to even small deviations from perfect symmetry.
At some point during the dozens of orbits of inspiral and mass transfer \citep{dan12}
a small numerical imperfection can kick-start the (asymmetric) disruption process. 
Consequently, the two spiral arms that form right after the merger have different sizes, 
with the smaller one quickly dissipating. Thus the core of the remnant is always a fraction 
of the star that is not fully disrupted and it is colder than the surrounding region, 
see Figure \ref{fig:Trho}. A perfectly symmetric system, however, is a completely academic
case and therefore such systems should be considered as only nearly perfectly symmetric.\\
Recent studies of \cite{aguilar09,pakmor10,pakmor11,zhu13} found that both components
are symmetrically disrupted and the higher temperature is located at the remnant's center. 
We were able to reproduce their results, but only when we started with their approximate initial 
conditions, neglecting the tidal deformation of the two stars, see Section \ref{sec:methods}.\\
The right panel of Figure \ref{fig:masstrends} shows the enclosed mass as a function of the radius from the center of 
the accretor with $1.2\,M_\odot$ and donor masses ranging 
from 0.3 to 1.0 $M_\odot$ three initial orbital periods after the merger. 
The solid black line shows the result from the beginning of mass transfer, at $t=0$ s.
Comparing the central density of the remnant with that at the moment when mass transfer sets in, 
we find that it changes by as much as 30\%. 
For low mass ratios the core experiences an expansion as the accretor is heated through shocks 
while for higher mass ratios the core is compressed by the massive outer material. \\
The hottest region of the remnant is the region between the central core and the 
surrounding disk. In this region matter is nearly virialized. Up to about 0.1 $M_\odot$ 
(decreasing with increasing mass ratio $q$) has been accreted prior to the actual 
merger \citep{dan11}. We distinguish between the disk and the hot envelope 
based on the angular velocity profile, as shown in the top-right of Figure \ref{fig:structure}. The peak temperature
is located in the envelope, corresponding fit formulae can be found in the appendix
(Eqs. \ref{eq:mtmax} and \ref{eq:mtmaxavg}).\\
While the envelope is supported against gravitational attraction mainly by the thermal pressure, the disk is 
centrifugally supported and much of its material has not been shocked. 
The disk plus envelope extent increases with the mass ratio for a constant total mass. Additionally, there is a
clear tendency to increase the disk plus envelope extent with the total mass for a fixed mass ratio.\\
The disk is fed by the matter in the trailing arm which is still bound to the central remnant.
The fallback material has a spiral shape and does not rotate at a Keplerian speed. The amount of matter flung out 
increases with decreasing mass ratio, see Figure \ref{fig:masstrends}, and with increasing total mass of the system.
Assuming this material is ballistically falling towards the central
remnant we calculate how much it takes for the matter to 
circularize \citep[as described in][]{rosswog07}. We find that the
fallback time for runs simulated here is ranging from days  
to years and most of the material will circularize inside the initial radius of the disk. 
For example for the system with $0.6+0.9\,M_\odot$ only $3\%$ of the
material falling back will circularize beyond the disk.

\begin{figure*}
\centerline{
 \includegraphics[height=4.0in]{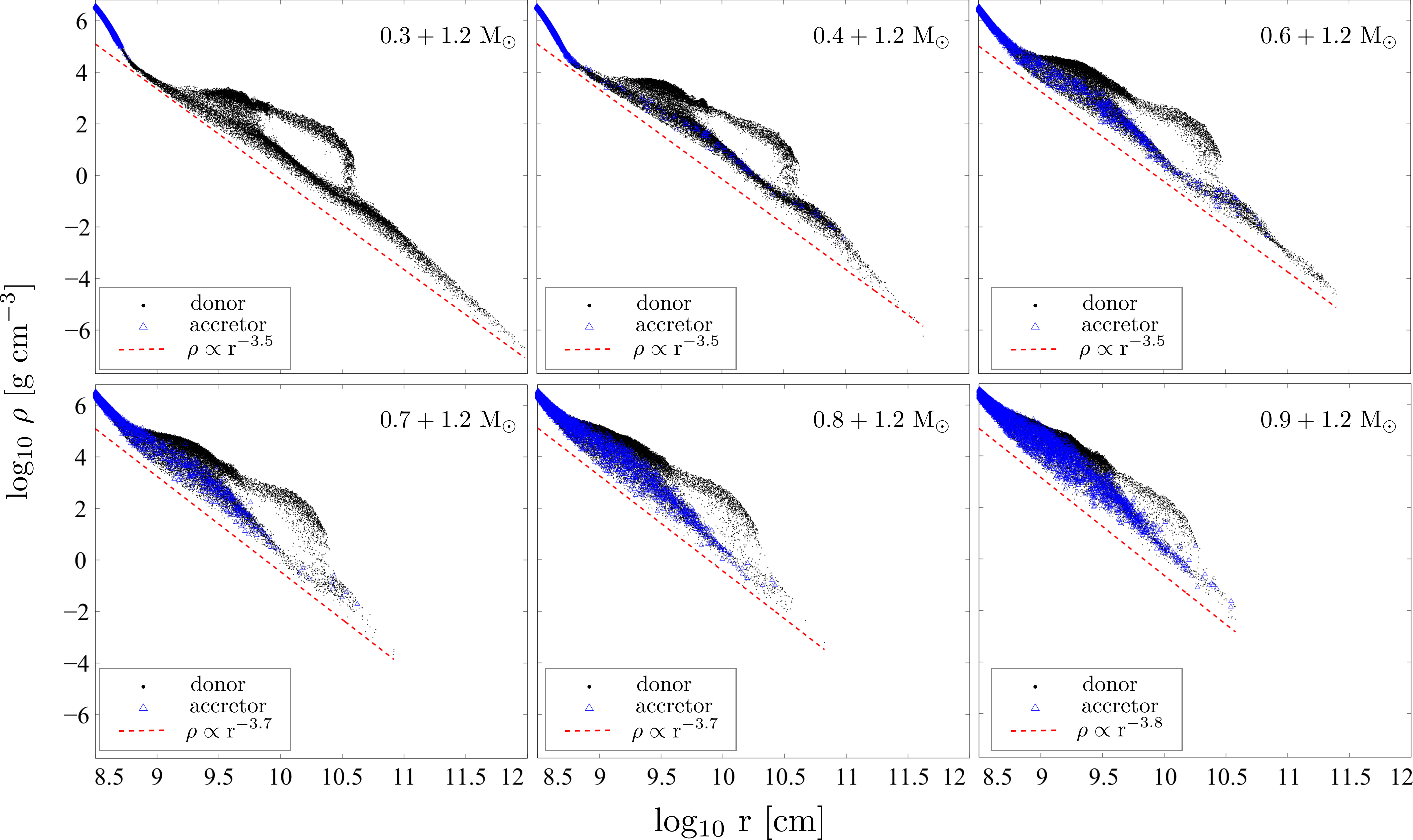}
}
\caption{Density profiles of the outer region of the remnant as a function of the radius with respect to the former 
accretor's center of mass for the systems with an accretor of $1.2\,M_\odot$ and donor masses between $0.3$ 
and $0.9\,M_\odot$. The blue triangles and black dots are the particles that belong to the former accretor and 
donor, respectively. The dashed red line is the linear fitted density profile and it ranges from a $r^{-3.8}$ to 
$r^{-3.5}$.}   
\label{fig:envelope}
\end{figure*}

\cite{fryer10} computed light-curves for explosions of Chandrasekhar mass WDs surrounded by a debris of 
material, left after the merger of two massive WDs. They used the merger of a $0.9+1.2\,M_\odot$ system to constrain 
the density profiles of the material surrounding the central remnant and found that the density follows an $ r^{-4}$ profile. 
To account for possible variations in the density profile they also did one calculation assuming a shallower 
$\propto r^{-3}$ profile. In Figure \ref{fig:envelope} we show the density profiles of the remnant's outer regions
for different systems. They all share the same accretor mass of $1.2\,M_\odot$ but have different donor masses of 0.3,  
0.4, 0.6 (upper panels), 0.7, 0.8 and 0.9 $M_\odot$ (bottom panels). We use two different colors to identify the 
particles formerly belonging to the accretor (blue triangles) and the donor (black dots). 
The red dashed-line shows the fitted density profile, as steep as $r^{-3.8}$ for the highest mass ratio binary, while 
the lowest mass ratio one has a $r^{-3.5}$. We observe the same trend over the entire WD-WD parameter space, 
the profile becomes shallower (ie. value of exponent increases) with decreasing mass ratio and the donor's density, 
ranging from $r^{-3.9}$ to $r^{-3.3}$.  As discussed by \cite{fryer10}, in a supernova explosion, a shallower density 
profile will cause the radiation to be trapped for a longer time and the shock breakout will have a lower, later X-ray 
luminosity peak and longer duration.\\
In the Appendix we provide fit formulae for the maximum angular velocity, $\omega_{\rm max}$, and the 
mass inclosed inside of it. 

\subsection{Spin dependence}
\label{sec:spin_dep}
\begin{figure}
\centerline{
 \includegraphics[height=3.3in]{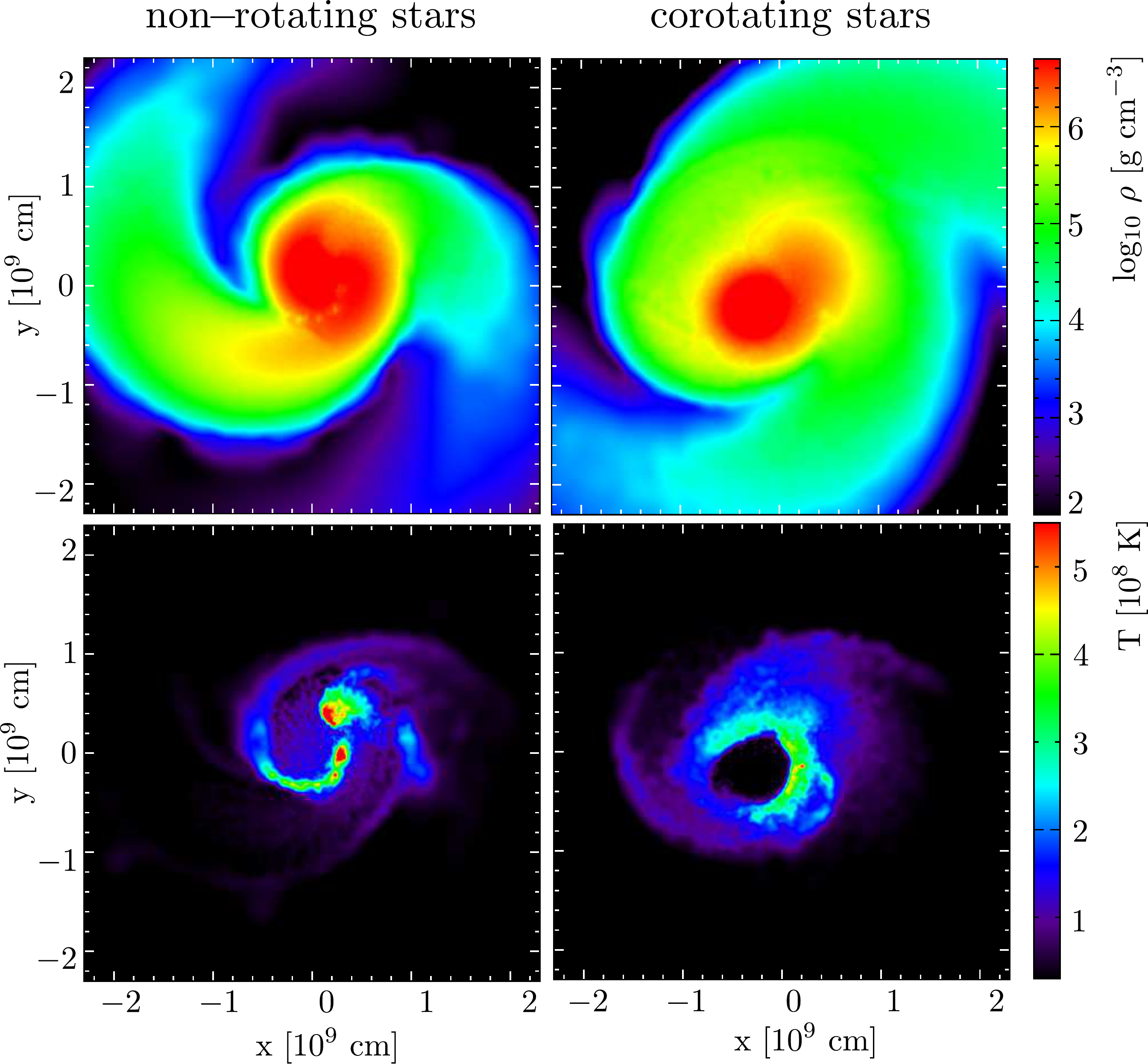}
}
\caption{Comparison of the outcome of different spin states. Each time a binary with 0.89 \Msun is
prepared, once with non-rotating (left column) and once with tidally locked stars (right column).
Both cases are shown at the moment when they are most prone to detonation ($t= 130$ s, left, 
and $t= 719$ s, right column.}   
\label{fig:pakmor_comparison_irrot_corot}
\end{figure}
\begin{figure}
\centerline{
 \includegraphics[height=4.6in]{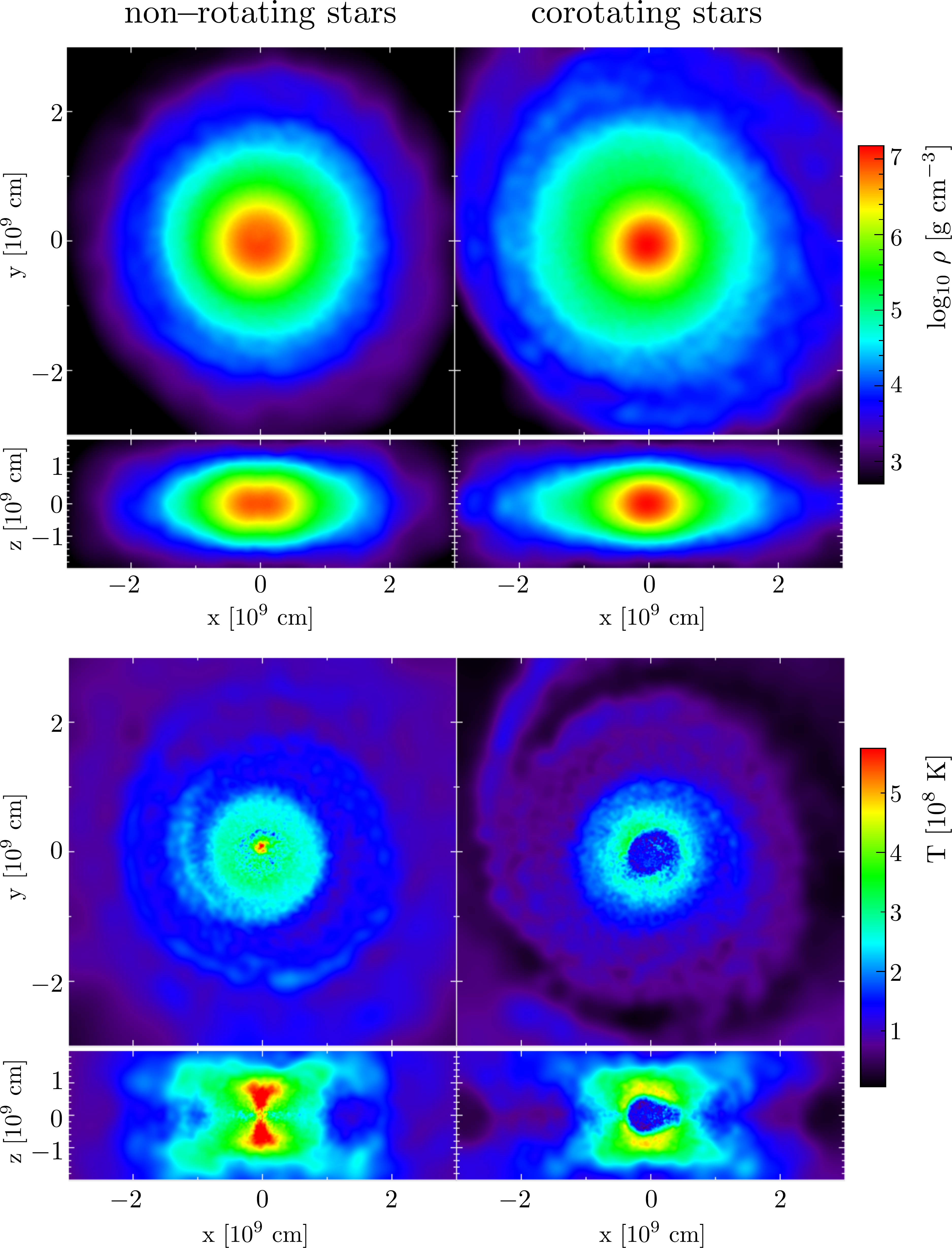}
}
\caption{Dependence of the remnant structure on the initial white dwarf spin. The left column 
shows the results for the initially non-rotating stars, the right column for an initially tidally locked
binary, both stars have 0.89 \msun. The non-spinning stars merge quickly and develop the
hottest temperatures in their core, while the corotating system orbits for much longer until
a numerical perturbation finally breaks the symmetry. The hottest regions in this case
are the surface layers of the resulting merger core.
}   
\label{fig:spin_dependence}
\end{figure}

Since the initial rotational state of a white dwarf binary is not well known and this study
is investigating the initially tidally locked systems, it is interesting to ask to which extent the results
would change for a different initial spin state. This is of particular relevance with respect to
the question where in the remnant a possible thermonuclear ignition could be triggered.\\
The major difference between a tidally locked and a non-rotating binary comes from the fact
that in the latter case the stars appear in the binary frame to rotate against the orbital direction 
and therefore at contact produce a shear layer with a velocity discontinuity of $\Delta v \sim 2 
R_\ast \omega_{\rm orb}$. In the perfectly co-rotating case, in contrast, the velocities vanish
in the binary frame, in reality small velocities occur due to the stars dynamically adapting to
the changing gravitational field of the companion. As a consequence, non-rotating stars with
their substantially larger shear produce a more pronounced string of Kelvin-Helmholtz vortices 
that can subsequently merge. This behaviour has been explored in detail in the context of 
neutron star mergers, see e.g. 
\cite{ruffert96,rosswog99,rasio99,rosswog02,price06,rosswog13,zrake13}, but it occurs in a
qualitatively similar way also for WD binaries.\\
To explore the spin dependence we set up a binary system with twice 0.89 \Msun and $2 \times 10^5$ 
particles exactly like in \cite{pakmor10} and another such system with the same particle number,
but in a tidally locked state prepared as described in \cite{dan11}. 
The non-spinning system merges quickly and does not have much time to develop large asymmetries 
(this may be partially aided by the approximate initial conditions that we apply
here for comparison reasons). We find that in carefully relaxed co-rotating systems 
the merger occurs on a much longer time scale and the final merger is initiated once a small 
numerical perturbation breaks the perfect initial symmetry. In Figure~\ref{fig:pakmor_comparison_irrot_corot}
we compare the two systems at the moments when they are most prone to detonation 
($t= 130$ s for 
the non-rotating, left panels, and at $t= 719$ s for the tidally locked case, right panels). Apart from the different 
morphology one sees the result of the much stronger Kelvin-Helmholtz instabilities for 
the non-rotating case: pronounced hot vortices form in the high-density core at 
the interface between the stars. They are also visible as density troughs in the 
upper-left panel. The corotating case, in contrast, shows that the locations most prone 
to detonation are at lower densities and temperatures, at the surface rather than 
the centre of the high-density core. The final density and temperature distributions 
are shown in Figure~\ref{fig:spin_dependence}, both for XY- and the XZ-plane. Once the 
Kelvin-Helmholtz vortices have merged, they result in a `hour-glass' shaped temperature 
distribution. This is agreement with the findings of \cite{zhu13} \citep[and with those 
for neutron star mergers;][]{rosswog13}.\\
This comparison illustrates that the spin state has a serious impact on the location 
where an explosion is likely triggered. Co-rotating systems likely ignite in the 
surface layers of the remnant core, while in irrotational systems explosions likely 
come from the central and highest density regions of the core. Therefore, corotating 
and irrotational systems should differ in the amount of unburnt material, their ejecta 
velocities and their degree of asymmetry.

\subsection{Mixing of matter between the binary components}
\label{sec:mixing}
\begin{figure}
\centerline{
 \includegraphics[height=2.7in]{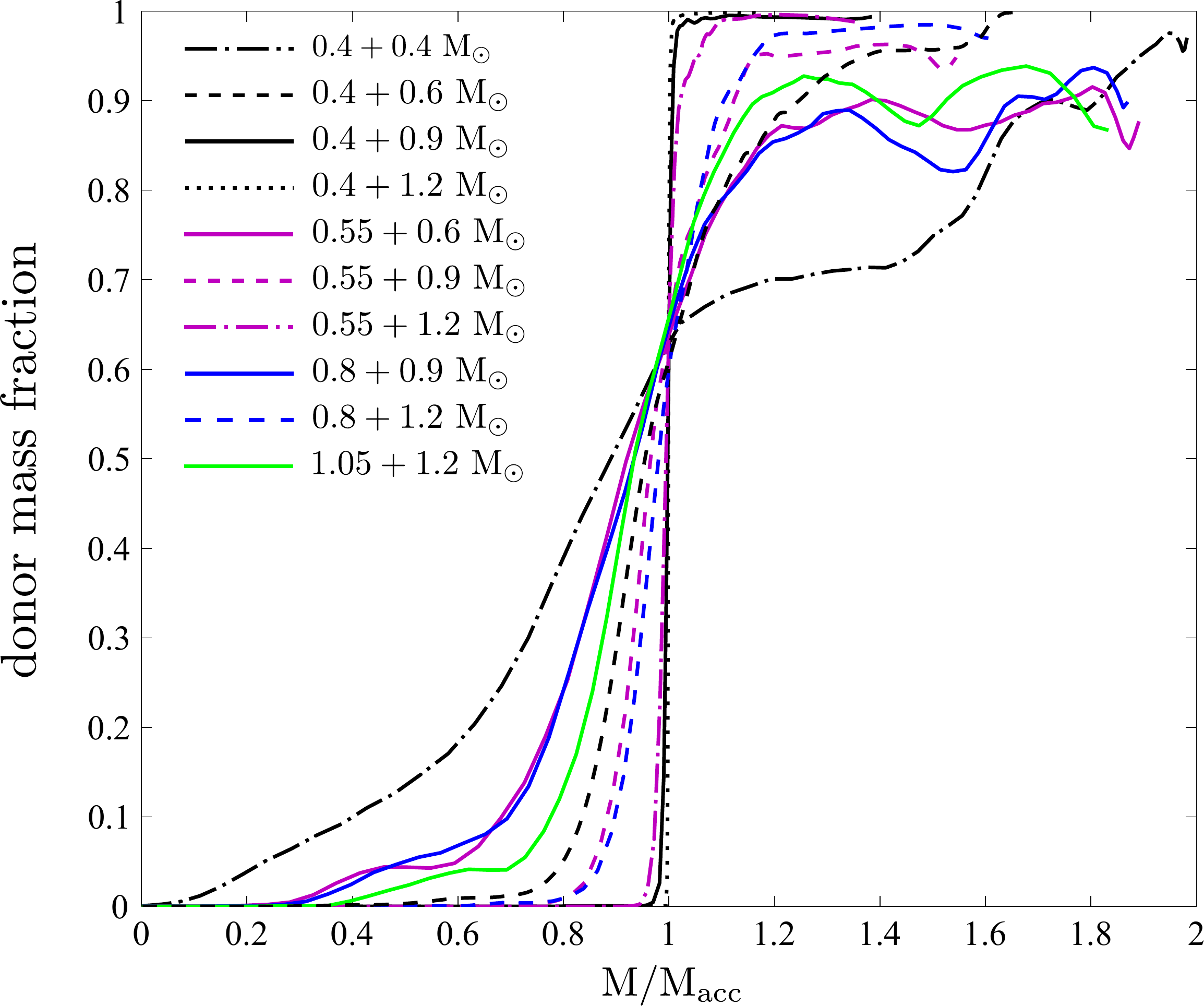}
}
\caption[Mixing of the two stars during the dynamical merger
  phase. Shown is the donor mass fraction as a function of the mass
  enclosed by given equipotential surfaces (scaled to the accretor's
  initial mass $M_{\rm acc}$). Ten systems covering the full parameter
  space of mass and chemical composition (same systems as in Figure
  \ref{fig:remnants}) are shown. The mixing between the binary
  components increases with the mass ratio, but it never becomes
  complete even for systems with a mass ratio of unity. This is due to
  the initially tidally locked spin state (Zhu et al. (2013) find for
  their non-spinning white dwarfs complete mixing for a mass ratio
  $q=1$).]{Mixing of the two stars during the dynamical merger
  phase. Shown is the donor mass fraction as a function of the mass
  enclosed by given equipotential surfaces (scaled to the accretor's
  initial mass $M_{\rm acc}$). Ten systems covering the full parameter
  space of mass and chemical composition (same systems as in Figure
  \ref{fig:remnants}) are shown. The mixing between the binary
  components increases with the mass ratio, but it never becomes
  complete even for systems with a mass ratio of unity. This is due to
  the initially tidally locked spin state (in contrast, \cite{zhu13} find for
  their non-spinning white dwarfs complete mixing for a mass ratio
  $q=1$).  }
\label{fig:mixing_mass_ratio}
\end{figure}

\begin{figure*}
\centerline{
 \includegraphics[height=5.4in]{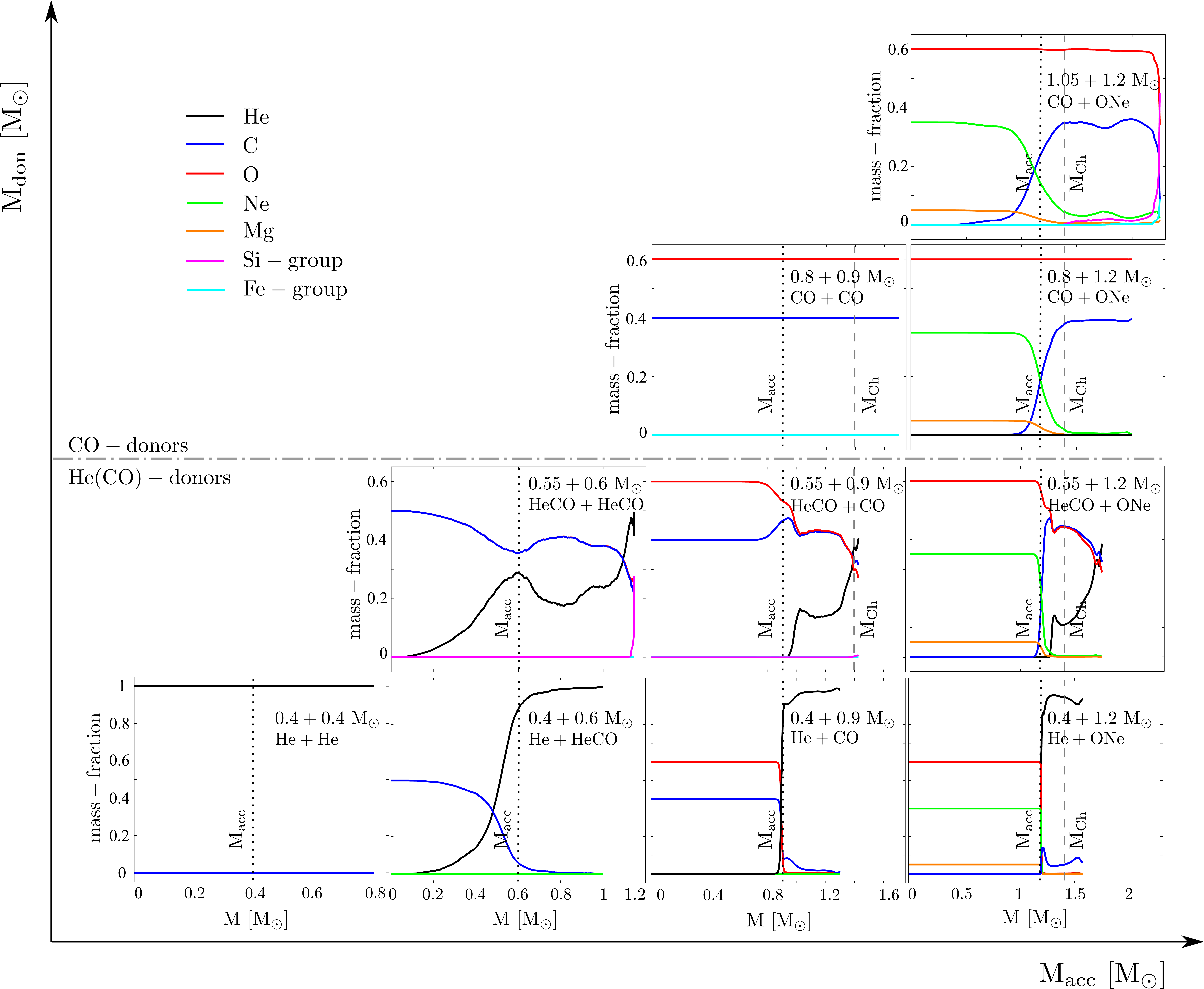}
}
\caption{Mixing of the two stars during the dynamical merger phase. Shown are the mass fractions of
different nuclear species for the same systems as above. The mass fractions are averages over 
equipotential surfaces and plotted against the enclosed mass. 
The horizontal dot-dashed line shows the transition from CO to He(CO) mass-transferring systems, the vertical dashed lines show the Chandrasekhar mass limit for a white dwarf and the dotted lines show the mass of the accretor.
}   
\label{fig:mixing_mass_fractions}
\end{figure*}

In Figure~\ref{fig:mixing_mass_ratio} we show the
fraction of the mass that originally belonged to the donor star as a function of the mass enclosed
by a common equipotential surface. For mass ratios below $q\approx 0.45$, the stars hardly mix at all,
while for a mass ratio close to unity maximum mixing occurs. Contrary to \cite{zhu13}, we never find
a complete mixing between the two stars. This is due to the different initial spin states (no spin in 
their study, tidal locking in ours) and it is consistent with the above findings of the spin dependence.\\
In Figure~\ref{fig:mixing_mass_fractions} we investigate how the stars mix chemically. To this end we 
plot averages of the mass fractions of the 7 species of our nuclear network over equipotential surfaces
for ten selected binary systems. For systems below $\sim\!\! 1.0$ \Msun burning is negligible and only mixing occurs, see for example the shown $0.4+0.6$ \Msun
system. This is consistent with our later discussion about dynamical burning, see Figure~\ref{fig:thermo}, 
left panel. Some burning occurs in the outer layers of the $0.55+0.6$ \Msun system which transforms
He into Si-group elements (magenta line). This matter is finally ejected. As shown below, we find several 
cases where nuclear energy production contributes substantially to the mass ejection process.
For He transferring systems with larger mass substantial burning may occur, the $0.4+1.2$ \Msun system, 
for example, burns roughly 0.1 \Msun of He into C.

\subsection{Dynamical burning and possible detonations}
\label{sec:burning}
\begin{figure*}
\centerline{
 \includegraphics[height=2.15in]{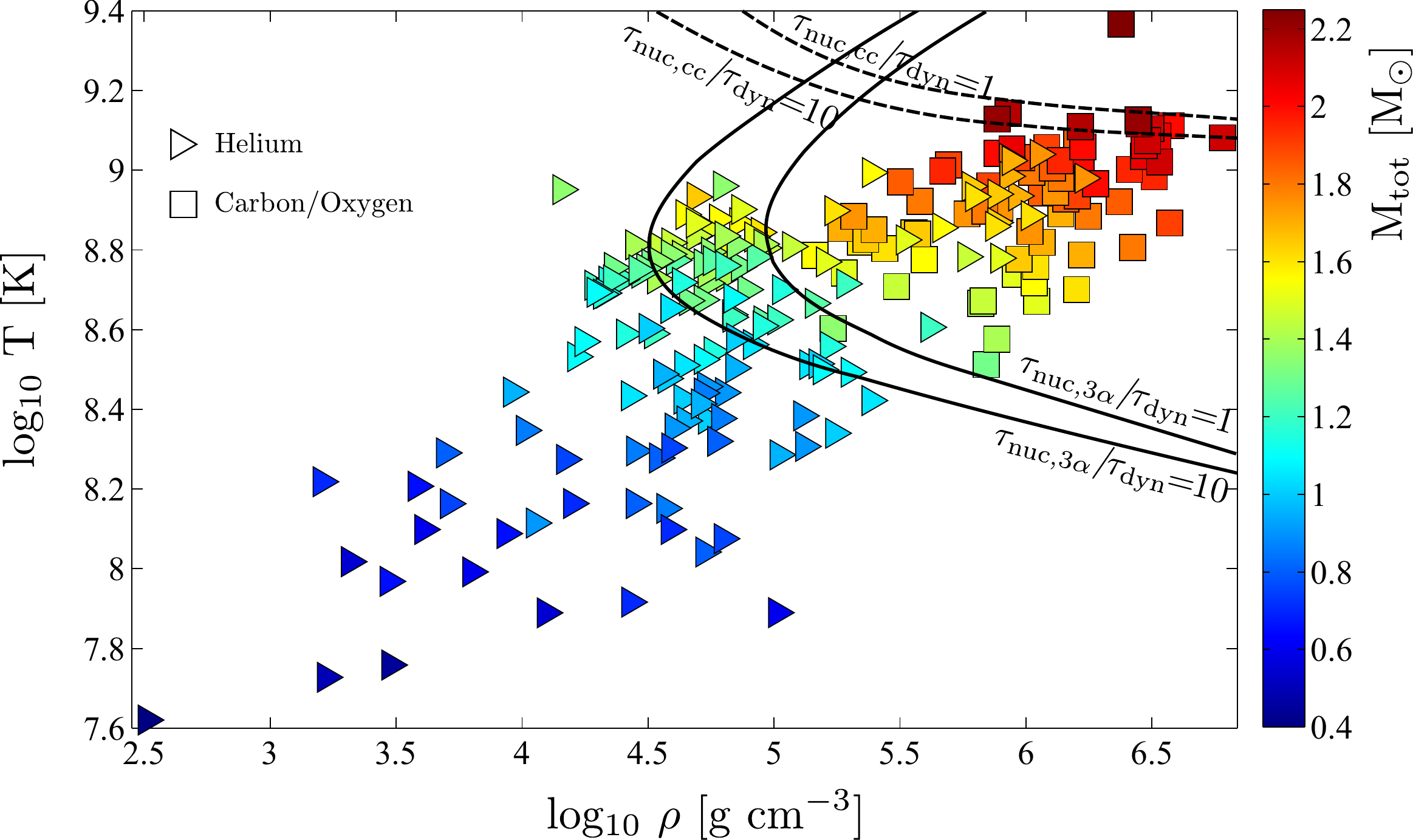} 
\includegraphics[height=2.15in]{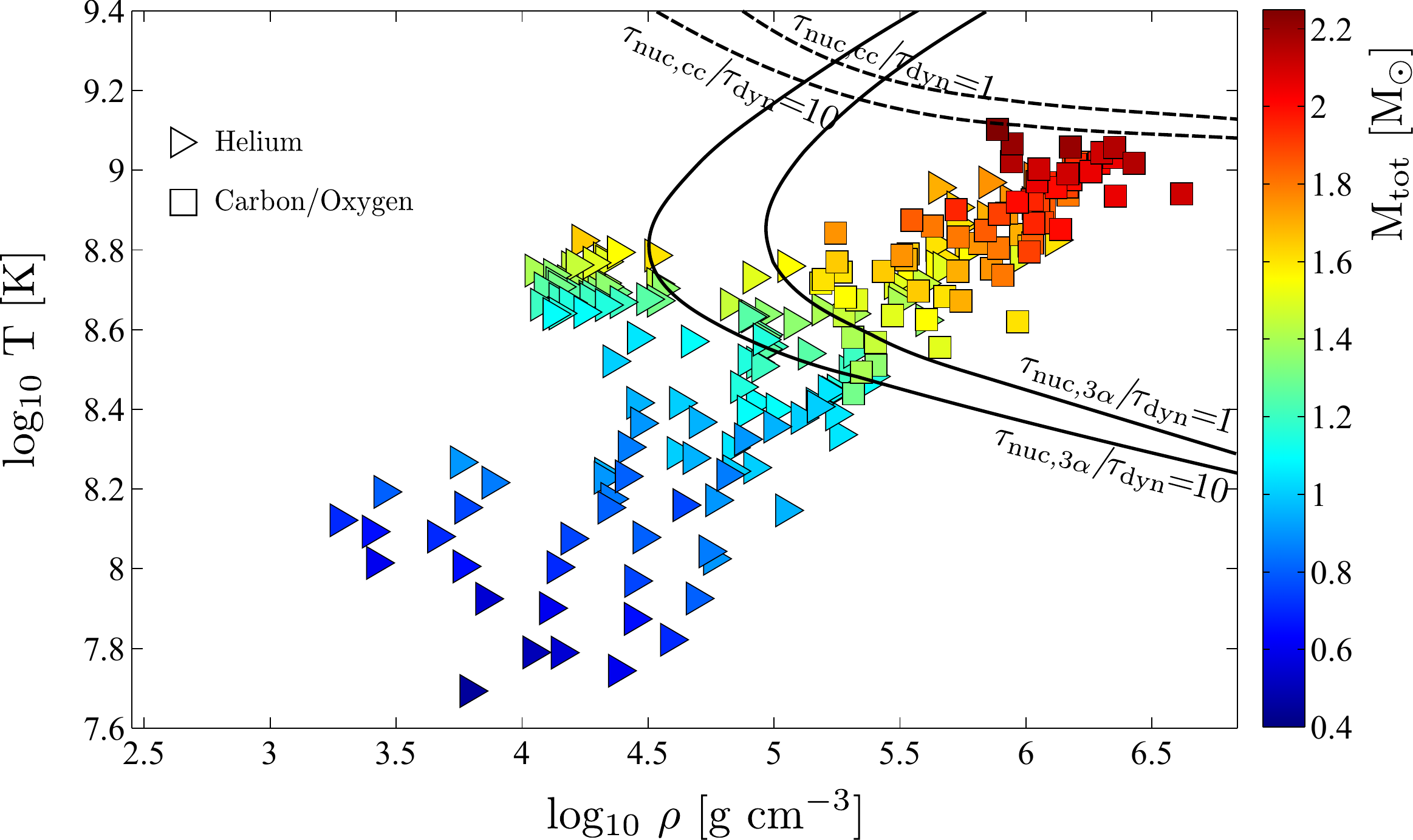} 
}
\caption{Left: maximum temperature $T_{\rm max}$ and density at maximum temperature $\rho(T_{\rm max})$ for
  both He (triangles) and CO (squares) mass-transferring systems. Right: maximum of SPH-smoothed temperatures 
  $\langle T\rangle_{\rm max} $ and  averaged density at the
  location of averaged maximum temperature $\rho(\langle T\rangle_{\rm max})$, again for both He (triangles) 
  and CO (squares) mass-transferring systems. On each panel color coded is the total mass of the system,
  $M_{\rm tot}$. Also shown are the contours for which the timescale for 
  triple-$\alpha$ reactions $\tau_{\rm nuc,3\alpha}$ and the nuclear timescale of carbon burning $\tau_{\rm nuc,cc}$ 
  are equal or ten times larger than the dynamical timescale $\tau_{\rm dyn}$. $\tau_{\rm nuc}/\tau_{\rm dyn}\leq 1$ 
  is the condition required to initiate the dynamical burning and potentially trigger a detonation. 
} 
\label{fig:thermo}
\end{figure*}

Figure \ref{fig:thermo} shows the maximum temperatures at the end of the simulations and 
the densities at which they occur. In the left panel we show the quantities directly as 
extracted from the simulation, the right panel is similar, but now we use more 
conservatively the maximum of the SPH-interpolated values of the temperatures, 
$\langle T \rangle$. For an explicit discussion of this difference,  we refer to Sec. 4.1 in
\cite{dan12}.
The figure shows the clear trend that both the maximum temperatures $T_{\rm max}$ 
and the densities where they occur increase with the total system mass $M_{\rm tot}$. 
We have not found such a trend with respect to the mass ratio $q$, though. The 
temperatures of the matter in the hot envelope region are typically a good fraction of 
the virial temperature $T_{\rm vir}= GM_{\rm acc} m_{\rm p}/(3R_{\rm acc}k_{\rm B})$, 
where $G, m_{\rm p}$ and $k_{\rm B}$ are gravitational constant, proton mass and 
Boltzmann constant, respectively. Typically, we find values of $T_{\rm max}\approx 
0.77 T_{\rm vir}$ and $\langle T\rangle_{\rm max} \approx 0.62 T_{\rm vir}$.\\
Dynamical burning and possibly a detonation sets in when matter heats up more rapidly than it can 
expand, cool and quench the burning processes, i.e. when $\tau_{\rm nuc}\leq \tau_{\rm dyn}$.
Here,  $\tau_{\rm nuc}=c_p T /\dot{\epsilon}_{\rm nuc}$ \citep[e.g.][]{taam80,nomoto82} is the nuclear, 
$\tau_{\rm dyn}\approx 1/\sqrt{G\rho}$ the dynamical timescale and $c_p$ is the specific heat at 
constant pressure (taken from the Helmholtz equation of state). The nuclear energy generation rate
$\dot{\epsilon}_{\rm nuc}$ is calculated analytically.  For CO composition we use \citep{blinnikov87}
\begin{equation}
\dot{\epsilon}_{\rm nuc,cc}= \rho q_{\rm c} A_{T9}Y_{\rm C}^2\ {\rm exp}(Q/T_{9a}^{1/3}),
\end{equation}
where $q_{\rm c}= 4.48\times 10^{18}\ {\rm erg\ mol^{-1}}$, $A_{T9}=8.54\times 10^{26}T_{9a}^{5/6}T_9^{-3/2}$ 
\citep{fowler75}, $Y_{\rm C}$ is the ${}^{12}{\rm C}$ abundance, $T_{9a}=T_9/(1+0.067T_9)$, $Q=84.165$ and $T_9=T/10^9\ {\rm K}$. $\tau_{\rm nuc,cc}$ is multiplied by the  factor  $\Theta={3T_9}/({QT_{9a}^{2/3}})$ to account 
for the shortening of the reaction time scale due to self-acceleration of burning \citep{frank67}.
The energy generation rate due to the triple-alpha reaction is \citep[e.g.][]{kippenhahn90}
\begin{equation}\label{eq:3a}
\dot{\epsilon}_{\rm nuc,3\alpha}=5.09\times 10^{11} f_{3\alpha} \rho^2X_4^3T_8^{-3}\ {\rm exp}(-44.027/T_8),
\end{equation}
where $f_{3\alpha}= {\rm exp}(2.76\times 10^{-3}\rho^{1/2} T_8^{-3/2})$ is the weak electron 
screening factor \citep{salpeter54,clayton68}, $X_4$ is the mass fraction of ${}^{4}{\rm He}$ (we assume pure He) 
and $T_8= T/10^8$ K. 

Focussing on the more optimistic estimates in the left panel of Figure \ref{fig:thermo}, we see that all
He accreting systems with a total mass beyond 1.1 $M_\odot$ undergo He detonations. 
These are roughly the same systems that are expected to produce surface detonations either
from instabilities within the accretion stream and/or at the moment first dynamical contact  \citep{guillochon10,dan12}.  
For He mass-transferring systems, neutrino cooling does not become important in comparison to
burning unless temperatures in excess of  $5\times 10^9$ K are reached \citep[see Figure 2 in][]{rosswog08}.

For the CO systems, above a temperature of about $6\times 10^8$ K \citep[e.g.][]{yoon07} the energy released from 
C-burning is larger than neutrino losses, but these systems never reach the necessary criterion for dynamical burning 
$\tau_{\rm nuc,cc}\leq \tau_{\rm dyn}$.  However, the conclusion may be different if the WDs carry a thin He surface
layer ($\sim 1\%$ by mass). \cite{raskin11} found He detonations at the surface of the accretor 
($\sim 1.0\,M_\odot$) but those were not sufficiently energetic to trigger C ignition. First calculations of the expected
electromagnetic display of a SN Ia explosion caused by CO-CO mergers have recently been calculated \citep{pakmor10,fryer10}.
In both approaches the explosion had been artificially initiated at heuristically chosen locations, the resulting lightcurves
and spectra did not match those of the most common SN Ia. Nevertheless, this topic warrants further exploration.\\
Binaries with total masses beyond 2.1 $M_\odot$ seem prone to trigger an CO explosion. Those would be natural
candidates for the progenitors of supra-chandrasekhar SN Ia \citep{howell06,howell11}, provided they are consistent
with producing $\sim 1\%$ of the observed type Ia systems.

\subsection{Ejected mass}
\label{sec:ejecta}

\begin{figure*}
\centerline{
 \includegraphics[height=2.5in]{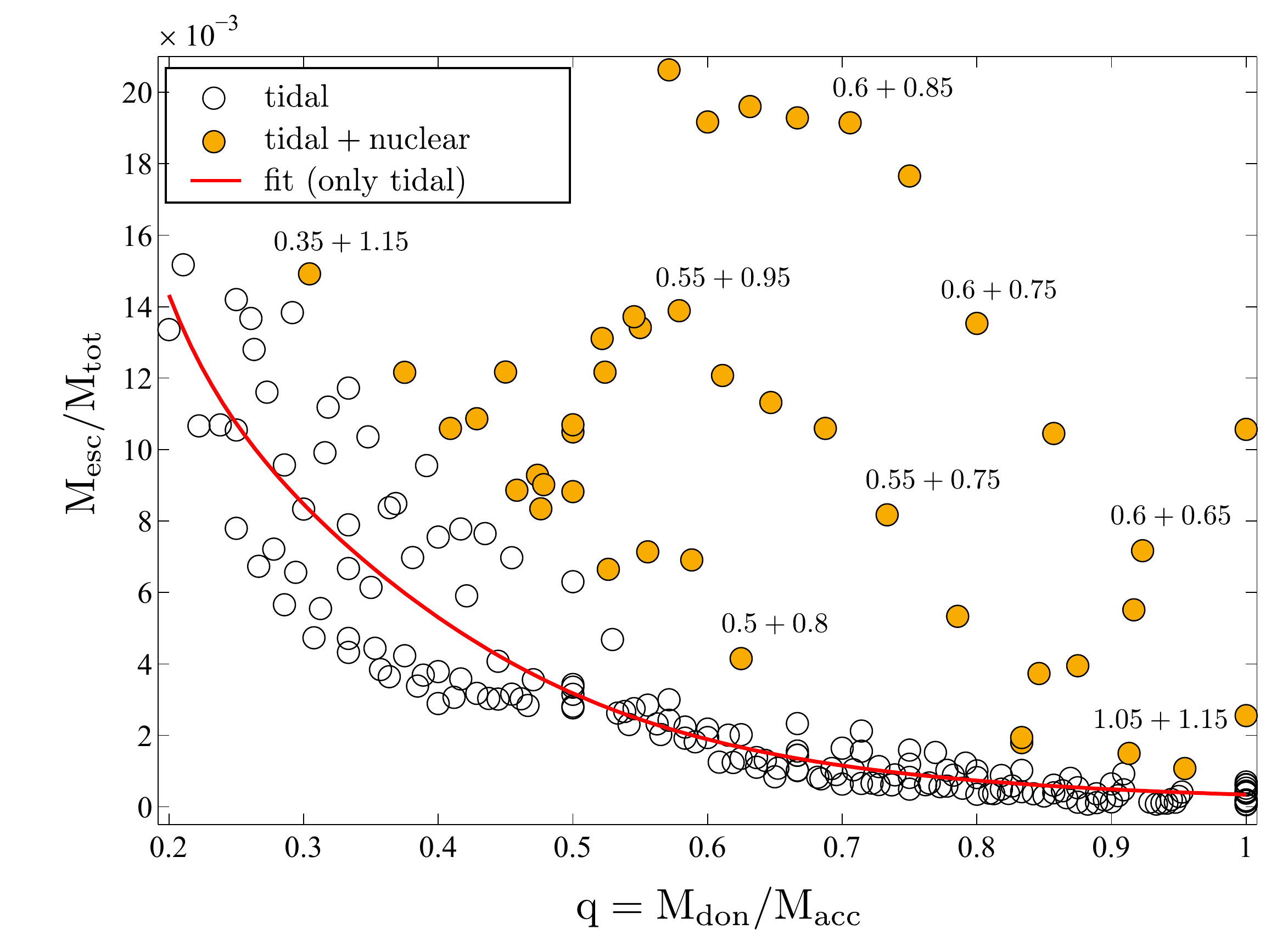} \hspace{-0.3cm}
\includegraphics[height=2.5in]{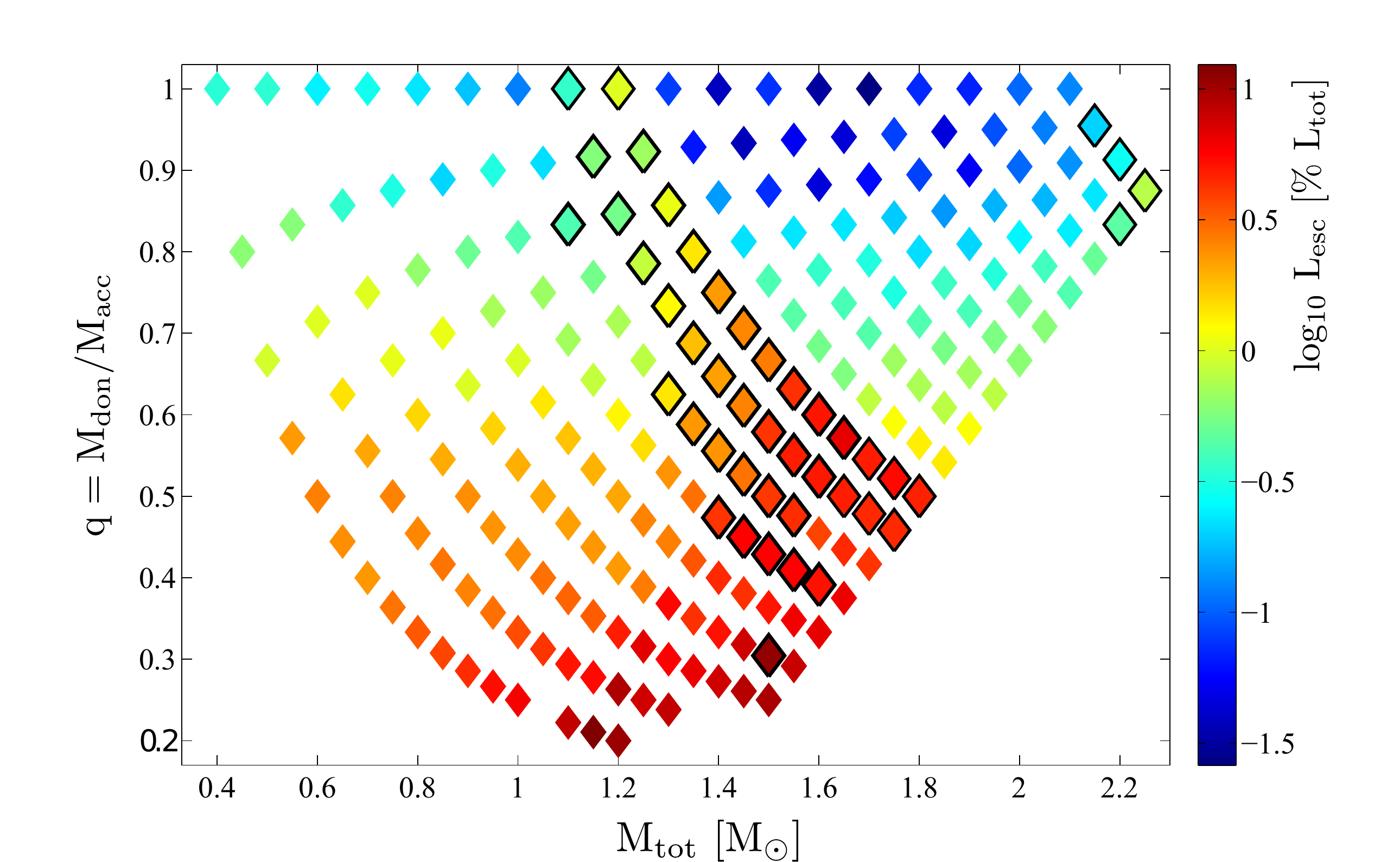} 
}
\caption{Left: ejected mass fraction as a function of the mass ratio $q=M_{\rm don}/M_{\rm acc}$. 
        In a number of cases, nuclear burning has a substantial impact on ejecting mass. 
        Such systems are marked by filled, yellow circles. Matter that is predominantly ejected
        by gravitational torques can be fit by Equation \ref{eq:mesc}. This fit is shown by the
        red line. To provide a visual aid, few of these systems have been labeled.
        Right: the fraction of angular momentum contained in material that is unbound in the 
        $q-M_{\rm tot}$ plane. The amount of ejected mass is tiny, up to $2\%$ of the total mass, 
        but this can take up to $10\%$ of the total angular momentum. The systems that have 
        not been used in the fit $M_{\rm esc}(q)$ are marked with slightly larger symbols and with a 
        black stroke. Most of the mass is ejected during the tidal disruption, although a small 
        fraction leaves the system through the outer Lagrange points during the mass transfer phase. 
}   
\label{fig:mesc}
\end{figure*}

Previous studies of \cite{benz90,segretain97,guerrero04,aguilar09} found ejected masses of $\sim 10^{-3} \; M_\odot$. 
Our large parameter space study is ideal to identify how $M_{\rm esc}$ depends on $q$ and $M_{\rm tot}$.
The fraction of ejected mass $M_{\rm esc}$, as a function of the mass ratio $q$, is shown in the
left-side of Figure \ref{fig:mesc}. We consider a particle to be unbound if the sum of its potential, 
kinetic and internal energy is positive. 
The fraction of ejecta tends to decrease with the increasing mass ratio $q$, 
and ranges between $10^{-4}$ and $3.4\times 10^{-2}\,M_\odot$. \cite{guerrero04} also found a similar 
trend, though they only investigated a small fraction of the parameter space.

Very little mass leaves the system
through the outer Lagrange points during the mass transfer phase, the majority 
becomes gravitationally unbound during the tidal disruption. At the stage shown in Figure 
\ref{fig:remnants}, the unbound fraction does 
not vary anymore. It is interesting to note that in an appreciable number of cases
nuclear burning substantially helps in unbinding material. These systems
either go through a hydrodynamical burning phase, such as those with massive He(CO) 
donors, or undergo substantial nuclear burning as the very massive CO mass-transferring 
systems with $M_{\rm tot}\gtrsim 2.0\,M_\odot$ \citep{dan12}. Such systems are marked by
filled, yellow circles, open circles refer to ejection purely by gravitational torques.
The escaping mass from these latter systems can be estimated by means
of Eq.~\ref{eq:mesc}, see Appendix.
Although the fraction of unbound mass is relatively small, it can take up to $12\%$ of 
the system's total angular momentum, see right-side of Figure \ref{fig:mesc}, having 
an impact on the further evolution of the remnant. 

\subsection{Resolution-dependence}
\label{sec:convergence}

We compared our results obtained with $4\times 10^4$ SPH particles with the runs 
from \cite{dan11} obtained using $2\times 10^5$ SPH particles, which corresponds 
to a reduction of the average resolution length by a factor of $5^{1/3}\approx 1.7$.

\begin{figure}
\centerline{
 \includegraphics[height=3.3in]{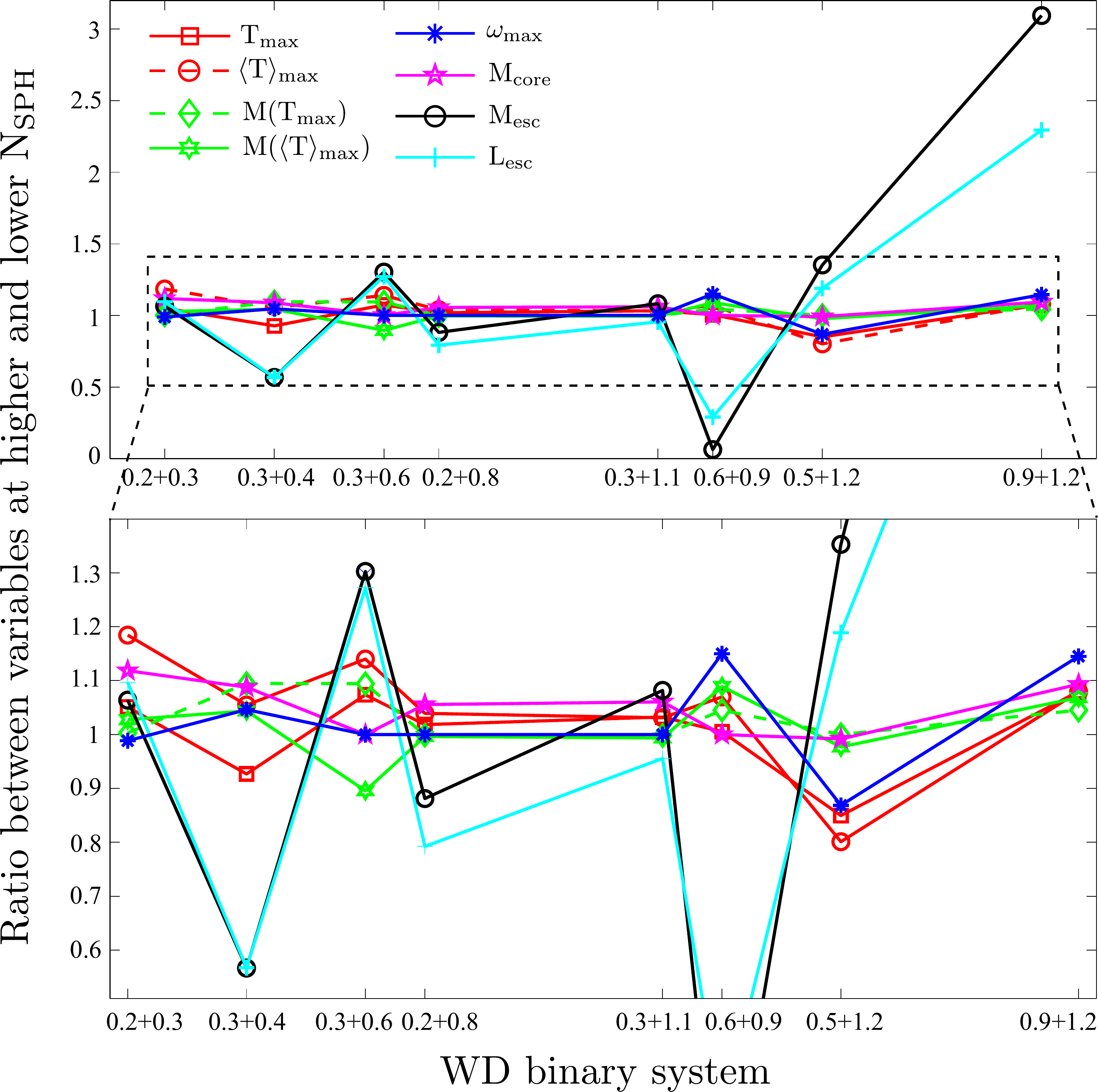}
}
\caption
[Ratio between different variables at $2\times 10^5$ and $4\times 10^4$ SPH particles runs. 
The runs with $2\times 10^5$ are those from Dan et al. (2011). Most of
the variables vary within $20\%$ between higher and lower numerical
resolution runs, only $M_{\rm esc}$ and $L_{\rm esc}$ are showing larger variations. ]   
{Ratio between different variables at $2\times 10^5$ and $4\times 10^4$ SPH particles runs. 
The runs with $2\times 10^5$ are those from \cite{dan11}. Most of the variables vary within $20\%$ 
between higher and lower numerical resolution runs, only $M_{\rm esc}$
and $L_{\rm esc}$ are showing larger variations. }   
\label{fig:convergence}
\end{figure}

\begin{figure*}
\centerline{
\includegraphics[height=2.25in]{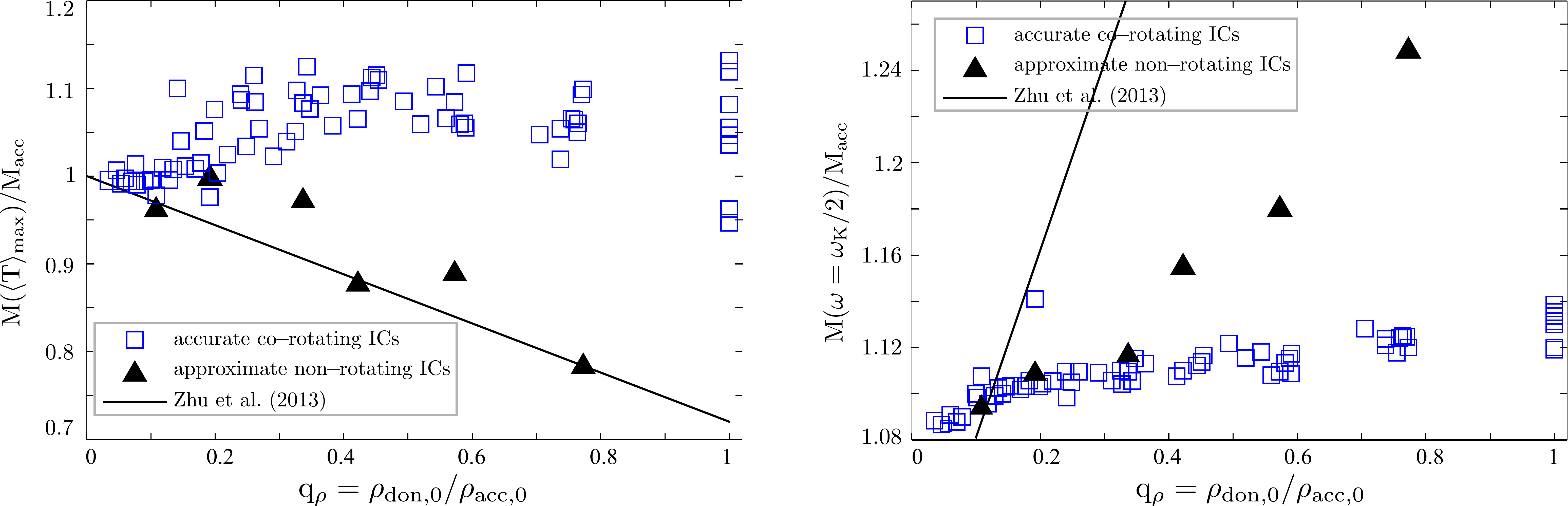}
}
\caption[Comparison with Zhu et al. (2013) for the CO mass-transferring systems: (left) 
enclosed mass at the location of maximum temperature $M(\langle T \rangle _{\rm max})$; 
(right) location in mass coordinate where $\omega=\omega_{\rm K}/2$, $M(\omega=\omega_{\rm K}/2)$. 
The black triangles mark the results obtained using the ``approximate'' non rotating ICs.
The masses are scaled in terms of the accretor's mass and plotted against the density ratio 
$q_\rho=\rho_{\rm don,0}/\rho_{\rm acc,0}$. The continuous lines are fits to the data and 
the dashed lines are the fits found by Zhu et al. (2012). 
]  
{Comparison with \cite{zhu13} for the CO mass-transferring systems: (left) enclosed mass 
at the location of maximum temperature $M(\langle T \rangle_{\rm max})$; (right) location 
in mass coordinate where $\omega=\omega_{\rm K}/2$, $M(\omega=\omega_{\rm K}/2)$. The 
black triangles mark the results obtained using the ``approximate'' non rotating ICs.
The masses are scaled in terms of the accretor's mass and plotted against the density 
ratio $q_\rho=\rho_{\rm don,0}/\rho_{\rm acc,0}$. The continuous lines are fits to 
the data and the dashed lines are the fits found by \cite{zhu13}. 
}   
\label{fig:compzhu}
\end{figure*}

The variation between higher and lower numerical resolution runs does not exceed $20\%$ 
for most of the variables with the exception of $M_{\rm esc}$ and $L_{\rm
  esc}$. 
The difference noted in the escape fraction of mass and angular momentum ($M_{\rm esc}$ and $L_{\rm esc}$, 
respectively) for the $0.6+0.9$ and $0.9+1.2\,M_\odot$ systems is due
the different initial compositions that affect the nuclear burning.
 For the runs using $2\times 10^5$ SPH particles presented in \cite{dan11} we used a different chemical 
composition. WDs below $0.6\,M_\odot$ are made of pure He. For
$M>0.6\,M_\odot$ we have chosen mass fractions  
X(${}^{12}{\rm C}$) = 0.5 and X(${}^{16}{\rm O}$) = 0.5 uniformly distributed throughout the star, with 
the exception of the $0.5+1.2\,M_\odot$ system which has an ONeMg composition, as in the current study.
For example, if we take the run with $0.9+1.2\,M_\odot$ from \cite{dan11} we find that the nuclear energy 
produced by C-burning is about 5 times larger compared to the current study. As discussed in \S 
\ref{sec:ejecta}, the fraction of unbound mass and angular momentum increases for the runs that 
undergo dynamical burning or substantial burning.

We therefore conclude that our results are not fully converged, but
the trends we found should be robust. Effects related to nuclear
burning may be seriously underestimated due to the  low numerical resolutions. 

\subsection{Comparison with other work}
\label{sec:comparison}

While finishing this paper, a preprint by \cite{zhu13} appeared, which
explores the parameter space of the CO WD mergers.  
They discuss the structure of similar- and dissimilar-mass mergers and provide detailed information on the 
remnants' main properties.  The differences between our and their calculations are the following: (i) we are using 
tidally locked stars while they start with non-spinning stars at an initial 
separation obtained from an estimate of the size of the Roche lobe \citep{eggleton83};  (ii) we are 
including feedback from the nuclear burning and (iii) we cover the full parameter 
space of both He(CO) and CO donors while they focus on the CO donors. 

As a proxy to indicate where ignition is most likely, we compare
the enclosed mass at the location where $T=\langle T \rangle_{\rm max}$, $M(\langle T \rangle_{\rm max})$, see Figure 
\ref{fig:compzhu} (left panel) and the location where the angular velocity is half the Keplerian value, $M(\omega=\omega_K/2)$. This 
location corresponds to the inner/outer edge of the disk/core in \cite{zhu13}, see Figure \ref{fig:compzhu} (right panel). 
As in \cite{zhu13} we are using the density ratio $q_{\rho}=\rho_{\rm don,0}/\rho_{\rm acc,0}$, 
where $\rho_{\rm don,0}$ and $\rho_{\rm acc,0}$ are the central densities at the moment when mass transfer sets in for 
the donor and the accretor, respectively. 

In accordance with \cite{zhu13}, we found that the fraction of the core that is heated up increases with the mass ratio. 
However, when comparing the location in mass coordinate of the peak temperature our results do not agree.
As we have shown in Section \ref{sec:spin_dep}, the difference between our and \cite{zhu13} is caused by the different ICs. 
In the non-rotating case, the peak temperature occurs closer to the center of the remnant as the mass ratio increases 
and that for nearly equal-mass components both stars are symmetrically
disrupted. In contrast, in the corotating case the locations of
maximum temperature are located at the surface rather than the centre of the
high-density core. 
To test the trends obtained by 
\cite{zhu13}, we followed the evolution of six systems with density ratios ranging between 0.1 and 0.8 using non-rotating 
ICs (represented with black triangles in Figure \ref{fig:compzhu}). Our results from the runs with non-rotating ICs confirm 
the results of \cite{zhu13}. 

The location in mass coordinate where $\omega=\omega_K/2$ is more similar with \cite{zhu13}, increasing with the mass 
(density) ratio, although the function obtained by \cite{zhu13} has a steeper slope. This is again attributed to the different 
ICs as the corotating ICs start with more angular momentum compared to
non-rotating ICs. The results for the non-rotating ICs are
over-plotted with black  triangles and match better the fitted
formula of \cite{zhu13}.  Note that the angular
velocity and temperature are averaged differently. In our study they
are averaged over equipotential surfaces, while in \cite{zhu13}
they are cylindrically averaged. Moreover, the enclosed mass is
defined differently, in \cite{zhu13} within spherical radii, while in
our case within equipotential surfaces.

As stated before, it is not well known whether the binary systems
realized in nature are synchronized or not. If they are,  
off centre ignitions at moderate densities ($\sim 10^5\ {\rm g\
  cm^{-3}}$) are likely. If instead they are not synchronized,  
dynamical burning may start deep in the remnant core, which, due to
faster reaction rate, may lead to a more robust  
explosion mechanism.

\section{Applications}
\label{sec:applications}

Our simulations have direct applications to a number of observed stellar
systems and events, such as hot subdwarfs, RCB stars, single massive white
dwarfs, supernovae and other transient events.

\subsection{Hot subdwarfs ($M_{\rm tot}\la 0.8\,M_\odot$)}

Mergers that contain at least one He WD and whose combined mass is
larger than $\sim 0.38 - 0.45\,M_\odot$ \citep[depending on the mass of the
more massive WD;][]{han03} will ignite He, generally in a
shell, and experience a series of He shell flashes that propagate
inwards (in the case of two He WD) until helium burns in the core
\citep{saio00}.  
During these He shell flashes, the envelope
expands and may attain giant dimensions, temporarily resembling an RCB
star \citep{saio00,justham11}. Once helium burns
steadily in the core (for $\sim 10^8$\,yr), the merged object will be
relatively compact (with a radius $\sim 0.1\,R_\odot$) and have the
appearance of a hot subdwarf. If the mass of the merger product is
$\la 0.8\,M_\odot$ \citep[see, e.g.,][]{paczynski71,iben85}, 
the object will remain compact after He core burning and ultimately
become a WD, a hybrid HeCO WD with a substantial He envelope \citep{iben85}.

The merger of two He WDs is believed to be the main channel to produce
single sdB stars with a typical Galactic formation rate of
$0.1\,$yr$^{-1}$ \citep[e.g.][]{tutukov90,nelemans01a,han03}. 
If the merger involves one (or possibly
two) HeCO WDs, the merger product will generally be more massive and
hotter. These merger products are excellent candidates for He-rich sdO
stars \citep{stroeer07,justham11} and may make up up
to $\sim 10$\,\% of the subdwarf population
\citep{nelemans01a,justham11}. 
Interestingly, \cite{stroeer07} find that
He-rich sdO stars come in different varieties and can be C and/or
N-rich, perhaps indicating different merger histories.

\subsection{RCB and EHe stars ($0.8\la M_{\rm tot}/M_\odot \le 1.4$)}

Merger products that contain a CO core and a He-rich layer and have a
mass larger than $\sim 0.8\,M_\odot$ will expand and become He giants
\citep{paczynski71,iben85}. This is believed to be the
main channel to produce extreme He stars (EHe stars) and RCB stars
\citep{webbink84,iben85}. Our simulations show that there
are at least two different channels to produce such objects with very
different chemical profiles and surface abundances. The merger of a He
+ CO WD \citep[the ``classical'' channel to produce RCB stars with an
estimated Galactic birth rate of $\sim 0.01\,$yr$^{-1}$;
e.g.,][]{iben96,han98} leads to a He-shell burning star
with a massive He envelope but low C and O abundance. In contrast, the
merger of a CO WD and a HeCO WD (and possibly the merger of two HeCO
WDs) produces an object with a large surface overabundance of C and O
(see Figure~\ref{fig:mixing_mass_fractions}). This may explain the large
observed abundances of C
and O in EHe stars \citep[and to a similar, but somewhat lower extent in RCB
stars;]{pandey01,saio02} without requiring
extreme mixing in the progenitor of the CO WD on the asymptotic giant
branch (cf. \citealt{herwig00}; also see \citealt{menon13}). Indeed,
different merger histories could account, at least in part, for the
apparent differences in the surface abundances of EHe and RCB stars
\citep[see the discussions in][]{pandey01,saio02}.

\subsection{Massive single white dwarfs ($M_{\rm tot} \le 1.4\,M_\odot$)}
\label{sec:massive}

The merger of two CO WDs with a combined mass below the Chandrasekhar
limit is expected to lead to the formation of a single CO white dwarf,
(assuming that carbon is not ignited in the merger event; see
\S~\ref{sec:COmergers}). The resulting WD may be substantially more
massive than any CO WD that can form from a single star ($\sim
1.1\,M_\odot$).  An excellent candidate is the rapidly rotating WD RE
J0317-853 \citep{barstow95} which appears to have a mass close to
the Chandrasekhar mass \citep{kuelebi10}.  RE J0317-853 is also
one of the most magnetic WDs known. This could be a direct consequence
of the merger process in which differential rotation may drive a
magnetic dynamo amplifying any seed magnetic field \citep[also see][]{tout08}.

\subsection{Helium detonations and sub-Chandrasekhar explosions
($M_{\rm tot} \le 1.4\,M_\odot$)}
\label{sec:helium}

Many of our simulations that contain a He-rich WD achieve conditions
in the $\rho$\,--$T$ plane where helium may be expected to detonate
(see Fig.~12). This can lead to a rather bright explosion and could
easily be mistaken for a faint supernova \citep{nomoto82a}. Indeed, as
such detonations produce elements similar to a SN Ia, they may
resemble a faint SN Ia and are therefore sometimes referred to as SNe .Ia
\citep{bildsten07}. The main difference is that the underlying CO
white dwarf is likely to survive. On the other hand, if the helium
detonation drives a sufficiently strong shock into the CO core, this
could trigger a second detonation near the core, a carbon detonation,
which would then destroy the whole white dwarf and produce a SN
Ia-like event. Such double-detonation models are a possible scenario
for sub-Chandrasekhar SNe Ia \citep{woosley94,fink10}. Even though
such double detonations are only likely to
occur for the more massive merger products (see Figure~\ref{fig:thermo}), they are
sufficiently common that they could easily produce a large fraction,
if not all, SNe Ia. One problem with this scenario is that the
simulated spectra \citep{kromer10}, using standard assumptions,
are too red to be consistent with observed SNe Ia. However,
\cite{kromer10} 
also show that this problem could be solved if the He
layer contained a large amount of carbon ($\sim 34$\,\% by
mass). Interestingly, this is close to the C abundance found in our
simulations in the outer He-rich layer if the merger involves a hybrid
HeCO WD (see Figure~\ref{fig:mixing_mass_fractions}), perhaps
providing some support for the \cite{kromer10} suggestion.

\subsection{CO + CO WD mergers: SN Ia or AIC ($M_{\rm tot} > 1.4\,M_\odot$)}
\label{sec:COmergers}

The mergers of two CO WDs with a total mass above the Chandrasekhar
mass limit is one of the main channels to produce a Type Ia supernova
(SN Ia) \citep{iben84,webbink84}.

However, it has been argued that in the WD merger scenario off-center
C burning transforms the CO WD into an ONeMg WD that will undergo
accretion induced collapse (AIC) to form a neutron star
\citep{nomoto85}. \cite{yoon07} found (cf.\ \S~\ref{sec:burning}) that
there may be some parameter range for which neutrino cooling prevents
the conversion of the CO WD into an ONeMg WD; in this case, carbon
will be ignited in the center, leading to a SN Ia, $\sim 10^5$ yr
after the merger itself.

In the case of nearly equal mass CO WDs, where both WDs are disrupted
and merge, C may detonate in the merging process \citep{pakmor10}.  In
our simulations, unlike those by \cite{pakmor10}, this is only likely
to happen for the most massive WDs (see Figure~\ref{fig:thermo})
because of the different initial assumed WD spins (see the discussion
in \S~\ref{sec:spin_dep}). The most massive mergers (with $M_{\rm
  tot}\ge 2.1 \,M_\odot$) may detonate during the merger and therefore
are potential candidates to explain the so-called super-Chandrasekhar
SN Ia explosions, which are much brighter than normal SNe Ia
\citep[e.g., SNLS-03D3bb,][]{howell06}. As such massive systems will
be rare, this would naturally explain their low observed frequency.

\subsection{Hybrid supernovae  ($M_{\rm tot} > 1.4\,M_\odot$)}
\label{sec:hybrid}

Our simulations also include cases where the combined mass exceeds the
Chandrasekhar mass but does not involve two CO WDs. In cases, where at
least one of the WDs contains a He-rich layer, it is possible, perhaps
even likely, that the He detonates leading to the ejection of this
layer (see Figure~\ref{fig:thermo} and \S~\ref{sec:helium}). In cases where the
post-merger object remains above the Chandrasekhar mass, it will
ultimately experience a SN \citep[e.g.][]{iben85}, either a
thermonuclear explosion (if C is ignited in the centre) or core
collapse (in the case of an ONeMg core). This may give rise to new
types of supernova events. For example, in the case of the collapse of
an ONeMg core to a neutron star with a significant He layer, the
explosion would technically be a core-collapse supernova, but as such
explosions \citep[initiated by the capture of electrons onto Ne and
Mg][]{nomoto84} are expected to be weak
\citep{podsiadlowski04,kitaura06}, most of the explosion energy may 
come from He burning, i.e., be thermonuclear in nature. Such hybrid
supernovae (observationally classified as a weak SN Ib) are potential
candidates for the class of ``Ca-rich'' SNe Ib \citep[such as SN
2005E][]{perets10} as the burning conditions are likely to favor 
the production of intermediate-mass elements
\citep[e.g.][]{waldman11}\footnote{He detonations on a CO or ONeMg
  core themselves 
 provide an alternative scenario for this supernova type
 \citep{waldman11}.}.  Ca-rich SNe as a class are preferentially found
in the outer parts of galaxies, far away from star-forming regions, implying
old and possibly metal-poor progenitor systems \citep{yuan13}.
This could be consistent with a merger origin if the merger requires a
long time delay to bring the systems into contact.

The merger of a CO and ONeMg WD could be another hybrid SN,
where the ONeMg core may collapse but carbon may or may not
detonate. Depending on the details, this could lead to either an
unusual SN Ia or a (faint?) SN Ic. Such mergers will be rare but
should exist in nature.

\section{Summary}
\label{sec:summary}

In this paper, we have investigated the merger properties of more than
200 WD-WD systems covering a large range of WD mass and chemical
compositions. 
We have determined the structure of the WD remnants as a function of
the total mass and the mass ratio of the binary. We distinguish
between four main regions: a cold core surrounded by  
a hot, thermally supported envelope, a Keplerian disk and a tidal
tail. The degree of heating experienced by the accretor is closely
related to the degree of mixing of the two stars, both increasing with
the mass ratio. Consequently, the fraction of the accretor that remains
cold (core) tends to decrease with increasing mass ratio
while the fraction of mass outside this region (combined mass of
the hot envelope and the
centrifugally supported disk) tends to increase with the mass ratio. 
Systems with lower mass ratio have more total angular momentum and 
as a result more mass is flung out in a tidal tail. Most of this
material will fall back at a later time onto the disk. 

However, the remnant's structure is strongly dependent on the nature
of  the initial conditions. We made a detailed comparison with the previous parameter
study of CO mass-transferring systems of \cite{zhu13} and found that the
differences in the remnants' structure are due to the different
initial conditions. When using the non-rotating initial stars as in \cite{zhu13}, 
we found that for nearly equal mass mergers both stars are symmetrically
disrupted and the location of peak temperature is at the remnant's
center. With our corotating initial conditions the location of peak
temperature is close (within 15\,\%)  to the surface of the former accretor. The 
initial conditions are also responsible for the difference found in the rotational profiles, the
corotating the initial conditions lead to more angular momentum in the central remnant
compared to the non-rotating the initial conditions.  

Little mass is ejected, between $10^{-4}$ and $3.4\times
10^{-2}\,M_\odot$ as the mass ratio decreases. However, this can take
up to $12\,\%$ of the system's total angular momentum having an impact
on the post-merger viscous evolution. 
In an appreciable number of cases, nuclear burning has a substantial
impact on ejected mass. These systems undergo substantial nuclear burning, such
as those with massive He(CO) donors or the very massive CO mass-transferring  
systems with $M_{\rm tot}\gtrsim 2.0\,M_\odot$.

We investigated whether hydrodynamical burning can occur after
the merger moment. Many of our simulations ($M_{\rm tot}\ga
1.1\,M_\odot$) that contain a He-rich WD 
achieve the conditions
required to trigger a He detonation.
These are roughly the same systems that are 
expected to produce surface detonations either from instabilities
within the accretion stream and/or at the moment first dynamical
contact \citep{guillochon10,dan12}. Further viscous evolution could
possibly increase the region of parameter space in which one expects a
detonation to occur \citep{schwab12}.  
The He-detonations may resemble a
SN Ia, but in this case the underlying CO white dwarf can survive. If
the He shock can trigger a second detonation in the CO core, then the
whole white dwarf is disrupted and will produce a SN Ia-like event. 

For our CO mass-transferring systems only the  most massive mergers
($M_{\rm tot}\ga 2.1\, M_\odot$) may detonate. These
systems are potential candidates to explain the so-called
super-Chandrasekhar SN Ia explosions, which are much brighter than
normal SNe Ia.
However, even systems with lower masses may still be prone to an
explosion long after the merger if the post-merger object remains
above the Chandrasekhar mass \citep[e.g.][]{iben85,yoon07} and C is
ignited in the centre. Alternatively, off center ignition of C burning
transforms the CO WD into an ONeMg WD that will undergo accretion
induced collapse (AIC) to form a neutron star \citep{nomoto85}.

Finally, we provide to the astrophysical community a large
database containing thermodynamic and rotational profiles to be used
for subsequent stellar evolution studies. The data from our large
simulation set can be downloaded from 
\verb+www.hs.uni-hamburg.de/DE/Ins/Per/Dan/wdwd_remnants+. 
In order to aid further comparisons, the general remnants properties
are tabulated, see Table \ref{tab:bigtable1} in the Appendix.

\section*{Acknowledgments}
 We thank Enrico Ramirez-Ruiz and James Guillochon for very useful
 discussions. M.D. and S.R. are supported by Deutsche
 Forschungsgemeinschaft under grants RO-3399/4-1 and RO-3399/4-2. 
M.B. acknowledges computing on the JUROPA supercomputer
at the Forschungszentrum J\"ulich under grant NIC5984 and 5056. 
Some of the figures in this paper were produced using the
visualization software SPLASH developed by Daniel \cite{price07}. 

\bibliographystyle{mn2e}
\bibliography{wd_remnants}

\begin{thebibliography}{103}
\expandafter\ifx\csname natexlab\endcsname\relax\def\natexlab#1{#1}\fi

\bibitem[{{Amaro-Seoane} {et~al}\mbox{.}(2012){Amaro-Seoane}, {Aoudia},
  {Babak}, {Bin{\'e}truy}, {Berti}, {Boh{\'e}}, {Caprini}, {Colpi}, {Cornish},
  {Danzmann}, {Dufaux}, {Gair}, {Jennrich}, {Jetzer}, {Klein}, {Lang}, {Lobo},
  {Littenberg}, {McWilliams}, {Nelemans}, {Petiteau}, {Porter}, {Schutz},
  {Sesana}, {Stebbins}, {Sumner}, {Vallisneri}, {Vitale}, {Volonteri}, \&
  {Ward}}]{amaro12}
{Amaro-Seoane} P. {et~al.}, 2012, Classical and Quantum Gravity, 29, 124016

\bibitem[{{Balsara}(1995)}]{Balsara95}
{Balsara} D.~S., 1995, Journal of Computational Physics, 121, 357

\bibitem[{{Barstow} {et~al}\mbox{.}(1995){Barstow}, {Jordan}, {O'Donoghue},
  {Burleigh}, {Napiwotzki}, \& {Harrop-Allin}}]{barstow95}
{Barstow} M.~A., {Jordan} S., {O'Donoghue} D., {Burleigh} M.~R., {Napiwotzki}
  R., {Harrop-Allin} M.~K., 1995, MNRAS, 277, 971

\bibitem[{{Benz} {et~al}\mbox{.}(1990){Benz}, {Cameron}, {Press}, \&
  {Bowers}}]{benz90}
{Benz} W., {Cameron} A.~G.~W., {Press} W.~H., {Bowers} R.~L., 1990, ApJ, 348,
  647

\bibitem[{{Bergeron} {et~al}\mbox{.}(1991){Bergeron}, {Kidder}, {Holberg},
  {Liebert}, {Wesemael}, \& {Saffer}}]{bergeron91}
{Bergeron} P., {Kidder} K.~M., {Holberg} J.~B., {Liebert} J., {Wesemael} F.,
  {Saffer} R.~A., 1991, ApJ, 372, 267

\bibitem[{{Bildsten} {et~al}\mbox{.}(2007){Bildsten}, {Shen}, {Weinberg}, \&
  {Nelemans}}]{bildsten07}
{Bildsten} L., {Shen} K.~J., {Weinberg} N.~N., {Nelemans} G., 2007, ApJL, 662,
  L95

\bibitem[{{Blinnikov} \& {Khokhlov}(1987)}]{blinnikov87}
{Blinnikov} S.~I., {Khokhlov} A.~M., 1987, Soviet Astronomy Letters, 13, 364

\bibitem[{{Brown} {et~al}\mbox{.}(2011){Brown}, {Kilic}, {Hermes}, {Allende
  Prieto}, {Kenyon}, \& {Winget}}]{brown11b}
{Brown} W.~R., {Kilic} M., {Hermes} J.~J., {Allende Prieto} C., {Kenyon} S.~J.,
  {Winget} D.~E., 2011, ApJL, 737, L23

\bibitem[{{Burkart} {et~al}\mbox{.}(2012){Burkart}, {Quataert}, {Arras}, \&
  {Weinberg}}]{burkart12}
{Burkart} J., {Quataert} E., {Arras} P., {Weinberg} N.~N., 2012, ArXiv e-prints

\bibitem[{{Clayton}(1968)}]{clayton68}
{Clayton} D.~D., 1968, {Principles of stellar evolution and nucleosynthesis}

\bibitem[{{Clayton}(2012)}]{clayton12}
{Clayton} G.~C., 2012, Journal of the American Association of Variable Star
  Observers (JAAVSO), 40, 539

\bibitem[{{Dan} {et~al}\mbox{.}(2011){Dan}, {Rosswog}, {Guillochon}, \&
  {Ramirez-Ruiz}}]{dan11}
{Dan} M., {Rosswog} S., {Guillochon} J., {Ramirez-Ruiz} E., 2011, ApJ, 737, 89

\bibitem[{{Dan} {et~al}\mbox{.}(2012){Dan}, {Rosswog}, {Guillochon}, \&
  {Ramirez-Ruiz}}]{dan12}
{Dan} M., {Rosswog} S., {Guillochon} J., {Ramirez-Ruiz} E., 2012, MNRAS, 422,
  2417

\bibitem[{{Eggleton}(1983)}]{eggleton83}
{Eggleton} P.~P., 1983, ApJ, 268, 368

\bibitem[{{Fink} {et~al}\mbox{.}(2010){Fink}, {R{\"o}pke}, {Hillebrandt},
  {Seitenzahl}, {Sim}, \& {Kromer}}]{fink10}
{Fink} M., {R{\"o}pke} F.~K., {Hillebrandt} W., {Seitenzahl} I.~R., {Sim}
  S.~A., {Kromer} M., 2010, A\&A, 514, A53

\bibitem[{{Fowler} {et~al}\mbox{.}(1975){Fowler}, {Caughlan}, \&
  {Zimmerman}}]{fowler75}
{Fowler} W.~A., {Caughlan} G.~R., {Zimmerman} B.~A., 1975, ARA\&A, 13, 69

\bibitem[{{Frank-Kamenetskii}(1967)}]{frank67}
{Frank-Kamenetskii} D., 1967, {Diffusion and Heat Transfer in Chemical
  Kinetic}. Nauka, Moscow

\bibitem[{{Fryer} {et~al}\mbox{.}(2010){Fryer}, {Ruiter}, {Belczynski},
  {Brown}, {Bufano}, {Diehl}, {Fontes}, {Frey}, {Holland}, {Hungerford},
  {Immler}, {Mazzali}, {Meakin}, {Milne}, {Raskin}, \& {Timmes}}]{fryer10}
{Fryer} C.~L. {et~al.}, 2010, ApJ, 725, 296

\bibitem[{{Fuller} \& {Lai}(2012)}]{fuller12}
{Fuller} J., {Lai} D., 2012, MNRAS, 421, 426

\bibitem[{{Gokhale} {et~al}\mbox{.}(2007){Gokhale}, {Peng}, \&
  {Frank}}]{gokhale07}
{Gokhale} V., {Peng} X.~M., {Frank} J., 2007, ApJ, 655, 1010

\bibitem[{{Guerrero} {et~al}\mbox{.}(2004){Guerrero}, {Garc{\'{\i}}a-Berro}, \&
  {Isern}}]{guerrero04}
{Guerrero} J., {Garc{\'{\i}}a-Berro} E., {Isern} J., 2004, A\&A, 413, 257

\bibitem[{{Guillochon} {et~al}\mbox{.}(2010){Guillochon}, {Dan},
  {Ramirez-Ruiz}, \& {Rosswog}}]{guillochon10}
{Guillochon} J., {Dan} M., {Ramirez-Ruiz} E., {Rosswog} S., 2010, ApJL, 709,
  L64

\bibitem[{{Han}(1998)}]{han98}
{Han} Z., 1998, MNRAS, 296, 1019

\bibitem[{{Han} {et~al}\mbox{.}(2003){Han}, {Podsiadlowski}, {Maxted}, \&
  {Marsh}}]{han03}
{Han} Z., {Podsiadlowski} P., {Maxted} P.~F.~L., {Marsh} T.~R., 2003, MNRAS,
  341, 669

\bibitem[{{Han} {et~al}\mbox{.}(2002){Han}, {Podsiadlowski}, {Maxted}, {Marsh},
  \& {Ivanova}}]{han02}
{Han} Z., {Podsiadlowski} P., {Maxted} P.~F.~L., {Marsh} T.~R., {Ivanova} N.,
  2002, MNRAS, 336, 449

\bibitem[{{Heber}(2009)}]{heber09}
{Heber} U., 2009, ARA\&A, 47, 211

\bibitem[{{Herwig}(2000)}]{herwig00}
{Herwig} F., 2000, A\&A, 360, 952

\bibitem[{{Hix} {et~al}\mbox{.}(1998){Hix}, {Khokhlov}, {Wheeler}, \&
  {Thielemann}}]{hix98}
{Hix} W.~R., {Khokhlov} A.~M., {Wheeler} J.~C., {Thielemann} F.-K., 1998, ApJ,
  503, 332

\bibitem[{{Howell}(2011)}]{howell11}
{Howell} D.~A., 2011, Nature Communications, 2

\bibitem[{{Howell} {et~al}\mbox{.}(2006){Howell}, {Sullivan}, {Nugent},
  {Ellis}, {Conley}, {Le Borgne}, {Carlberg}, {Guy}, {Balam}, {Basa},
  {Fouchez}, {Hook}, {Hsiao}, {Neill}, {Pain}, {Perrett}, \&
  {Pritchet}}]{howell06}
{Howell} D.~A. {et~al.}, 2006, Nature, 443, 308

\bibitem[{{Iben} \& {Tutukov}(1984)}]{iben84}
{Iben}, Jr. I., {Tutukov} A.~V., 1984, ApJS, 54, 335

\bibitem[{{Iben} \& {Tutukov}(1985)}]{iben85}
{Iben}, Jr. I., {Tutukov} A.~V., 1985, ApJS, 58, 661

\bibitem[{{Iben} {et~al}\mbox{.}(1996){Iben}, {Tutukov}, \&
  {Yungelson}}]{iben96}
{Iben}, Jr. I., {Tutukov} A.~V., {Yungelson} L.~R., 1996, ApJ, 456, 750

\bibitem[{{Jeffery} {et~al}\mbox{.}(2011){Jeffery}, {Karakas}, \&
  {Saio}}]{jeffery11}
{Jeffery} C.~S., {Karakas} A.~I., {Saio} H., 2011, MNRAS, 414, 3599

\bibitem[{{Justham} {et~al}\mbox{.}(2011){Justham}, {Podsiadlowski}, \&
  {Han}}]{justham11}
{Justham} S., {Podsiadlowski} P., {Han} Z., 2011, MNRAS, 410, 984

\bibitem[{{Kilic} {et~al}\mbox{.}(2011){Kilic}, {Brown}, {Allende Prieto},
  {Ag{\"u}eros}, {Heinke}, \& {Kenyon}}]{kilic11}
{Kilic} M., {Brown} W.~R., {Allende Prieto} C., {Ag{\"u}eros} M.~A., {Heinke}
  C., {Kenyon} S.~J., 2011, ApJ, 727, 3

\bibitem[{{Kippenhahn} \& {Weigert}(1990)}]{kippenhahn90}
{Kippenhahn} R., {Weigert} A., 1990, {Stellar Structure and Evolution}

\bibitem[{{Kitaura} {et~al}\mbox{.}(2006){Kitaura}, {Janka}, \&
  {Hillebrandt}}]{kitaura06}
{Kitaura} F.~S., {Janka} H.-T., {Hillebrandt} W., 2006, A\&A, 450, 345

\bibitem[{{Kromer} {et~al}\mbox{.}(2010){Kromer}, {Sim}, {Fink}, {R{\"o}pke},
  {Seitenzahl}, \& {Hillebrandt}}]{kromer10}
{Kromer} M., {Sim} S.~A., {Fink} M., {R{\"o}pke} F.~K., {Seitenzahl} I.~R.,
  {Hillebrandt} W., 2010, ApJ, 719, 1067

\bibitem[{{K{\"u}lebi} {et~al}\mbox{.}(2010){K{\"u}lebi}, {Jordan}, {Nelan},
  {Bastian}, \& {Altmann}}]{kuelebi10}
{K{\"u}lebi} B., {Jordan} S., {Nelan} E., {Bastian} U., {Altmann} M., 2010,
  A\&A, 524, A36

\bibitem[{{Lor{\'e}n-Aguilar} {et~al}\mbox{.}(2009){Lor{\'e}n-Aguilar},
  {Isern}, \& {Garc{\'i}a-Berro}}]{aguilar09}
{Lor{\'e}n-Aguilar} P., {Isern} J., {Garc{\'i}a-Berro} E., 2009, A\&A, 500,
  1193

\bibitem[{{Marsh}(2011)}]{marsh11}
{Marsh} T.~R., 2011, Classical and Quantum Gravity, 28, 094019

\bibitem[{{Marsh} {et~al}\mbox{.}(2004){Marsh}, {Nelemans}, \&
  {Steeghs}}]{marsh04}
{Marsh} T.~R., {Nelemans} G., {Steeghs} D., 2004, MNRAS, 350, 113

\bibitem[{{Menon} {et~al}\mbox{.}(2013){Menon}, {Herwig}, {Denissenkov},
  {Clayton}, {Staff}, {Pignatari}, \& {Paxton}}]{menon13}
{Menon} A., {Herwig} F., {Denissenkov} P.~A., {Clayton} G.~C., {Staff} J.,
  {Pignatari} M., {Paxton} B., 2013, ApJ, 772, 59

\bibitem[{{Monaghan}(2005)}]{monaghan05}
{Monaghan} J.~J., 2005, Reports on Progress in Physics, 68, 1703

\bibitem[{{Morris} \& {Monaghan}(1997)}]{morris97}
{Morris} J.~P., {Monaghan} J.~J., 1997, Journal of Computational Physics, 136,
  41

\bibitem[{{Napiwotzki}(2009)}]{napiwotzki09}
{Napiwotzki} R., 2009, Journal of Physics Conference Series, 172, 012004

\bibitem[{{Napiwotzki} {et~al}\mbox{.}(2004){Napiwotzki}, {Yungelson},
  {Nelemans}, {Marsh}, {Leibundgut}, {Renzini}, {Homeier}, {Koester},
  {Moehler}, {Christlieb}, {Reimers}, {Drechsel}, {Heber}, {Karl}, \&
  {Pauli}}]{napiwotzki04}
{Napiwotzki} R. {et~al.}, 2004, in Astronomical Society of the Pacific
  Conference Series, Vol. 318, Spectroscopically and Spatially Resolving the
  Components of the Close Binary Stars, {Hilditch} R.~W., {Hensberge} H.,
  {Pavlovski} K., eds., pp. 402--410

\bibitem[{{Nelemans}(2005)}]{nelemans05}
{Nelemans} G., 2005, in Astronomical Society of the Pacific Conference Series,
  Vol. 330, The Astrophysics of Cataclysmic Variables and Related Objects,
  {J.-M.~Hameury \& J.-P.~Lasota}, ed., p.~27

\bibitem[{{Nelemans}(2009)}]{nelemans09}
{Nelemans} G., 2009, Classical and Quantum Gravity, 26, 094030

\bibitem[{{Nelemans} {et~al}\mbox{.}(2001){Nelemans}, {Yungelson}, {Portegies
  Zwart}, \& {Verbunt}}]{nelemans01a}
{Nelemans} G., {Yungelson} L.~R., {Portegies Zwart} S.~F., {Verbunt} F., 2001,
  A\&A, 365, 491

\bibitem[{{Nomoto}(1982{\natexlab{a}})}]{nomoto82}
{Nomoto} K., 1982{\natexlab{a}}, ApJ, 257, 780

\bibitem[{{Nomoto}(1982{\natexlab{b}})}]{nomoto82a}
{Nomoto} K., 1982{\natexlab{b}}, ApJ, 253, 798

\bibitem[{{Nomoto}(1984)}]{nomoto84}
{Nomoto} K., 1984, ApJ, 277, 791

\bibitem[{{Nomoto} \& {Iben}(1985)}]{nomoto85}
{Nomoto} K., {Iben}, Jr. I., 1985, ApJ, 297, 531

\bibitem[{{Nomoto} \& {Kondo}(1991)}]{nomoto91}
{Nomoto} K., {Kondo} Y., 1991, ApJL, 367, L19

\bibitem[{{Paczy{\'n}ski}(1971)}]{paczynski71}
{Paczy{\'n}ski} B., 1971, Acta Astronomica, 21, 1

\bibitem[{{Pakmor} {et~al}\mbox{.}(2011){Pakmor}, {Hachinger}, {R{\"o}pke}, \&
  {Hillebrandt}}]{pakmor11}
{Pakmor} R., {Hachinger} S., {R{\"o}pke} F.~K., {Hillebrandt} W., 2011, A\&A,
  528, A117+

\bibitem[{{Pakmor} {et~al}\mbox{.}(2010){Pakmor}, {Kromer}, {R{\"o}pke}, {Sim},
  {Ruiter}, \& {Hillebrandt}}]{pakmor10}
{Pakmor} R., {Kromer} M., {R{\"o}pke} F.~K., {Sim} S.~A., {Ruiter} A.~J.,
  {Hillebrandt} W., 2010, Nature, 463, 61

\bibitem[{{Pandey} {et~al}\mbox{.}(2001){Pandey}, {Kameswara Rao}, {Lambert},
  {Jeffery}, \& {Asplund}}]{pandey01}
{Pandey} G., {Kameswara Rao} N., {Lambert} D.~L., {Jeffery} C.~S., {Asplund}
  M., 2001, MNRAS, 324, 937

\bibitem[{{Parsons} {et~al}\mbox{.}(2011){Parsons}, {Marsh}, {G{\"a}nsicke},
  {Drake}, \& {Koester}}]{parsons11}
{Parsons} S.~G., {Marsh} T.~R., {G{\"a}nsicke} B.~T., {Drake} A.~J., {Koester}
  D., 2011, ApJL, 735, L30

\bibitem[{{Perets} {et~al}\mbox{.}(2010){Perets}, {Gal-Yam}, {Mazzali},
  {Arnett}, {Kagan}, {Filippenko}, {Li}, {Arcavi}, {Cenko}, {Fox}, {Leonard},
  {Moon}, {Sand}, {Soderberg}, {Anderson}, {James}, {Foley}, {Ganeshalingam},
  {Ofek}, {Bildsten}, {Nelemans}, {Shen}, {Weinberg}, {Metzger}, {Piro},
  {Quataert}, {Kiewe}, \& {Poznanski}}]{perets10}
{Perets} H.~B. {et~al.}, 2010, Nature, 465, 322

\bibitem[{{Podsiadlowski} {et~al}\mbox{.}(2004){Podsiadlowski}, {Langer},
  {Poelarends}, {Rappaport}, {Heger}, \& {Pfahl}}]{podsiadlowski04}
{Podsiadlowski} P., {Langer} N., {Poelarends} A.~J.~T., {Rappaport} S., {Heger}
  A., {Pfahl} E., 2004, ApJ, 612, 1044

\bibitem[{{Price}(2007)}]{price07}
{Price} D.~J., 2007, PASA, 24, 159

\bibitem[{{Price} \& {Rosswog}(2006)}]{price06}
{Price} D.~J., {Rosswog} S., 2006, Science, 312, 719

\bibitem[{{Rasio} \& {Shapiro}(1999)}]{rasio99}
{Rasio} F.~A., {Shapiro} S.~L., 1999, Classical and Quantum Gravity, 16, 1

\bibitem[{{Raskin} {et~al}\mbox{.}(2011){Raskin}, {Scannapieco}, {Fryer},
  {Rockefeller\ }, \& {Timmes}}]{raskin11}
{Raskin} C., {Scannapieco} E., {Fryer} C., {Rockefeller\ } G., {Timmes} F.~X.,
  2011, ArXiv e-prints

\bibitem[{{Rosswog}(2007)}]{rosswog07}
{Rosswog} S., 2007, MNRAS, 376, L48

\bibitem[{{Rosswog}(2009)}]{rosswog09}
{Rosswog} S., 2009, NewAR, 53, 78

\bibitem[{{Rosswog} \& {Davies}(2002)}]{rosswog02}
{Rosswog} S., {Davies} M.~B., 2002, MNRAS, 334, 481

\bibitem[{{Rosswog} {et~al}\mbox{.}(2000){Rosswog}, {Davies}, {Thielemann}, \&
  {Piran}}]{rosswog00}
{Rosswog} S., {Davies} M.~B., {Thielemann} F.-K., {Piran} T., 2000, A\&A, 360,
  171

\bibitem[{{Rosswog} {et~al}\mbox{.}(2009{\natexlab{a}}){Rosswog}, {Kasen},
  {Guillochon}, \& {Ramirez-Ruiz}}]{rosswog09c}
{Rosswog} S., {Kasen} D., {Guillochon} J., {Ramirez-Ruiz} E.,
  2009{\natexlab{a}}, ApJL, 705, L128

\bibitem[{{Rosswog} {et~al}\mbox{.}(1999){Rosswog}, {Liebend{\"o}rfer},
  {Thielemann}, {Davies}, {Benz}, \& {Piran}}]{rosswog99}
{Rosswog} S., {Liebend{\"o}rfer} M., {Thielemann} F., {Davies} M.~B., {Benz}
  W., {Piran} T., 1999, A\&A, 341, 499

\bibitem[{{Rosswog} {et~al}\mbox{.}(2013){Rosswog}, {Piran}, \&
  {Nakar}}]{rosswog13}
{Rosswog} S., {Piran} T., {Nakar} E., 2013, MNRAS, 430, 2585

\bibitem[{{Rosswog} {et~al}\mbox{.}(2009{\natexlab{b}}){Rosswog},
  {Ramirez-Ruiz}, \& {Hix}}]{rosswog09b}
{Rosswog} S., {Ramirez-Ruiz} E., {Hix} W.~R., 2009{\natexlab{b}}, ApJ, 695, 404

\bibitem[{{Rosswog} {et~al}\mbox{.}(2008){Rosswog}, {Ramirez-Ruiz}, {Hix}, \&
  {Dan}}]{rosswog08}
{Rosswog} S., {Ramirez-Ruiz} E., {Hix} W.~R., {Dan} M., 2008, Computer Physics
  Communications, 179, 184

\bibitem[{{Rosswog} {et~al}\mbox{.}(2004){Rosswog}, {Speith}, \&
  {Wynn}}]{rosswog04}
{Rosswog} S., {Speith} R., {Wynn} G.~A., 2004, MNRAS, 351, 1121

\bibitem[{{Ruffert} {et~al}\mbox{.}(1996){Ruffert}, {Janka}, \&
  {Schaefer}}]{ruffert96}
{Ruffert} M., {Janka} H.-T., {Schaefer} G., 1996, A\&A, 311, 532

\bibitem[{{Saio} \& {Jeffery}(2000)}]{saio00}
{Saio} H., {Jeffery} C.~S., 2000, MNRAS, 313, 671

\bibitem[{{Saio} \& {Jeffery}(2002)}]{saio02}
{Saio} H., {Jeffery} C.~S., 2002, MNRAS, 333, 121

\bibitem[{{Saio} \& {Nomoto}(1985)}]{saio85}
{Saio} H., {Nomoto} K., 1985, A\&A, 150, L21

\bibitem[{{Saio} \& {Nomoto}(2004)}]{saio04}
{Saio} H., {Nomoto} K., 2004, ApJ, 615, 444

\bibitem[{{Salpeter}(1954)}]{salpeter54}
{Salpeter} E.~E., 1954, Australian Journal of Physics, 7, 373

\bibitem[{{Schwab} {et~al}\mbox{.}(2012){Schwab}, {Shen}, {Quataert}, {Dan}, \&
  {Rosswog}}]{schwab12}
{Schwab} J., {Shen} K.~J., {Quataert} E., {Dan} M., {Rosswog} S., 2012, MNRAS,
  427, 190

\bibitem[{{Segretain} {et~al}\mbox{.}(1997){Segretain}, {Chabrier}, \&
  {Mochkovitch}}]{segretain97}
{Segretain} L., {Chabrier} G., {Mochkovitch} R., 1997, ApJ, 481, 355

\bibitem[{{Shen} {et~al}\mbox{.}(2010){Shen}, {Kasen}, {Weinberg}, {Bildsten},
  \& {Scannapieco}}]{shen10}
{Shen} K.~J., {Kasen} D., {Weinberg} N.~N., {Bildsten} L., {Scannapieco} E.,
  2010, ApJ, 715, 767

\bibitem[{{Sim} {et~al}\mbox{.}(2010){Sim}, {R{\"o}pke}, {Hillebrandt},
  {Kromer}, {Pakmor}, {Fink}, {Ruiter}, \& {Seitenzahl}}]{sim10}
{Sim} S.~A., {R{\"o}pke} F.~K., {Hillebrandt} W., {Kromer} M., {Pakmor} R.,
  {Fink} M., {Ruiter} A.~J., {Seitenzahl} I.~R., 2010, ApJL, 714, L52

\bibitem[{{Steinfadt} {et~al}\mbox{.}(2010){Steinfadt}, {Kaplan}, {Shporer},
  {Bildsten}, \& {Howell}}]{steinfadt10}
{Steinfadt} J.~D.~R., {Kaplan} D.~L., {Shporer} A., {Bildsten} L., {Howell}
  S.~B., 2010, ApJL, 716, L146

\bibitem[{{Stroeer} {et~al}\mbox{.}(2007){Stroeer}, {Heber}, {Lisker},
  {Napiwotzki}, {Dreizler}, {Christlieb}, \& {Reimers}}]{stroeer07}
{Stroeer} A., {Heber} U., {Lisker} T., {Napiwotzki} R., {Dreizler} S.,
  {Christlieb} N., {Reimers} D., 2007, A\&A, 462, 269

\bibitem[{{Taam}(1980)}]{taam80}
{Taam} R.~E., 1980, ApJ, 237, 142

\bibitem[{{Timmes} \& {Swesty}(2000)}]{timmes00}
{Timmes} F.~X., {Swesty} F.~D., 2000, ApJS, 126, 501

\bibitem[{{Tout} {et~al}\mbox{.}(2008){Tout}, {Wickramasinghe}, {Liebert},
  {Ferrario}, \& {Pringle}}]{tout08}
{Tout} C.~A., {Wickramasinghe} D.~T., {Liebert} J., {Ferrario} L., {Pringle}
  J.~E., 2008, MNRAS, 387, 897

\bibitem[{{Tutukov} \& {Yungelson}(1990)}]{tutukov90}
{Tutukov} A.~V., {Yungelson} L.~R., 1990, A. Zh., 67, 109

\bibitem[{{van Kerkwijk} {et~al}\mbox{.}(2010){van Kerkwijk}, {Chang}, \&
  {Justham}}]{vankerkwijk10}
{van Kerkwijk} M.~H., {Chang} P., {Justham} S., 2010, ApJL, 722, L157

\bibitem[{{Vennes} {et~al}\mbox{.}(2011){Vennes}, {Thorstensen}, {Kawka},
  {N{\'e}meth}, {Skinner}, {Pigulski}, {St{\c e}{\' s}licki},
  {Ko{\l}aczkowski}, \& {{\'S}r{\'o}dka}}]{vennes11}
{Vennes} S. {et~al.}, 2011, ApJL, 737, L16

\bibitem[{{Waldman} {et~al}\mbox{.}(2011){Waldman}, {Sauer}, {Livne}, {Perets},
  {Glasner}, {Mazzali}, {Truran}, \& {Gal-Yam}}]{waldman11}
{Waldman} R., {Sauer} D., {Livne} E., {Perets} H., {Glasner} A., {Mazzali} P.,
  {Truran} J.~W., {Gal-Yam} A., 2011, ApJ, 738, 21

\bibitem[{{Webbink}(1984)}]{webbink84}
{Webbink} R.~F., 1984, ApJ, 277, 355

\bibitem[{{Woosley} \& {Kasen}(2011)}]{woosley11}
{Woosley} S.~E., {Kasen} D., 2011, ApJ, 734, 38

\bibitem[{{Woosley} \& {Weaver}(1994)}]{woosley94}
{Woosley} S.~E., {Weaver} T.~A., 1994, ApJ, 423, 371

\bibitem[{{Yoon} {et~al}\mbox{.}(2007){Yoon}, {Podsiadlowski}, \&
  {Rosswog}}]{yoon07}
{Yoon} S., {Podsiadlowski} P., {Rosswog} S., 2007, MNRAS, 380, 933

\bibitem[{{Yuan} {et~al}\mbox{.}(2013){Yuan}, {Kobayashi}, {Schmidt},
  {Podsiadlowski}, {Sim}, \& {Scalzo}}]{yuan13}
{Yuan} F., {Kobayashi} C., {Schmidt} B.~P., {Podsiadlowski} P., {Sim} S.~A.,
  {Scalzo} R.~A., 2013, MNRAS, 432, 1680

\bibitem[{{Zhu} {et~al}\mbox{.}(2013){Zhu}, {Chang}, {van Kerkwijk}, \&
  {Wadsley}}]{zhu13}
{Zhu} C., {Chang} P., {van Kerkwijk} M.~H., {Wadsley} J., 2013, ApJ, 767, 164

\bibitem[{{Zrake} \& {MacFadyen}(2013)}]{zrake13}
{Zrake} J., {MacFadyen} A.~I., 2013, ArXiv e-prints

\end{thebibliography}

\appendix

\setcounter{table}{0} \renewcommand{\thetable}{A\arabic{table}}

\section{Polynomial fitting functions}
\label{sec:fits}

We provide approximate formulae for $M(T_{\rm max})$, $M(\langle
T\rangle_{\rm max})$, $M(\omega_{\rm max})$, $M(\omega_k/2)$, $M_{\rm core}$, $M_{\rm env}$,
$M_{\rm disk}$, $M_{\rm fb}$, 
and $M_{\rm esc}$ as a function of $q$ and $M_{\rm tot}$, valid for
the entire parameter space under study, ie. both He and CO
mass-transferring systems. 

To measure the goodness of the fit of the estimated equations we use
the $R^2$ statistic, defined as the ratio of the sum of squares of the
residuals and the total sum of squares. $R^2$ can take values between
0 and 1 and the closer is to 1 the better the fit. For
example, an $R^2$ value of $0.7$ means that the fit explains 70\% of 
the variance.

\begin{eqnarray}
M(T_{\rm max}) &=&  M_{\rm tot}( 0.863 - 0.3335q) \label{eq:mtmax}\\
M(\langle T\rangle_{\rm max}) &=&  M_{\rm tot}(0.851 -0.319 q) \label{eq:mtmaxavg},
\end{eqnarray}
having an $R^2$ goodness-of-fit of 0.75 and 0.85, respectively. 

\begin{equation}
\omega_{\rm max} =  \omega_{\rm orb,0}(-3.476 q + 6.148)\label{eq:wmax},
\end{equation}
with $R^2$ equal to 0.6012. 

\begin{equation}
M(\omega_{\rm max}) = M_{\rm tot} (0.8684 - 0.3284 q) \label{eq:mwmax}.
\end{equation}
The $R^2$ goodness-of-fit is equal to 0.92 if we exclude seven runs
($0.2+0.7\,M_\odot$, $0.2+0.9\,M_\odot$,$0.2+0.95\,M_\odot$, $0.25+0.7\,M_\odot$,
$0.25+0.8\,M_\odot$, $0.3+0.85\,M_\odot$, $0.7+1\,M_\odot$) that have
$\omega_{\max}$ close to the accretor's center of mass. 

\begin{eqnarray}
M_{\rm core} & = &  M_{\rm tot} (0.7786 - 0.5114 q) \label{eq:mcore}\\
M_{\rm env} & =  &  M_{\rm tot} (0.2779 -0.464 q + 0.7161 q^2) \label{eq:menv}\\
M_{\rm disk} & =  &  M_{\rm tot} (- 0.1185 + 0.9763 q -0.6559 q^2) \label{eq:mdisk}\\
M_{\rm fb} & =  &  M_{\rm tot} (0.07064 -0.0648 q) \label{eq:mfb},
\end{eqnarray}
with $R^2$ equal to 0.97, 0.88, 0.78 and 0.8, respectively.

In order to compare with \cite{zhu13} we also provide the fitting
formula for  $M(T_{\rm max})$ and $M(\omega=\omega_{\rm K}/2)$ with
respect to $q_\rho=\rho_{\rm don,0}/\rho_{\rm acc,0}$ and $M_{\rm acc}$,
where $\rho_{\rm don,0}$ and $\rho_{\rm acc,0}$ are the initial central
density of the donor and accretor, respectively, for the systems with
CO donors. 
For additional information about our comparison with \cite{zhu13}, see Section \ref{sec:comparison}. 
\begin{equation}
M(\langle T_{\rm max}\rangle) =  M_{\rm acc}(0.952 + 0.65 q_\rho -
0.9945 q_\rho^2 + 0.4378 q_\rho^3) \label{eq:mtmaxzhu}
\end{equation}
\begin{equation}
M(\omega=\omega_{\rm K}/2) = M_{\rm tot}(1.095 + 0.03784 q_\rho), \label{eq:mwk2}
\end{equation}
with $R^2$ equal to 0.45 and 0.73, respectively.

This rational polynomial for the unbound fraction of mass $M_{\rm
  esc}$ as a function of $M_{\rm tot}$ and $q$ was obtained by
excluding most of the systems that reach, or are very  
close to the conditions for dynamical burning \citep[as obtained in][]{dan12}, see 
Figure \ref{fig:mesc} where we marked these systems.
\begin{equation}\label{eq:mesc}
M_{\rm esc} =  M_{\rm tot} \frac{0.0001807}{-0.01672+0.2463 q - 0.6982 q^2+q^3}.
\end{equation}

\section{Summary of white dwarf merger remnants }
\label{sec:appendix}

Detailed instructions how to read the data in the remnant structure
profiles made available through 
\verb+www.hs.uni-hamburg.de/DE/Ins/Per/Dan/wdwd_remnants+ are included
on the website. 

The quantities presented in the following table are: $P_0$ is the initial orbital 
period in seconds; $T_{\rm max}$ is the peak temperature and $\langle T\rangle_{\rm max}$ 
is the SPH-smoothed peak temperature, both in units of $10^8$ K; 
$\rho(T_{\rm max})$ and $\rho(\langle T\rangle_{\rm max})$ are the densities at the
location of $T_{\rm max}$ and $\langle T\rangle_{\rm max}$,
respectively, in units of $10^3\ {\rm g\ cm^{-3}}$;
$\rho_{\rm max}$ is the peak density in units of $10^6\ {\rm g\ cm^{-3}}$;
 $M(T_{\rm max})$ and $M(\langle T\rangle_{\rm max})$ are the enclosed mass at the location
of peak temperature $T_{\rm max}$ and $\langle T\rangle_{\rm
  max}$, respectively, in units of
$M_\odot$; 
$\omega_{\rm max}$ is the peak value of the averaged (over equipotential surfaces) angular 
velocity and $M(\omega_{\rm max})$ is the corresponding location in
mass coordinate. $M_{\omega_k/2}$ is the location in mass coordinate
where the angular velocity equals half the Keplerian value. 
$M_{\rm core}$, $M_{\rm env}$ and $M_{\rm disk}$ are the masses 
of the central (isothermal) core, the hot envelope and the
(nearly) Keplerian disk, respectively, in units of $M_\odot$;  $M_{\rm fb}$ is the mass
of the tidal tail that is falling back onto the central remnant in
units of $10^{-2}\,M_\odot$; 
$M_{\rm esc}$ is the escape fraction of mass in units of $10^{-3}\,M_\odot$; 
$L_{\rm esc}$ is the fraction of angular momentum contained in 
material that is unbound.
Enclosed masses are computed with respect to the accretor’s center of mass 
and defined within equipotential surfaces.

\clearpage
\eject \pdfpagewidth=21cm \pdfpageheight=28cm
\onecolumn

\setlength{\tabcolsep}{2.5pt}
\def\Tiny{\fontsize{7pt}{7pt}\selectfont}

\begin{landscape}
\begin{center}
\Tiny
\renewcommand{\arraystretch}{1.2}
\begin{longtable}{cccccccccccccccccccc}
\caption{Summary of the runs performed for this paper.}
\label{tab:bigtable1}\\
\hline \hline \\[-2ex]
$M_{\rm don}$ & $M_{\rm acc}$ & $P_0$ & $\rho_{\rm max}$ & $T_{\rm max}$ &
$\rho(T_{\rm max})$ & $M(T_{\rm max})$ & $\langle T\rangle_{\rm max}$
& $\rho(\langle T\rangle_{\rm max})$ & $M(\langle T\rangle_{\rm max})$
& $\omega_{\rm max}$ & $M(\omega_{\rm max}$) & $M(\omega_{\rm k}/2)$ &
$M_{\rm core}$ & $M_{\rm env}$ & $M_{\rm € disk}$ & $M_{\rm fb}$ & $M_{\rm esc}$ & $L_{\rm esc}$ & Outcome\\
$\rm M_{\odot}$ & $\rm M_{\odot}$ & s & $10^6\ {\rm g/cm^{3}}$ &
$10^8$ K & $10^3\ {\rm g/cm^{3}}$ & $\rm M_{\odot}$ & $10^8$ K &
$10^3\ {\rm g/cm^{3}}$ & $\rm M_{\odot}$ & ${\rm rad/s}$ & $\rm
M_{\odot}$  & $\rm M_{\odot}$ & $\rm M_{\odot}$  & $\rm M_{\odot}$ &
$\rm M_{\odot}$ & $10^{-2} \rm M_{\odot}$ & $10^{-3} \rm M_{\odot}$ &
$\%L_{\rm  tot}$ & \\[0.5ex] \hline 
   \\[-1.8ex]
\endfirsthead

\multicolumn{20}{c}{{\tablename} \thetable{} -- Continued} \\[0.5ex]
\hline \hline \\[-2ex]
$M_{\rm don}$ & $M_{\rm acc}$ & $P_0$ & $\rho_{\rm max}$ & $T_{\rm max}$ &
$\rho(T_{\rm max})$ & $M(T_{\rm max})$ & $\langle T\rangle_{\rm max}$
& $\rho(\langle T\rangle_{\rm max})$ & $M(\langle T\rangle_{\rm max})$
& $\omega_{\rm max}$ & $M(\omega_{\rm max}$) & $M(\omega_{\rm k}/2)$ &
$M_{\rm core}$ & $M_{\rm env}$ & $M_{\rm disk}$ & $M_{\rm fb}$ &
$M_{\rm esc}$ & $L_{\rm esc}$ & Outcome\\ 
$\rm M_{\odot}$ & $\rm M_{\odot}$ & s & $10^6\ {\rm g/cm^{3}}$ &
$10^8$ K & $10^3\ {\rm g/cm^{3}}$ & $\rm M_{\odot}$ & $10^8$ K &
$10^3\ {\rm g/cm^{3}}$ & $\rm M_{\odot}$ & ${\rm rad/s}$ & $\rm
M_{\odot}$  & $\rm M_{\odot}$ & $\rm M_{\odot}$  & $\rm M_{\odot}$ &
$\rm M_{\odot}$ & $10^{-2} \rm M_{\odot}$ & $10^{-3} \rm M_{\odot}$ &
$\%L_{\rm  tot}$ & \\[0.5ex] \hline
   \\[-1.8ex]
\endhead

\hline
\multicolumn{20}{l}{{Continued on next page\ldots}} \\
\endfoot
\\[-1.8ex] \hline \hline
\endlastfoot

\multicolumn{20}{c}{{Helium\hspace{0.2cm}--\hspace{0.2cm}Helium}}\\[1.0ex]
\hline
0.2 &  0.2 &  215.249 &  0.203 &  0.414 &  0.326 &  0.322 &  0.285 &  6.394 &  0.269 &  0.076 &  0.210 &  0.230 &  0.107 &  0.225 &  0.059 &  0.873 &  0.244 &  0.331 &  \\ 
0.2 &  0.25 &  216.726 &  0.328 &  0.569 &  3.023 &  0.304 &  0.481 &  6.834 &  0.285 &  0.081 &  0.289 &  0.280 &  0.148 &  0.207 &  0.088 &  0.741 &  0.452 &  0.600 &  \\ 
0.2 &  0.3 &  207.567 &  0.484 &  0.785 &  4.571 &  0.347 &  0.602 &  12.051 &  0.327 &  0.087 &  0.336 &  0.334 &  0.211 &  0.181 &  0.098 &  0.965 &  0.720 &  0.942 &  \\ 
0.2 &  0.35 &  229.584 &  0.691 &  1.033 &  2.087 &  0.406 &  0.822 &  7.873 &  0.381 &  0.089 &  0.376 &  0.391 &  0.259 &  0.162 &  0.111 &  1.627 &  1.648 &  2.200 &  \\ 
0.2 &  0.4 &  226.102 &  0.951 &  1.245 &  4.078 &  0.438 &  1.012 &  2.944 &  0.445 &  0.096 &  0.431 &  0.440 &  0.324 &  0.147 &  0.104 &  2.226 &  2.054 &  2.626 &  \\ 
0.2 &  0.45 &  232.394 &  1.283 &  1.597 &  3.843 &  0.486 &  1.213 &  2.831 &  0.493 &  0.105 &  0.481 &  0.487 &  0.352 &  0.154 &  0.091 &  5.153 &  1.961 &  2.395 &  \\ 
0.25 &  0.25 &  169.393 &  0.334 &  0.530 &  1.675 &  0.351 &  0.395 &  22.371 &  0.300 &  0.097 &  0.265 &  0.286 &  0.125 &  0.278 &  0.090 &  0.615 &  0.308 &  0.339 &  \\ 
0.25 &  0.3 &  170.515 &  0.504 &  0.769 &  12.493 &  0.360 &  0.601 &  15.577 &  0.333 &  0.103 &  0.342 &  0.336 &  0.173 &  0.255 &  0.113 &  0.886 &  0.559 &  0.590 &  \\ 
0.25 &  0.35 &  170.383 &  0.721 &  0.974 &  6.324 &  0.411 &  0.778 &  14.021 &  0.390 &  0.109 &  0.398 &  0.389 &  0.231 &  0.211 &  0.143 &  1.376 &  0.934 &  0.993 &  \\ 
0.25 &  0.4 &  175.930 &  0.986 &  1.216 &  8.652 &  0.444 &  0.992 &  6.422 &  0.453 &  0.113 &  0.442 &  0.446 &  0.285 &  0.193 &  0.157 &  1.422 &  1.312 &  1.427 &  \\ 
0.25 &  0.45 &  174.924 &  1.321 &  1.446 &  15.918 &  0.485 &  1.180 &  5.106 &  0.503 &  0.119 &  0.487 &  0.497 &  0.348 &  0.187 &  0.138 &  2.506 &  1.992 &  2.057 &  \\ 
0.3 &  0.3 &  135.734 &  0.516 &  0.770 &  103.478 &  0.293 &  0.541 &  26.103 &  0.359 &  0.120 &  0.311 &  0.342 &  0.178 &  0.305 &  0.109 &  0.722 &  0.265 &  0.240 &  \\ 
0.3 &  0.35 &  137.012 &  0.736 &  0.922 &  2.973 &  0.458 &  0.731 &  30.134 &  0.376 &  0.127 &  0.391 &  0.391 &  0.194 &  0.293 &  0.152 &  1.065 &  0.392 &  0.346 &  \\ 
0.3 &  0.4 &  142.497 &  1.019 &  1.244 &  38.787 &  0.443 &  0.987 &  15.002 &  0.457 &  0.130 &  0.462 &  0.448 &  0.262 &  0.259 &  0.160 &  1.775 &  1.110 &  1.005 &  \\ 
0.3 &  0.45 &  141.926 &  1.359 &  1.445 &  5.162 &  0.530 &  1.165 &  16.997 &  0.504 &  0.137 &  0.508 &  0.501 &  0.307 &  0.239 &  0.181 &  2.159 &  1.170 &  1.055 &  \\ 
0.35 &  0.35 &  114.803 &  0.741 &  0.818 &  27.088 &  0.415 &  0.648 &  41.743 &  0.391 &  0.143 &  0.360 &  0.396 &  0.172 &  0.411 &  0.106 &  1.034 &  0.365 &  0.285 &  \\ 
0.35 &  0.4 &  118.403 &  1.053 &  1.180 &  63.018 &  0.449 &  0.911 &  30.182 &  0.468 &  0.149 &  0.460 &  0.452 &  0.230 &  0.377 &  0.132 &  0.981 &  0.402 &  0.309 &  \\ 
0.35 &  0.45 &  118.157 &  1.416 &  1.446 &  28.210 &  0.540 &  1.174 &  32.558 &  0.498 &  0.156 &  0.515 &  0.505 &  0.301 &  0.311 &  0.172 &  1.500 &  0.819 &  0.629 &  \\ 
0.4 &  0.4 &  96.566 &  1.047 &  1.093 &  53.206 &  0.478 &  0.823 &  52.636 &  0.450 &  0.169 &  0.391 &  0.456 &  0.205 &  0.417 &  0.168 &  0.945 &  0.335 &  0.223 &  \\ 
0.4 &  0.45 &  101.591 &  1.456 &  1.405 &  37.268 &  0.545 &  1.082 &  59.120 &  0.486 &  0.175 &  0.519 &  0.509 &  0.258 &  0.416 &  0.166 &  1.071 &  0.311 &  0.203 &  \\ 
0.45 &  0.45 &  86.252 &  1.466 &  1.292 &  11.308 &  0.624 &  1.037 &  61.555 &  0.532 &  0.195 &  0.449 &  0.514 &  0.291 &  0.415 &  0.186 &  0.787 &  0.383 &  0.179 &  \\ [0.5ex]
\hline
\\ [-1.5ex]
\multicolumn{20}{c}{Helium\hspace{0.2cm}--\hspace{0.2cm}Helium/Carbon/Oxygen}\\[1.0ex]
\hline\\
0.2 &  0.5 &  228.886 &  1.739 &  1.641 &  1.613 &  0.541 &  1.296 &  2.115 &  0.541 &  0.109 &  0.531 &  0.540 &  0.405 &  0.155 &  0.093 &  4.474 &  2.025 &  2.243 &  \\ 
0.2 &  0.55 &  229.339 &  2.289 &  1.993 &  39.066 &  0.555 &  1.416 &  46.742 &  0.557 &  0.111 &  0.577 &  0.593 &  0.402 &  0.208 &  0.092 &  4.477 &  2.732 &  2.846 &  \\ 
0.2 &  0.6 &  230.037 &  3.008 &  2.072 &  59.437 &  0.602 &  1.673 &  27.744 &  0.614 &  0.116 &  0.619 &  0.645 &  0.473 &  0.188 &  0.091 &  4.488 &  3.457 &  3.312 &  \\ 
0.25 &  0.5 &  177.730 &  1.766 &  1.866 &  14.913 &  0.539 &  1.395 &  6.520 &  0.547 &  0.126 &  0.539 &  0.548 &  0.378 &  0.209 &  0.134 &  2.697 &  2.511 &  2.539 &  \\ 
0.25 &  0.55 &  178.772 &  2.310 &  1.937 &  4.978 &  0.600 &  1.528 &  3.145 &  0.606 &  0.134 &  0.587 &  0.597 &  0.425 &  0.208 &  0.128 &  3.596 &  2.523 &  2.491 &  \\ 
0.25 &  0.6 &  179.892 &  3.002 &  2.364 &  61.484 &  0.614 &  1.688 &  69.397 &  0.606 &  0.136 &  0.635 &  0.654 &  0.477 &  0.208 &  0.130 &  3.234 &  3.044 &  2.792 &  \\ 
0.3 &  0.5 &  143.922 &  1.814 &  1.882 &  34.897 &  0.523 &  1.395 &  23.735 &  0.538 &  0.143 &  0.556 &  0.554 &  0.374 &  0.234 &  0.169 &  2.105 &  1.738 &  1.548 &  \\ 
0.3 &  0.55 &  146.260 &  2.335 &  2.206 &  10.328 &  0.614 &  1.613 &  8.333 &  0.614 &  0.150 &  0.595 &  0.607 &  0.413 &  0.238 &  0.163 &  3.408 &  2.335 &  2.017 &  \\ 
0.3 &  0.6 &  146.020 &  3.054 &  2.401 &  130.200 &  0.593 &  1.814 &  6.262 &  0.662 &  0.159 &  0.645 &  0.657 &  0.448 &  0.256 &  0.164 &  2.915 &  2.832 &  2.356 &  \\ 
0.35 &  0.5 &  120.603 &  1.874 &  1.959 &  28.288 &  0.560 &  1.468 &  24.337 &  0.565 &  0.162 &  0.565 &  0.560 &  0.351 &  0.285 &  0.194 &  1.889 &  1.394 &  1.069 &  \\ 
0.35 &  0.55 &  121.539 &  2.414 &  2.236 &  40.264 &  0.602 &  1.659 &  22.942 &  0.612 &  0.170 &  0.602 &  0.612 &  0.395 &  0.286 &  0.197 &  2.067 &  1.242 &  0.972 &  \\ 
0.35 &  0.6 &  120.954 &  3.114 &  2.573 &  51.150 &  0.635 &  1.861 &  51.150 &  0.635 &  0.179 &  0.654 &  0.661 &  0.445 &  0.259 &  0.219 &  2.503 &  2.123 &  1.586 &  \\ 
0.4 &  0.5 &  102.523 &  1.922 &  2.012 &  132.652 &  0.509 &  1.456 &  62.742 &  0.554 &  0.184 &  0.577 &  0.561 &  0.323 &  0.366 &  0.200 &  1.079 &  0.738 &  0.495 &  \\ 
0.4 &  0.55 &  103.503 &  2.488 &  2.331 &  44.860 &  0.618 &  1.697 &  23.024 &  0.640 &  0.189 &  0.628 &  0.618 &  0.372 &  0.335 &  0.228 &  1.309 &  1.068 &  0.712 &  \\ 
0.4 &  0.6 &  103.109 &  3.195 &  2.644 &  43.660 &  0.664 &  1.915 &  44.352 &  0.664 &  0.196 &  0.663 &  0.669 &  0.407 &  0.357 &  0.214 &  2.011 &  1.447 &  0.975 &  \\ 
0.45 &  0.5 &  88.221 &  2.008 &  1.915 &  105.100 &  0.554 &  1.373 &  117.948 &  0.528 &  0.202 &  0.576 &  0.565 &  0.291 &  0.477 &  0.174 &  0.707 &  0.609 &  0.311 &  \\ 
0.45 &  0.55 &  88.850 &  2.592 &  2.354 &  54.571 &  0.626 &  1.724 &  73.256 &  0.594 &  0.210 &  0.639 &  0.622 &  0.343 &  0.430 &  0.215 &  1.144 &  0.875 &  0.421 &  \\ 
0.45 &  0.6 &  89.581 &  3.303 &  2.694 &  26.982 &  0.693 &  1.975 &  72.952 &  0.654 &  0.218 &  0.685 &  0.674 &  0.391 &  0.391 &  0.247 &  1.927 &  1.252 &  0.662 &  \\ [0.5ex]
\hline
\\ [-1.5ex]
\multicolumn{20}{c}{Helium/Carbon/Oxygen\hspace{0.2cm}--\hspace{0.2cm}Helium/Carbon/Oxygen}\\[1.0ex]
\hline
0.5 &  0.5 &  74.362 &  2.029 &  2.170 &  174.519 &  0.503 &  1.760 &  88.536 &  0.563 &  0.216 &  0.544 &  0.576 &  0.261 &  0.567 &  0.165 &  0.596 &  0.624 &  0.120 &  \\ 
0.5 &  0.55 &  75.401 &  2.701 &  2.622 &  243.511 &  0.566 &  2.131 &  192.096 &  0.584 &  0.224 &  0.644 &  0.633 &  0.342 &  0.473 &  0.224 &  1.048 &  0.970 &  0.214 &  \\ 
0.5 &  0.6 &  74.288 &  3.438 &  3.133 &  155.623 &  0.663 &  2.467 &  108.393 &  0.649 &  0.235 &  0.678 &  0.673 &  0.362 &  0.471 &  0.248 &  1.639 &  1.978 &  0.415 &  \\ 
0.55 &  0.55 &  66.277 &  2.755 &  3.087 &  200.887 &  0.571 &  2.569 &  90.632 &  0.661 &  0.232 &  0.632 &  0.632 &  0.310 &  0.568 &  0.210 &  0.981 &  2.807 &  0.351 &  \\ 
0.55 &  0.6 &  66.680 &  3.456 &  4.032 &  90.061 &  0.647 &  3.281 &  84.085 &  0.663 &  0.229 &  0.672 &  0.682 &  0.375 &  0.521 &  0.228 &  2.002 &  6.341 &  0.575 &  \\ 
0.6 &  0.6 &  59.194 &  3.370 &  4.179 &  102.862 &  0.642 &  3.552 &  102.862 &  0.642 &  0.235 &  0.647 &  0.679 &  0.329 &  0.456 &  0.379 &  2.352 &  12.681 &  1.007 &  \\ [0.5ex]
\hline
\\ [1.5ex]
\multicolumn{20}{c}{{Helium\hspace{0.2cm}--\hspace{0.2cm}Carbon/Oxygen}}\\[0.5ex]
\hline
0.2 &  0.65 &  230.434 &  3.911 &  2.740 &  52.593 &  0.654 &  1.982 &  28.511 &  0.663 &  0.120 &  0.673 &  0.692 &  0.504 &  0.206 &  0.102 &  3.301 &  4.022 &  3.731 &  \\ 
0.2 &  0.7 &  229.756 &  5.097 &  2.841 &  54.098 &  0.706 &  2.277 &  31.882 &  0.711 &  0.126 &  0.477 &  0.742 &  0.567 &  0.191 &  0.100 &  3.750 &  5.090 &  4.331 &  \\ 
0.2 &  0.75 &  232.641 &  6.568 &  3.052 &  36.193 &  0.760 &  2.562 &  30.778 &  0.768 &  0.131 &  0.750 &  0.792 &  0.594 &  0.212 &  0.084 &  5.399 &  6.394 &  5.231 &  \\ 
0.2 &  0.8 &  231.157 &  8.586 &  3.976 &  31.743 &  0.806 &  3.262 &  24.823 &  0.812 &  0.135 &  0.795 &  0.841 &  0.664 &  0.191 &  0.100 &  3.708 &  7.794 &  5.969 &  \\ 
0.2 &  0.85 &  235.614 &  \multicolumn{17}{c}{does not merge after $86 \times P_0$}  \\ 
0.2 &  0.9 &  235.165 &  14.003 &  4.899 &  19.740 &  0.907 &  4.291 &  14.474 &  0.914 &  0.142 &  0.062 &  0.939 &  0.730 &  0.222 &  0.083 &  5.272 &  11.732 &  8.152 &  \\ 
0.2 &  0.95 &  235.215 &  16.594 &  4.847 &  20.894 &  0.954 &  4.310 &  15.272 &  0.959 &  0.150 &  0.051 &  0.989 &  0.704 &  0.298 &  0.057 &  7.357 &  17.446 &  12.408 &  \\ 
0.2 &  1 &  234.578 &  21.742 &  5.264 &  22.916 &  1.002 &  4.534 &  12.925 &  1.008 &  0.159 &  0.998 &  1.033 &  0.762 &  0.280 &  0.088 &  5.356 &  16.031 &  10.387 &  \\ 
0.2 &  1.05 &  235.761 & \multicolumn{17}{c}{does not merge after $77.5\times P_0$} \\ 
0.25 &  0.65 &  180.932 &  3.923 &  2.744 &  63.787 &  0.664 &  2.075 &  81.452 &  0.645 &  0.144 &  0.668 &  0.703 &  0.507 &  0.223 &  0.126 &  4.011 &  3.042 &  2.584 &  \\ 
0.25 &  0.7 &  181.024 &  5.102 &  3.164 &  70.063 &  0.711 &  2.292 &  54.088 &  0.717 &  0.151 &  0.153 &  0.754 &  0.554 &  0.218 &  0.121 &  5.268 &  3.650 &  2.878 &  \\ 
0.25 &  0.75 &  182.846 &  6.583 &  3.755 &  70.664 &  0.759 &  2.561 &  45.251 &  0.770 &  0.157 &  0.776 &  0.803 &  0.617 &  0.206 &  0.118 &  5.464 &  4.714 &  3.556 &  \\ 
0.25 &  0.8 &  182.183 &  8.434 &  4.463 &  38.723 &  0.804 &  3.737 &  31.002 &  0.816 &  0.160 &  0.090 &  0.852 &  0.654 &  0.215 &  0.121 &  5.437 &  5.825 &  4.141 &  \\ 
0.25 &  0.85 &  184.417 &  10.538 &  4.875 &  21.456 &  0.858 &  4.340 &  19.115 &  0.864 &  0.161 &  0.856 &  0.902 &  0.677 &  0.244 &  0.123 &  4.888 &  7.221 &  5.026 &  \\ 
0.25 &  0.9 &  183.236 &  13.566 &  5.224 &  21.967 &  0.902 &  4.582 &  16.745 &  0.908 &  0.181 &  0.901 &  0.948 &  0.740 &  0.222 &  0.117 &  6.258 &  8.297 &  5.344 &  \\ 
0.25 &  0.95 &  184.034 &  17.224 &  5.105 &  19.319 &  0.956 &  4.685 &  15.256 &  0.961 &  0.188 &  0.952 &  0.992 &  0.780 &  0.228 &  0.091 &  8.646 &  15.365 &  9.193 &  \\ 
0.25 &  1 &  184.290 &  23.082 &  5.583 &  28.908 &  1.001 &  4.983 &  13.359 &  1.010 &  0.211 &  1.004 &  1.038 &  0.823 &  0.224 &  0.127 &  6.373 &  13.183 &  7.536 &  \\ 
0.25 &  1.05 &  185.003 &  29.167 &  5.657 &  23.106 &  1.052 &  5.025 &  15.149 &  1.059 &  0.204 &  1.050 &  1.088 &  0.827 &  0.279 &  0.121 &  5.975 &  13.900 &  8.159 &  \\ 
0.3 &  0.65 &  146.813 &  3.964 &  2.752 &  9.190 &  0.711 &  2.232 &  105.983 &  0.641 &  0.171 &  0.689 &  0.708 &  0.497 &  0.248 &  0.154 &  4.749 &  2.874 &  2.339 &  \\ 
0.3 &  0.7 &  146.585 &  5.113 &  3.242 &  149.740 &  0.688 &  2.508 &  156.919 &  0.680 &  0.173 &  0.742 &  0.762 &  0.581 &  0.214 &  0.167 &  3.544 &  3.177 &  2.465 &  \\ 
0.3 &  0.75 &  147.840 &  6.603 &  3.621 &  82.347 &  0.773 &  2.745 &  173.410 &  0.738 &  0.182 &  0.790 &  0.811 &  0.595 &  0.247 &  0.142 &  6.191 &  3.974 &  2.876 &  \\ 
0.3 &  0.8 &  147.050 &  8.275 &  4.739 &  68.888 &  0.807 &  3.661 &  50.374 &  0.816 &  0.183 &  0.790 &  0.861 &  0.649 &  0.241 &  0.147 &  5.848 &  4.651 &  3.107 &  \\ 
0.3 &  0.85 &  149.691 &  10.368 &  5.188 &  42.981 &  0.856 &  4.523 &  23.206 &  0.865 &  0.188 &  0.002 &  0.911 &  0.707 &  0.228 &  0.160 &  4.918 &  5.105 &  3.233 &  \\ 
0.3 &  0.9 &  149.108 &  13.171 &  5.279 &  25.217 &  0.906 &  4.678 &  20.180 &  0.910 &  0.192 &  0.904 &  0.957 &  0.754 &  0.223 &  0.154 &  6.127 &  8.004 &  4.747 &  \\ 
0.3 &  0.95 &  149.422 &  16.915 &  5.475 &  29.697 &  0.953 &  4.814 &  17.958 &  0.961 &  0.206 &  0.950 &  1.000 &  0.767 &  0.250 &  0.146 &  7.545 &  12.391 &  6.690 &  \\ 
0.3 &  1 &  149.953 &  22.586 &  5.629 &  31.051 &  1.002 &  5.141 &  14.832 &  1.011 &  0.227 &  1.000 &  1.048 &  0.790 &  0.271 &  0.133 &  9.516 &  10.832 &  5.708 &  \\ 
0.3 &  1.05 &  150.690 &  30.706 &  5.864 &  34.484 &  1.051 &  5.382 &  14.531 &  1.059 &  0.265 &  1.051 &  1.091 &  0.885 &  0.223 &  0.148 &  8.043 &  12.919 &  6.469 &  \\ 
0.35 &  0.65 &  121.642 &  4.045 &  2.978 &  37.689 &  0.701 &  2.349 &  135.768 &  0.639 &  0.187 &  0.705 &  0.716 &  0.490 &  0.274 &  0.200 &  3.377 &  2.668 &  1.972 &  \\ 
0.35 &  0.7 &  122.430 &  5.156 &  3.224 &  137.712 &  0.715 &  2.549 &  155.470 &  0.690 &  0.199 &  0.745 &  0.764 &  0.524 &  0.282 &  0.201 &  3.977 &  2.909 &  2.060 &  \\ 
0.35 &  0.75 &  123.539 &  6.592 &  3.685 &  17.646 &  0.816 &  2.842 &  239.128 &  0.717 &  0.210 &  0.798 &  0.813 &  0.584 &  0.276 &  0.208 &  2.981 &  3.123 &  2.175 &  \\ 
0.35 &  0.8 &  122.852 &  8.287 &  4.978 &  107.360 &  0.806 &  3.699 &  96.419 &  0.811 &  0.202 &  0.803 &  0.866 &  0.651 &  0.254 &  0.200 &  4.170 &  3.490 &  2.279 &  \\ 
0.35 &  0.85 &  124.890 &  10.152 &  5.715 &  63.928 &  0.854 &  4.599 &  26.507 &  0.869 &  0.219 &  0.854 &  0.920 &  0.676 &  0.287 &  0.188 &  4.502 &  3.675 &  2.235 &  \\ 
0.35 &  0.9 &  125.082 &  12.975 &  5.884 &  57.009 &  0.902 &  4.775 &  20.711 &  0.917 &  0.219 &  0.900 &  0.962 &  0.720 &  0.268 &  0.212 &  4.531 &  4.608 &  2.669 &  \\ 
0.35 &  0.95 &  124.644 &  16.407 &  5.777 &  52.520 &  0.951 &  4.751 &  19.982 &  0.963 &  0.226 &  0.951 &  1.009 &  0.773 &  0.250 &  0.210 &  5.624 &  11.037 &  5.493 &  \\ 
0.35 &  1 &  125.596 &  22.122 &  6.101 &  38.868 &  1.002 &  5.029 &  23.317 &  1.006 &  0.255 &  1.000 &  1.055 &  0.793 &  0.277 &  0.211 &  6.042 &  8.292 &  4.218 &  \\ 
0.35 &  1.05 &  125.409 &  29.235 &  6.784 &  63.511 &  1.049 &  5.195 &  16.471 &  1.060 &  0.271 &  1.050 &  1.101 &  0.869 &  0.250 &  0.208 &  6.154 &  11.052 &  5.166 &  \\ 
0.4 &  0.65 &  103.817 &  4.141 &  3.212 &  43.898 &  0.713 &  2.397 &  185.441 &  0.630 &  0.207 &  0.724 &  0.721 &  0.462 &  0.337 &  0.224 &  2.466 &  2.108 &  1.407 &  \\ 
0.4 &  0.7 &  104.815 &  5.288 &  3.375 &  16.512 &  0.786 &  2.720 &  179.546 &  0.690 &  0.216 &  0.765 &  0.775 &  0.519 &  0.324 &  0.226 &  2.853 &  2.667 &  1.729 &  \\ 
0.4 &  0.75 &  104.650 &  6.735 &  3.847 &  25.780 &  0.824 &  2.988 &  258.232 &  0.722 &  0.226 &  0.788 &  0.822 &  0.562 &  0.311 &  0.239 &  3.396 &  3.018 &  1.923 &  \\ 
0.4 &  0.8 &  105.005 &  8.549 &  5.148 &  193.044 &  0.800 &  3.447 &  218.119 &  0.797 &  0.238 &  0.844 &  0.866 &  0.620 &  0.296 &  0.240 &  4.003 &  3.380 &  2.077 &  \\ 
0.4 &  0.85 &  105.398 &  10.219 &  5.896 &  40.741 &  0.863 &  4.679 &  33.674 &  0.870 &  0.245 &  0.853 &  0.921 &  0.673 &  0.296 &  0.233 &  4.328 &  4.447 &  2.562 &  \\ 
0.4 &  0.9 &  105.854 &  12.432 &  5.773 &  32.572 &  0.911 &  4.863 &  25.181 &  0.913 &  0.234 &  0.897 &  0.974 &  0.688 &  0.288 &  0.271 &  4.819 &  5.293 &  2.627 &  \\ 
0.4 &  0.95 &  106.466 &  15.976 &  6.448 &  70.841 &  0.948 &  4.885 &  21.718 &  0.965 &  0.247 &  0.947 &  1.017 &  0.729 &  0.307 &  0.249 &  5.651 &  7.975 &  3.454 &  \\ 
0.4 &  1 &  106.900 &  21.120 &  6.064 &  43.036 &  1.000 &  5.102 &  16.958 &  1.013 &  0.272 &  0.997 &  1.059 &  0.806 &  0.282 &  0.250 &  5.202 &  10.570 &  4.438 &  \\ 
0.4 &  1.05 &  107.339 &  28.328 &  6.518 &  39.046 &  1.050 &  5.399 &  19.819 &  1.057 &  0.281 &  1.048 &  1.106 &  0.897 &  0.221 &  0.257 &  6.527 &  10.118 &  4.287 &  \\ 
0.45 &  0.65 &  89.708 &  4.234 &  3.324 &  52.587 &  0.734 &  2.436 &  83.398 &  0.698 &  0.226 &  0.720 &  0.727 &  0.444 &  0.359 &  0.267 &  2.931 &  1.181 &  0.693 &  \\ 
0.45 &  0.7 &  89.157 &  5.402 &  3.584 &  166.312 &  0.711 &  2.751 &  188.196 &  0.696 &  0.237 &  0.765 &  0.778 &  0.500 &  0.350 &  0.278 &  2.075 &  1.495 &  0.852 &  \\ 
0.45 &  0.75 &  90.732 &  6.833 &  4.244 &  68.312 &  0.814 &  3.078 &  209.468 &  0.739 &  0.248 &  0.808 &  0.828 &  0.566 &  0.336 &  0.268 &  2.676 &  2.326 &  1.286 &  \\ 
0.45 &  0.8 &  90.876 &  8.145 &  6.011 &  85.187 &  0.815 &  4.227 &  84.626 &  0.817 &  0.246 &  0.807 &  0.882 &  0.620 &  0.313 &  0.272 &  4.163 &  2.900 &  1.468 &  \\ 
0.45 &  0.85 &  89.934 &  9.635 &  5.683 &  73.832 &  0.844 &  4.636 &  37.089 &  0.865 &  0.245 &  0.850 &  0.933 &  0.659 &  0.301 &  0.284 &  5.059 &  6.086 &  2.345 &  \\ 
0.45 &  0.9 &  91.636 &  11.785 &  6.186 &  56.092 &  0.901 &  4.983 &  38.794 &  0.901 &  0.256 &  0.893 &  0.968 &  0.689 &  0.310 &  0.284 &  5.866 &  8.509 &  2.918 &  \\ 
0.45 &  0.95 &  91.470 &  15.542 &  6.476 &  90.665 &  0.942 &  5.138 &  22.826 &  0.959 &  0.279 &  0.943 &  0.997 &  0.759 &  0.261 &  0.291 &  7.608 &  12.992 &  4.139 &  \\ 
0.45 &  1 &  92.022 &  20.037 &  6.343 &  35.357 &  0.995 &  5.113 &  36.875 &  0.993 &  0.278 &  0.991 &  1.052 &  0.789 &  0.279 &  0.291 &  7.303 &  17.651 &  5.735 &  \\ 
0.45 &  1.05 &  92.531 &  27.016 &  6.746 &  83.393 &  1.040 &  5.306 &  87.909 &  1.039 &  0.290 &  1.043 &  1.101 &  0.828 &  0.282 &  0.308 &  6.621 &  16.294 &  5.717 &  \\ [0.5ex]
\hline
\\ [-1.5ex]
\multicolumn{20}{c}{{Helium/Carbon/Oxygen\hspace{0.2cm}--\hspace{0.2cm}Carbon/Oxygen}}\\[1.0ex]
\hline
0.5 &  0.65 &  77.846 &  4.385 &  3.477 &  57.549 &  0.746 &  2.810 &  78.616 &  0.716 &  0.246 &  0.742 &  0.733 &  0.434 &  0.423 &  0.272 &  1.927 &  1.748 &  0.522 &  \\ 
0.5 &  0.7 &  77.677 &  5.563 &  3.857 &  33.316 &  0.794 &  3.172 &  95.099 &  0.760 &  0.258 &  0.776 &  0.780 &  0.562 &  0.367 &  0.290 &  -2.036 &  2.544 &  0.736 &  \\ 
0.5 &  0.75 &  77.792 &  6.995 &  4.153 &  90.687 &  0.798 &  3.412 &  144.790 &  0.776 &  0.269 &  0.817 &  0.834 &  0.544 &  0.392 &  0.282 &  2.927 &  2.913 &  0.796 &  \\ 
0.5 &  0.8 &  78.410 &  8.859 &  4.667 &  45.821 &  0.871 &  3.760 &  222.249 &  0.812 &  0.277 &  0.862 &  0.880 &  0.592 &  0.391 &  0.275 &  3.694 &  5.393 &  1.413 &  \\ 
0.5 &  0.85 &  78.621 &  11.127 &  8.869 &  14.421 &  1.023 &  4.141 &  418.228 &  0.843 &  0.289 &  0.903 &  0.927 &  0.653 &  0.342 &  0.306 &  4.057 &  9.329 &  2.294 &  \\ 
0.5 &  0.9 &  79.102 &  14.166 &  5.482 &  54.356 &  0.967 &  4.584 &  326.166 &  0.888 &  0.303 &  0.944 &  0.971 &  0.678 &  0.367 &  0.302 &  4.275 &  9.991 &  2.324 &  \\ 
0.5 &  0.95 &  79.346 &  18.514 &  6.015 &  583.849 &  0.920 &  5.128 &  408.402 &  0.953 &  0.323 &  1.001 &  1.025 &  0.739 &  0.356 &  0.299 &  4.602 &  9.638 &  2.809 &  \\ 
0.5 &  1 &  80.422 &  24.354 &  6.635 &  332.741 &  0.998 &  5.611 &  488.456 &  0.985 &  0.346 &  1.050 &  1.071 &  0.764 &  0.375 &  0.291 &  5.641 &  13.235 &  4.050 &  \\ 
0.5 &  1.05 &  80.310 &  32.680 &  7.113 &  464.021 &  1.046 &  6.228 &  617.862 &  1.034 &  0.370 &  1.100 &  1.119 &  0.842 &  0.327 &  0.305 &  6.354 &  12.928 &  4.309 &  \\ 
0.55 &  0.65 &  67.241 &  4.446 &  4.006 &  416.649 &  0.620 &  3.239 &  90.766 &  0.713 &  0.254 &  0.723 &  0.738 &  0.396 &  0.516 &  0.265 &  1.817 &  4.477 &  0.526 &  \\ 
0.55 &  0.7 &  68.252 &  5.642 &  4.304 &  67.964 &  0.772 &  3.501 &  97.451 &  0.754 &  0.265 &  0.772 &  0.785 &  0.479 &  0.449 &  0.285 &  3.060 &  6.665 &  0.852 &  \\ 
0.55 &  0.75 &  68.037 &  6.987 &  4.693 &  52.523 &  0.820 &  3.763 &  96.541 &  0.790 &  0.274 &  0.816 &  0.827 &  0.508 &  0.431 &  0.311 &  3.880 &  10.624 &  1.230 &  \\ 
0.55 &  0.8 &  69.026 &  8.705 &  5.169 &  55.063 &  0.864 &  4.064 &  127.112 &  0.837 &  0.288 &  0.859 &  0.873 &  0.557 &  0.411 &  0.325 &  4.200 &  14.302 &  1.796 &  \\ 
0.55 &  0.85 &  69.081 &  10.995 &  5.292 &  34.265 &  0.929 &  4.297 &  216.763 &  0.855 &  0.306 &  0.917 &  0.920 &  0.632 &  0.367 &  0.339 &  4.545 &  15.845 &  2.165 &  \\ 
0.55 &  0.9 &  69.554 &  13.928 &  6.329 &  92.468 &  0.978 &  4.741 &  328.682 &  0.884 &  0.321 &  0.958 &  0.969 &  0.653 &  0.385 &  0.340 &  5.509 &  17.503 &  2.585 &  \\ 
0.55 &  0.95 &  70.081 &  18.147 &  5.970 &  789.231 &  0.896 &  5.138 &  314.783 &  0.944 &  0.338 &  0.986 &  1.021 &  0.717 &  0.406 &  0.308 &  4.840 &  20.830 &  4.124 &  \\ 
0.55 &  1 &  70.446 &  24.158 &  9.787 &  244.069 &  1.033 &  5.665 &  459.437 &  0.989 &  0.366 &  1.042 &  1.065 &  0.760 &  0.373 &  0.329 &  6.691 &  20.797 &  4.830 &  \\ 
0.55 &  1.05 &  70.093 &  32.859 &  7.175 &  753.509 &  1.024 &  6.423 &  458.041 &  1.055 &  0.385 &  1.091 &  1.124 &  0.849 &  0.323 &  0.346 &  6.275 &  19.470 &  4.918 &  \\ 
0.6 &  0.65 &  59.832 &  4.514 &  4.597 &  144.793 &  0.729 &  3.806 &  97.853 &  0.712 &  0.262 &  0.723 &  0.733 &  0.421 &  0.428 &  0.369 &  2.240 &  8.959 &  0.686 &  \\ 
0.6 &  0.7 &  60.050 &  5.645 &  4.977 &  78.272 &  0.786 &  4.238 &  86.751 &  0.758 &  0.264 &  0.743 &  0.794 &  0.444 &  0.504 &  0.306 &  3.348 &  13.588 &  1.088 &  \\ 
0.6 &  0.75 &  59.984 &  6.863 &  9.048 &  62.639 &  0.913 &  4.291 &  98.371 &  0.816 &  0.282 &  0.799 &  0.829 &  0.466 &  0.475 &  0.349 &  4.224 &  18.269 &  1.408 &  \\ 
0.6 &  0.8 &  60.960 &  8.734 &  5.371 &  64.380 &  0.884 &  4.408 &  163.915 &  0.815 &  0.293 &  0.854 &  0.877 &  0.535 &  0.428 &  0.364 &  4.862 &  24.721 &  2.227 &  \\ 
0.6 &  0.85 &  61.237 &  10.809 &  5.484 &  66.283 &  0.916 &  4.537 &  71.821 &  0.923 &  0.312 &  0.909 &  0.920 &  0.602 &  0.427 &  0.334 &  5.929 &  27.764 &  2.487 &  \\ 
0.6 &  0.9 &  61.262 &  13.922 &  5.840 &  160.281 &  0.929 &  4.911 &  331.915 &  0.898 &  0.333 &  0.952 &  0.969 &  0.648 &  0.411 &  0.367 &  4.583 &  28.932 &  2.686 &  \\ 
0.6 &  0.95 &  62.184 &  18.030 &  6.256 &  63.335 &  1.010 &  5.274 &  355.850 &  0.940 &  0.354 &  0.986 &  1.019 &  0.715 &  0.388 &  0.359 &  5.781 &  30.384 &  4.233 &  \\ 
0.6 &  1 &  62.202 &  23.732 &  6.932 &  88.429 &  1.050 &  5.884 &  496.703 &  0.972 &  0.381 &  1.040 &  1.074 &  0.790 &  0.359 &  0.365 &  5.594 &  30.669 &  5.051 &  \\ 
0.6 &  1.05 &  62.032 &  31.703 &  7.333 &  768.514 &  1.014 &  6.524 &  596.270 &  1.029 &  0.407 &  1.079 &  1.120 &  0.805 &  0.390 &  0.364 &  5.652 &  34.037 &  6.606 &  \\ [0.5ex]
\hline
\\ [-1.5ex]
\multicolumn{20}{c}{{Carbon/Oxygen\hspace{0.2cm}--\hspace{0.2cm}Carbon/Oxygen}}\\[1.0ex]
\hline
0.65 &  0.65 &  51.865 &  4.498 &  3.229 &  700.736 &  0.559 &  2.704 &  218.810 &  0.703 &  0.333 &  0.599 &  0.738 &  0.289 &  0.797 &  0.200 &  1.319 &  0.205 &  0.080 &  \\ 
0.65 &  0.7 &  53.239 &  5.990 &  3.977 &  172.768 &  0.802 &  3.192 &  268.065 &  0.742 &  0.343 &  0.802 &  0.787 &  0.382 &  0.734 &  0.220 &  1.415 &  0.178 &  0.063 &  \\ 
0.65 &  0.75 &  53.749 &  7.712 &  4.591 &  666.019 &  0.753 &  3.750 &  217.223 &  0.838 &  0.358 &  0.864 &  0.838 &  0.468 &  0.601 &  0.311 &  2.004 &  0.354 &  0.143 &  \\ 
0.65 &  0.8 &  54.470 &  9.814 &  5.063 &  308.348 &  0.848 &  4.254 &  200.931 &  0.892 &  0.375 &  0.891 &  0.891 &  0.516 &  0.540 &  0.373 &  2.014 &  0.538 &  0.217 &  \\ 
0.65 &  0.85 &  54.133 &  12.626 &  5.482 &  191.333 &  0.915 &  4.660 &  225.467 &  0.915 &  0.391 &  0.948 &  0.948 &  0.573 &  0.519 &  0.384 &  2.312 &  1.002 &  0.414 &  \\ 
0.65 &  0.9 &  53.986 &  16.218 &  6.065 &  145.986 &  0.994 &  5.136 &  163.126 &  0.976 &  0.411 &  1.003 &  0.999 &  0.637 &  0.489 &  0.393 &  2.975 &  1.025 &  0.436 &  \\ 
0.65 &  0.95 &  54.432 &  21.360 &  6.552 &  221.599 &  1.027 &  5.477 &  195.902 &  1.022 &  0.434 &  1.040 &  1.048 &  0.705 &  0.463 &  0.395 &  3.581 &  1.252 &  0.527 &  \\ 
0.65 &  1 &  54.640 &  28.183 &  6.902 &  361.440 &  1.068 &  6.083 &  358.800 &  1.040 &  0.459 &  1.096 &  1.103 &  0.746 &  0.457 &  0.406 &  3.966 &  1.388 &  0.556 &  \\ 
0.65 &  1.05 &  55.136 &  38.207 &  7.707 &  1381.898 &  1.027 &  6.809 &  988.864 &  1.027 &  0.489 &  1.131 &  1.163 &  0.823 &  0.438 &  0.397 &  3.947 &  2.110 &  0.823 &  \\ 
0.7 &  0.7 &  46.842 &  5.796 &  3.732 &  771.280 &  0.655 &  3.064 &  234.975 &  0.792 &  0.374 &  0.666 &  0.792 &  0.340 &  0.813 &  0.231 &  1.582 &  0.109 &  0.038 &  \\ 
0.7 &  0.75 &  48.161 &  7.817 &  4.665 &  681.538 &  0.808 &  3.640 &  267.291 &  0.824 &  0.384 &  0.832 &  0.840 &  0.408 &  0.760 &  0.264 &  1.830 &  0.110 &  0.037 &  \\ 
0.7 &  0.8 &  47.824 &  10.220 &  5.438 &  913.532 &  0.788 &  4.253 &  311.627 &  0.844 &  0.410 &  0.900 &  0.887 &  0.481 &  0.687 &  0.309 &  2.277 &  0.193 &  0.072 &  \\ 
0.7 &  0.85 &  48.301 &  13.021 &  5.912 &  397.345 &  0.943 &  4.735 &  203.383 &  0.943 &  0.428 &  0.973 &  0.949 &  0.533 &  0.607 &  0.384 &  2.515 &  0.578 &  0.223 &  \\ 
0.7 &  0.9 &  48.243 &  16.723 &  6.313 &  276.838 &  0.990 &  5.244 &  166.696 &  1.012 &  0.449 &  1.017 &  0.995 &  0.593 &  0.579 &  0.399 &  2.772 &  0.922 &  0.351 &  \\ 
0.7 &  0.95 &  48.480 &  21.914 &  6.866 &  373.104 &  1.023 &  5.814 &  188.279 &  1.059 &  0.471 &  1.044 &  1.054 &  0.658 &  0.539 &  0.420 &  3.269 &  1.016 &  0.404 &  \\ 
0.7 &  1 &  84.987 &  29.338 &  7.173 &  787.639 &  1.028 &  6.374 &  1064.668 &  0.976 &  0.492 &  0.359 &  1.141 &  0.731 &  0.527 &  0.415 &  2.570 &  1.085 &  0.427 &  \\ 
0.7 &  1.05 &  48.875 &  38.881 &  8.005 &  792.237 &  1.058 &  6.868 &  186.278 &  1.155 &  0.532 &  1.135 &  1.155 &  0.798 &  0.456 &  0.456 &  3.769 &  1.778 &  0.674 &  \\ 
0.75 &  0.75 &  42.406 &  7.664 &  4.654 &  1111.741 &  0.669 &  3.536 &  477.711 &  0.785 &  0.420 &  0.753 &  0.854 &  0.346 &  0.883 &  0.263 &  0.759 &  0.231 &  0.074 &  \\ 
0.75 &  0.8 &  47.046 &  10.249 &  5.175 &  1084.536 &  0.812 &  4.158 &  423.421 &  0.840 &  0.434 &  0.866 &  0.900 &  0.443 &  0.797 &  0.293 &  1.660 &  0.166 &  0.052 &  \\ 
0.75 &  0.85 &  42.771 &  13.332 &  5.689 &  1100.392 &  0.851 &  4.751 &  499.338 &  0.901 &  0.459 &  0.970 &  0.948 &  0.525 &  0.696 &  0.363 &  1.508 &  0.121 &  0.045 &  \\ 
0.75 &  0.9 &  43.153 &  17.344 &  6.656 &  232.610 &  1.007 &  5.496 &  286.758 &  1.001 &  0.482 &  1.001 &  1.001 &  0.575 &  0.643 &  0.412 &  1.892 &  0.694 &  0.222 &  \\ 
0.75 &  0.95 &  43.452 &  22.299 &  7.037 &  250.783 &  1.089 &  5.886 &  351.801 &  1.029 &  0.507 &  1.072 &  1.054 &  0.637 &  0.584 &  0.455 &  2.202 &  0.884 &  0.312 &  \\ 
0.75 &  1 &  43.607 &  29.622 &  7.598 &  208.763 &  1.101 &  6.463 &  683.181 &  1.034 &  0.539 &  1.095 &  1.105 &  0.733 &  0.536 &  0.452 &  2.792 &  0.864 &  0.308 &  \\ 
0.75 &  1.05 &  44.025 &  39.943 &  8.285 &  382.368 &  1.152 &  7.157 &  446.014 &  1.104 &  0.572 &  1.137 &  1.161 &  0.777 &  0.511 &  0.476 &  3.516 &  1.174 &  0.420 &  \\ 
0.8 &  0.8 &  38.085 &  10.045 &  4.978 &  1597.373 &  0.747 &  4.101 &  963.585 &  0.770 &  0.472 &  0.725 &  0.919 &  0.375 &  0.920 &  0.294 &  1.108 &  0.108 &  0.031 &  \\ 
0.8 &  0.85 &  38.448 &  13.646 &  5.933 &  943.778 &  0.902 &  4.905 &  392.917 &  0.929 &  0.490 &  0.901 &  0.956 &  0.455 &  0.853 &  0.326 &  1.571 &  0.170 &  0.044 &  \\ 
0.8 &  0.9 &  38.854 &  17.806 &  6.540 &  1168.151 &  0.901 &  5.500 &  561.438 &  0.953 &  0.521 &  0.994 &  1.002 &  0.562 &  0.705 &  0.405 &  2.751 &  0.196 &  0.058 &  \\ 
0.8 &  0.95 &  38.854 &  23.346 &  7.706 &  586.806 &  1.007 &  6.031 &  338.403 &  1.042 &  0.550 &  1.078 &  1.057 &  0.617 &  0.644 &  0.467 &  2.070 &  0.641 &  0.191 &  \\ 
0.8 &  1 &  39.187 &  30.463 &  7.842 &  690.128 &  1.078 &  6.717 &  563.213 &  1.051 &  0.581 &  1.117 &  1.110 &  0.693 &  0.636 &  0.440 &  3.050 &  0.630 &  0.213 &  \\ 
0.8 &  1.05 &  39.058 &  41.284 &  8.694 &  1160.689 &  1.105 &  7.404 &  370.513 &  1.148 &  0.616 &  1.148 &  1.165 &  0.766 &  0.559 &  0.492 &  3.059 &  1.147 &  0.377 &  \\ 
0.85 &  0.85 &  33.992 &  13.196 &  6.076 &  1670.837 &  0.772 &  4.635 &  576.422 &  0.951 &  0.533 &  0.817 &  0.972 &  0.392 &  0.954 &  0.343 &  1.100 &  0.099 &  0.026 &  \\ 
0.85 &  0.9 &  34.660 &  18.203 &  6.961 &  1045.649 &  0.959 &  5.581 &  760.399 &  0.959 &  0.553 &  1.006 &  1.006 &  0.487 &  0.827 &  0.417 &  1.749 &  0.358 &  0.082 &  \\ 
0.85 &  0.95 &  34.501 &  24.142 &  7.599 &  1779.133 &  0.912 &  6.293 &  809.584 &  1.030 &  0.589 &  1.062 &  1.054 &  0.572 &  0.788 &  0.414 &  2.651 &  0.349 &  0.081 &  \\ 
0.85 &  1 &  34.727 &  31.535 &  8.395 &  1450.772 &  1.056 &  6.967 &  720.451 &  1.065 &  0.625 &  1.117 &  1.110 &  0.653 &  0.689 &  0.483 &  2.437 &  0.561 &  0.139 &  \\ 
0.85 &  1.05 &  34.826 &  42.932 &  9.071 &  724.831 &  1.135 &  7.671 &  814.835 &  1.091 &  0.668 &  1.167 &  1.161 &  0.702 &  0.677 &  0.486 &  3.387 &  0.723 &  0.204 &  \\ 
0.9 &  0.9 &  30.310 &  17.425 &  6.352 &  2669.223 &  0.805 &  5.376 &  841.261 &  0.950 &  0.601 &  0.853 &  1.017 &  0.449 &  1.046 &  0.293 &  1.267 &  0.342 &  0.071 &  \\ 
0.9 &  0.95 &  31.222 &  24.741 &  8.271 &  2358.052 &  0.899 &  6.480 &  1102.244 &  1.011 &  0.628 &  1.058 &  1.068 &  0.551 &  0.889 &  0.392 &  1.741 &  0.240 &  0.046 &  \\ 
0.9 &  1 &  31.100 &  32.846 &  8.560 &  1609.054 &  0.991 &  7.291 &  1207.070 &  1.066 &  0.674 &  1.108 &  1.108 &  0.636 &  0.770 &  0.472 &  2.200 &  0.262 &  0.053 &  \\ 
0.9 &  1.05 &  31.184 &  44.074 &  9.554 &  3248.604 &  0.980 &  7.868 &  553.130 &  1.148 &  0.718 &  1.155 &  1.163 &  0.679 &  0.757 &  0.488 &  2.546 &  0.766 &  0.164 &  \\ 
0.95 &  0.95 &  27.414 &  23.920 &  7.315 &  3748.350 &  0.800 &  6.153 &  1069.803 &  0.984 &  0.672 &  0.948 &  1.077 &  0.472 &  1.042 &  0.369 &  1.628 &  0.366 &  0.068 &  \\ 
0.95 &  1 &  27.583 &  34.590 &  9.961 &  2591.980 &  0.970 &  7.262 &  1114.504 &  1.054 &  0.713 &  1.102 &  1.121 &  0.605 &  0.881 &  0.447 &  1.646 &  0.562 &  0.091 &  \\ 
0.95 &  1.05 &  27.714 &  46.075 &  10.191 &  1668.466 &  1.130 &  8.219 &  961.232 &  1.157 &  0.766 &  1.157 &  1.174 &  0.640 &  0.829 &  0.510 &  1.982 &  0.627 &  0.103 &  \\ 
1 &  1 &  24.311 &  31.069 &  9.165 &  1899.998 &  1.059 &  7.012 &  1415.971 &  1.035 &  0.765 &  0.972 &  1.119 &  0.484 &  1.078 &  0.422 &  1.514 &  0.764 &  0.104 &  \\ 
1 &  1.05 &  24.931 &  48.777 &  10.758 &  2942.135 &  1.090 &  8.512 &  2329.601 &  1.070 &  0.819 &  1.160 &  1.180 &  0.600 &  0.961 &  0.470 &  1.780 &  0.850 &  0.119 &  \\ 
1.05 &  1.05 &  21.361 &  46.198 &  10.407 &  3413.441 &  1.106 &  8.620 &  4231.055 &  0.994 &  0.871 &  1.081 &  1.176 &  0.513 &  1.020 &  0.547 &  1.795 &  1.453 &  0.125 &  \\ [0.5ex]
\hline
\\ [-1.5ex]
\multicolumn{20}{c}{{Helium\hspace{0.2cm}--\hspace{0.2cm}Oxygen/Neon}}\\[1.0ex]
\hline
0.2 &  1.1 &  236.693 &  \multicolumn{17}{c}{does not merge after $84 \times P_0$}\\
0.2 &  1.15 &  238.214 &  \multicolumn{17}{c}{does not merge after $80 \times P_0$}\\
0.2 &  1.2 &  235.168 &  \multicolumn{17}{c}{does not merge after $80 \times P_0$}\\
0.25 &  1.1 &  186.210 &  \multicolumn{17}{c}{does not merge after $75 \times P_0$}\\
0.25 &  1.15 &  185.373 &  \multicolumn{17}{c}{does not merge after $75 \times P_0$}\\
0.25 &  1.2 &  185.471 &  \multicolumn{17}{c}{does not merge after $80 \times P_0$}\\
0.3 &  1.1 &  151.295 &  40.873 &  6.449 &  28.069 &  1.103 &  5.525 &  12.290 &  1.110 &  0.261 &  1.100 &  1.138 &  0.907 &  0.238 &  0.142 &  9.562 &  16.247 &  7.654 &  \\ 
0.3 &  1.15 &  150.134 &  57.305 &  6.384 &  117.765 &  1.149 &  5.704 &  17.799 &  1.154 &  0.320 &  1.150 &  1.185 &  0.930 &  0.265 &  0.138 &  9.668 &  19.818 &  8.831 &  \\ 
0.3 &  1.2 &  151.476 &  79.510 &  7.318 &  48.858 &  1.204 &  5.780 &  20.762 &  1.208 &  0.296 &  1.205 &  1.232 &  0.952 &  0.289 &  0.151 &  8.617 &  21.295 &  9.421 &  \\ 
0.35 &  1.1 &  124.643 &  38.839 &  6.798 &  49.035 &  1.100 &  5.630 &  17.097 &  1.107 &  0.282 &  1.099 &  1.144 &  0.900 &  0.256 &  0.206 &  7.158 &  16.222 &  7.540 &  \\ 
0.35 &  1.15 &  125.184 &  50.143 &  7.904 &  73.854 &  1.151 &  5.963 &  980.481 &  1.087 &  0.281 &  1.155 &  1.193 &  0.880 &  0.326 &  0.204 &  6.730 &  22.376 &  10.792 &  \\ 
0.35 &  1.2 &  126.520 &  81.018 &  7.674 &  44.376 &  1.200 &  6.140 &  25.883 &  1.205 &  0.342 &  1.200 &  1.234 &  0.964 &  0.279 &  0.197 &  8.837 &  21.441 &  7.998 &  \\ 
0.4 &  1.1 &  106.581 &  39.735 &  7.046 &  54.698 &  1.098 &  5.815 &  23.236 &  1.108 &  0.301 &  1.099 &  1.145 &  0.905 &  0.249 &  0.265 &  6.815 &  12.563 &  5.051 &  \\ 
0.4 &  1.15 &  107.435 &  53.919 &  7.486 &  60.355 &  1.148 &  5.927 &  22.787 &  1.154 &  0.340 &  1.149 &  1.191 &  0.933 &  0.262 &  0.254 &  8.498 &  16.053 &  6.120 &  \\ 
0.4 &  1.2 &  107.690 &  81.457 &  7.631 &  1043.333 &  1.151 &  6.595 &  1346.704 &  1.132 &  0.376 &  1.198 &  1.234 &  0.994 &  0.244 &  0.257 &  8.702 &  18.754 &  6.514 &  \\ 
0.45 &  1.1 &  91.509 &  37.159 &  7.328 &  72.652 &  1.095 &  5.669 &  120.311 &  1.091 &  0.320 &  1.095 &  1.146 &  0.921 &  0.249 &  0.277 &  8.683 &  16.421 &  5.736 &  \\ 
0.45 &  1.15 &  91.732 &  53.949 &  7.848 &  173.361 &  1.140 &  6.033 &  36.254 &  1.150 &  0.338 &  1.147 &  1.191 &  0.937 &  0.278 &  0.291 &  7.885 &  15.283 &  5.139 &  \\ 
0.45 &  1.2 &  92.617 &  78.080 &  8.501 &  49.249 &  1.196 &  6.572 &  18.771 &  1.205 &  0.395 &  1.196 &  1.235 &  0.970 &  0.280 &  0.303 &  7.640 &  20.070 &  6.293 &  \\ [0.5ex]
\hline
\\ [-1.5ex]
\multicolumn{20}{c}{{Helium/Carbon/Oxygen\hspace{0.2cm}--\hspace{0.2cm}Oxygen/Neon}}\\[1.0ex]
\hline
0.5 &  1.1 &  79.774 &  46.507 &  8.512 &  627.041 &  1.092 &  7.254 &  553.186 &  1.096 &  0.394 &  1.152 &  1.173 &  0.881 &  0.337 &  0.304 &  6.701 &  11.159 &  3.767 &  \\ 
0.5 &  1.15 &  80.598 &  64.642 &  9.044 &  607.305 &  1.147 &  7.964 &  554.041 &  1.150 &  0.431 &  1.185 &  1.219 &  0.947 &  0.310 &  0.320 &  6.055 &  12.617 &  4.445 &  \\ 
0.5 &  1.2 &  81.316 &  96.250 &  10.487 &  876.008 &  1.188 &  8.932 &  469.217 &  1.219 &  0.472 &  1.235 &  1.266 &  0.982 &  0.322 &  0.310 &  7.206 &  13.220 &  4.189 &  \\ 
0.55 &  1.1 &  70.267 &  46.091 &  8.637 &  770.185 &  1.078 &  7.163 &  569.397 &  1.096 &  0.421 &  1.141 &  1.171 &  0.905 &  0.327 &  0.339 &  6.111 &  17.319 &  4.800 &  \\ 
0.55 &  1.15 &  70.343 &  64.596 &  9.583 &  857.534 &  1.141 &  8.175 &  1092.033 &  1.128 &  0.465 &  1.197 &  1.221 &  0.976 &  0.302 &  0.337 &  7.014 &  15.330 &  4.488 &  \\ 
0.55 &  1.2 &  71.280 &  96.559 &  10.874 &  1132.422 &  1.183 &  9.215 &  738.608 &  1.203 &  0.503 &  1.242 &  1.272 &  0.990 &  0.322 &  0.357 &  6.541 &  15.508 &  4.462 &  \\ 
0.6 &  1.1 &  62.111 &  45.826 &  8.548 &  928.747 &  1.075 &  7.539 &  729.351 &  1.080 &  0.445 &  1.160 &  1.176 &  0.914 &  0.319 &  0.382 &  6.138 &  23.325 &  4.832 &  \\ 
0.6 &  1.15 &  62.454 &  63.549 &  9.473 &  1705.963 &  1.095 &  8.310 &  895.597 &  1.135 &  0.486 &  1.196 &  1.226 &  0.939 &  0.346 &  0.377 &  6.511 &  22.944 &  5.373 &  \\ 
0.6 &  1.2 &  62.510 &  95.710 &  10.631 &  855.484 &  1.201 &  9.446 &  1109.502 &  1.184 &  0.538 &  1.231 &  1.274 &  1.016 &  0.316 &  0.381 &  6.772 &  19.251 &  4.645 &  \\ [0.5ex]
\hline
\\ [-1.5ex]
\multicolumn{20}{c}{{Carbon/Oxygen\hspace{0.2cm}--\hspace{0.2cm}Oxygen/Neon}}\\[1.0ex]
\hline
0.65 &  1.1 &  55.281 &  52.503 &  8.736 &  1317.569 &  1.088 &  7.706 &  872.186 &  1.115 &  0.525 &  1.179 &  1.206 &  0.852 &  0.434 &  0.405 &  5.581 &  3.180 &  1.194 &  \\ 
0.65 &  1.15 &  55.270 &  73.902 &  9.801 &  1060.712 &  1.166 &  8.564 &  1518.323 &  1.140 &  0.578 &  1.224 &  1.249 &  0.933 &  0.388 &  0.409 &  6.573 &  3.644 &  1.324 &  \\ 
0.65 &  1.2 &  53.732 &  112.982 &  11.610 &  1602.232 &  1.188 &  10.346 &  1806.667 &  1.194 &  0.635 &  1.251 &  1.306 &  0.975 &  0.398 &  0.417 &  5.627 &  4.258 &  1.384 &  \\ 
0.7 &  1.1 &  49.191 &  54.180 &  9.459 &  1388.710 &  1.102 &  7.784 &  1137.366 &  1.095 &  0.569 &  1.181 &  1.210 &  0.889 &  0.401 &  0.465 &  4.367 &  1.995 &  0.726 &  \\ 
0.7 &  1.15 &  49.330 &  75.394 &  10.036 &  1282.083 &  1.159 &  8.765 &  1195.404 &  1.143 &  0.623 &  1.214 &  1.251 &  0.907 &  0.423 &  0.461 &  5.630 &  2.312 &  0.804 &  \\ 
0.7 &  1.2 &  49.286 &  113.915 &  11.315 &  1217.284 &  1.232 &  10.081 &  1590.375 &  1.208 &  0.680 &  1.264 &  1.304 &  0.973 &  0.407 &  0.454 &  6.196 &  3.652 &  1.172 &  \\ 
0.75 &  1.1 &  44.169 &  55.788 &  9.268 &  319.270 &  1.193 &  7.887 &  1228.021 &  1.095 &  0.613 &  1.188 &  1.209 &  0.829 &  0.478 &  0.502 &  3.978 &  1.534 &  0.531 &  \\ 
0.75 &  1.15 &  44.409 &  77.064 &  10.670 &  969.644 &  1.169 &  8.973 &  1504.169 &  1.143 &  0.662 &  1.244 &  1.263 &  0.922 &  0.435 &  0.489 &  5.177 &  2.053 &  0.698 &  \\ 
0.75 &  1.2 &  44.041 &  116.054 &  12.092 &  3088.380 &  1.166 &  10.234 &  1629.991 &  1.197 &  0.731 &  1.274 &  1.309 &  0.988 &  0.425 &  0.478 &  5.675 &  2.642 &  0.813 &  \\ 
0.8 &  1.1 &  39.297 &  57.230 &  10.230 &  496.682 &  1.235 &  8.185 &  1163.873 &  1.109 &  0.665 &  1.193 &  1.212 &  0.878 &  0.468 &  0.510 &  4.301 &  1.185 &  0.399 &  \\ 
0.8 &  1.15 &  39.287 &  79.925 &  10.473 &  1374.399 &  1.181 &  9.195 &  1237.723 &  1.161 &  0.721 &  1.234 &  1.260 &  0.896 &  0.496 &  0.507 &  4.974 &  1.764 &  0.545 &  \\ 
0.8 &  1.2 &  39.492 &  119.181 &  12.826 &  3798.235 &  1.153 &  10.314 &  2009.456 &  1.188 &  0.790 &  1.282 &  1.308 &  0.962 &  0.440 &  0.534 &  6.097 &  2.054 &  0.609 &  \\ 
0.85 &  1.1 &  35.004 &  58.494 &  9.878 &  470.637 &  1.222 &  8.327 &  1098.367 &  1.127 &  0.715 &  1.199 &  1.216 &  0.782 &  0.613 &  0.517 &  3.720 &  1.100 &  0.319 &  \\ 
0.85 &  1.15 &  35.235 &  83.566 &  11.350 &  1692.525 &  1.156 &  9.368 &  1553.895 &  1.163 &  0.776 &  1.259 &  1.269 &  0.843 &  0.554 &  0.559 &  4.236 &  1.786 &  0.507 &  \\ 
0.85 &  1.2 &  35.324 &  124.731 &  12.091 &  3358.091 &  1.166 &  10.653 &  2287.650 &  1.195 &  0.848 &  1.282 &  1.318 &  0.936 &  0.490 &  0.569 &  5.210 &  2.123 &  0.589 &  \\ 
0.9 &  1.1 &  31.372 &  61.506 &  10.939 &  773.097 &  1.201 &  8.938 &  1361.269 &  1.125 &  0.771 &  1.220 &  1.220 &  0.756 &  0.717 &  0.493 &  3.280 &  0.987 &  0.246 &  \\ 
0.9 &  1.15 &  31.134 &  86.424 &  11.210 &  885.785 &  1.252 &  9.820 &  1864.311 &  1.154 &  0.838 &  1.270 &  1.270 &  0.836 &  0.617 &  0.560 &  3.535 &  1.818 &  0.381 &  \\ 
0.9 &  1.2 &  31.579 &  126.927 &  12.613 &  3144.213 &  1.217 &  10.945 &  2054.908 &  1.209 &  0.916 &  1.302 &  1.323 &  0.846 &  0.612 &  0.599 &  4.085 &  1.731 &  0.417 &  \\ 
0.95 &  1.1 &  27.877 &  63.855 &  11.437 &  2962.060 &  1.103 &  9.254 &  1144.728 &  1.163 &  0.829 &  1.226 &  1.226 &  0.713 &  0.734 &  0.569 &  3.206 &  0.920 &  0.167 &  \\ 
0.95 &  1.15 &  31.135 &  88.035 &  11.775 &  3054.531 &  1.155 &  9.950 &  1165.923 &  1.212 &  0.903 &  1.262 &  1.276 &  0.781 &  0.724 &  0.565 &  2.911 &  1.221 &  0.238 &  \\ 
0.95 &  1.2 &  28.166 &  131.369 &  13.791 &  854.627 &  1.294 &  11.285 &  2306.283 &  1.218 &  0.979 &  1.312 &  1.324 &  0.872 &  0.615 &  0.617 &  4.394 &  2.626 &  0.490 &  \\ 
1 &  1.1 &  24.681 &  66.239 &  11.959 &  6072.826 &  0.967 &  9.691 &  1512.814 &  1.165 &  0.882 &  1.210 &  1.227 &  0.672 &  0.851 &  0.552 &  2.350 &  1.002 &  0.136 &  \\ 
1 &  1.15 &  24.957 &  91.897 &  12.753 &  1658.527 &  1.200 &  10.396 &  913.077 &  1.256 &  0.962 &  1.248 &  1.280 &  0.832 &  0.657 &  0.634 &  2.488 &  1.711 &  0.225 &  \\ 
1 &  1.2 &  24.999 &  133.209 &  13.322 &  773.319 &  1.332 &  11.525 &  916.374 &  1.304 &  1.047 &  1.325 &  1.318 &  0.840 &  0.632 &  0.684 &  3.943 &  4.269 &  0.452 &  \\ 
1.05 &  1.1 &  21.875 &  70.602 &  12.960 &  2830.360 &  1.152 &  10.302 &  2751.701 &  1.152 &  0.943 &  1.231 &  1.241 &  0.671 &  0.888 &  0.571 &  1.748 &  2.309 &  0.199 &  \\ 
1.05 &  1.15 &  21.990 &  98.120 &  13.257 &  2803.448 &  1.193 &  11.314 &  1549.808 &  1.248 &  1.022 &  1.238 &  1.290 &  0.749 &  0.811 &  0.617 &  2.033 &  3.290 &  0.281 &  \\ 
1.05 &  1.2 &  22.009 &  130.896 &  23.068 &  2403.125 &  1.317 &  12.535 &  803.453 &  1.317 &  1.063 &  1.279 &  1.325 &  0.804 &  0.729 &  0.656 &  5.116 &  8.890 &  0.786 &  \\ 
\end{longtable}
\label{lastpage}
\end{center}
\normalsize
\end{landscape}

\end{document}